\RequirePackage{fix-cm}
\documentclass[smallextended]{svjour3wide} 
\smartqed  
\usepackage[numbers,sort&compress]{natbib}
\usepackage{graphicx}
\usepackage{scalefnt}
\usepackage{amsmath}
\usepackage{amssymb}
\usepackage{mathrsfs}
\usepackage{color}
\usepackage{bm}
\usepackage{ulem}
\usepackage{comment}

\journalname{Journal of Statistical Physics}
\begin{document}

\title{Statistical mechanical expressions of slip length}

\author{Hiroyoshi Nakano \and Shin-ichi Sasa}

\institute{
H. Nakano \and S. Sasa \at
Department of Physics, Kyoto University, Kyoto 606-8502, Japan \\
\email{h.takagi@scphys.kyoto-u.ac.jp, sasa@scphys.kyoto-u.ac.jp}
}

\date{Received: date / Accepted: date}

\maketitle

\begin{abstract}
We provide general derivations of the partial slip boundary condition from microscopic dynamics and linearized fluctuating hydrodynamics. The derivations are based on the assumption of separation of scales between microscopic behavior, such as collision of particles, and macroscopic behavior, such as relaxation of fluid to global equilibrium. The derivations lead to several statistical mechanical expressions of the slip length, which are classified into two types. The expression in the first type is given as a local transport coefficient, which is related to the linear response theory that describes the relaxation process of the fluid. The second type is related to the linear response theory that describes the non-equilibrium steady state and the slip length is given as combination of global transport coefficients, which are dependent on macroscopic lengths such as a system size. Our derivations clarify that the separation of scales must be seriously considered in order to distinguish the expressions belonging to two types. Based on these linear response theories, we organize the relationship among the statistical mechanical expressions of the slip length suggested in previous studies.

\end{abstract}

\setcounter{secnumdepth}{3}
\setcounter{tocdepth}{3}
\tableofcontents

\section{Introduction} \label{sec:intro}
\subsection{Motivation of the paper}
Over the past two decades, interest in applying hydrodynamics to nano- and micro- scale system have increased, because of the remarkable developments in experimental techniques for small-scale systems~\cite{neto2005boundary,lauga2007microfluidics,cao2009molecular,bocquet2010nanofluidics}. From a large number of studies on this topic, it is confirmed that certain specific fluids confined in such systems slip on the solid surface but behave accurately as Newtonian liquids in the bulk region~\cite{pit2000direct,zhu2001rate,zhu2002limits,cottin2005boundary,maali2008measurement,vinogradova2009direct,thompson1997general,gupta1997shear,barrat1999large,cieplak2001boundary}. This result indicated the breakdown of macroscopic hydrodynamics that the behavior of fluids are described by the solutions of the Navier--Stokes equation with stick boundary condition. Here, the stick boundary condition means that a fluid has no relative velocity when in contact with a solid surface~\cite{landau1959course}. Determining the correct boundary conditions to replace the stick boundary condition, is one of the important challenges in hydrodynamics.

Slipping at the solid surface is often characterized by the partial slip boundary condition~\cite{vinogradova1995drainage}. Specifically, the slip velocity $v_s$ of the fluid at the wall is taken to be linearly proportional to the shear rate $\dot{\gamma}$ at the wall as~\cite{navier1823,lamb1993hydrodynamics,happel2012low}
\begin{eqnarray}
v_s = b \dot{\gamma},
\end{eqnarray}
where the slip length $b$ represents the distance beyond the surface of the wall at which the fluid velocity extrapolates to zero. Much effort has been spent in investigating factors that affect the slip length such as surface roughness~\cite{cottin2003low,priezjev2006influence,lee2014interfacial} and wettability~\cite{barrat1999influence,huang2008water,voronov2008review}.

In this paper, we shall try to describe the partial slip boundary condition by underlying microscopic theories and linearized fluctuating hydrodynamics. The derivation of hydrodynamics from the microscopic theories is one of the main problem in nonequilibrium statistical mechanics~\cite{kirkwood1946statistical,kirkwood1949statistical,irving1950statistical,green1954markoff,mori1958statistical,kawasaki1973theory,zubarev1996statistical,sasa2014derivation,hongo2018nonrelativistic}. Its principle objectives consist of two aspects. The first one is to determine the macroscopic equations and investigate the limits of their validity. The second one is to provide microscopic descriptions of the transport coefficients and investigate the relationship between the transport coefficients and the microscopic parameters. Regarding the partial slip boundary condition, several research groups attempted to derive its microscopic description and proposed some different statistical mechanical expressions of the slip length~\cite{bocquet1994hydrodynamic,fuchs2002statistical,petravic2007equilibrium,kobryn2008molecular,hansen2011prediction,huang2014green,ramos2016hydrodynamic,nakamura2014perturbation,nakamura2018stick}.  However, these expressions are all different in form and no relationships among each other are known. To make matters worse, some of them appear to be inconsistent with each other~\cite{petravic2007equilibrium,bocquet2013green,ramos2016hydrodynamic}. Then, this proliferation of the expressions remains a controversial point. Therefore, the goal of this paper are: i) to provide the complete derivation of the partial slip boundary condition; and ii) to organize the expressions of the slip length found in the literatures based on the unified theory.

\subsubsection{confusion between two linear response theories: local vs global}
\label{sec:Confusion between two linear response theories: local vs non-local}
\begin{figure}[htbp]
\centering
\parbox{5cm}{
\includegraphics[width=1.0\linewidth, bb=0 0 505 462]{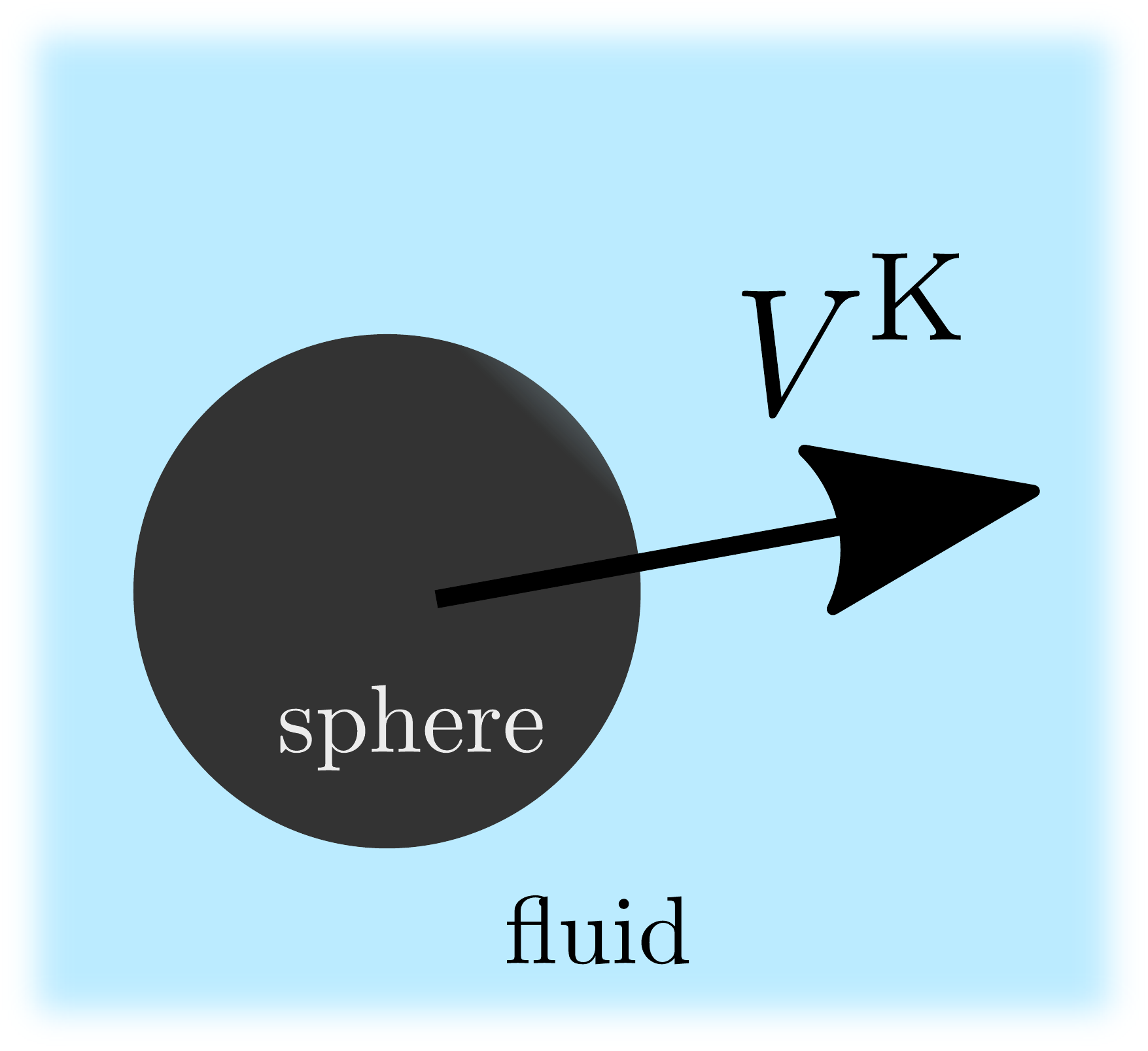}
\caption{model using by Kirkwood}
\label{fig:intro1}}
\hspace{1cm}
\begin{minipage}{5cm}
\includegraphics[width=1.0\linewidth, bb=0 0 365 376]{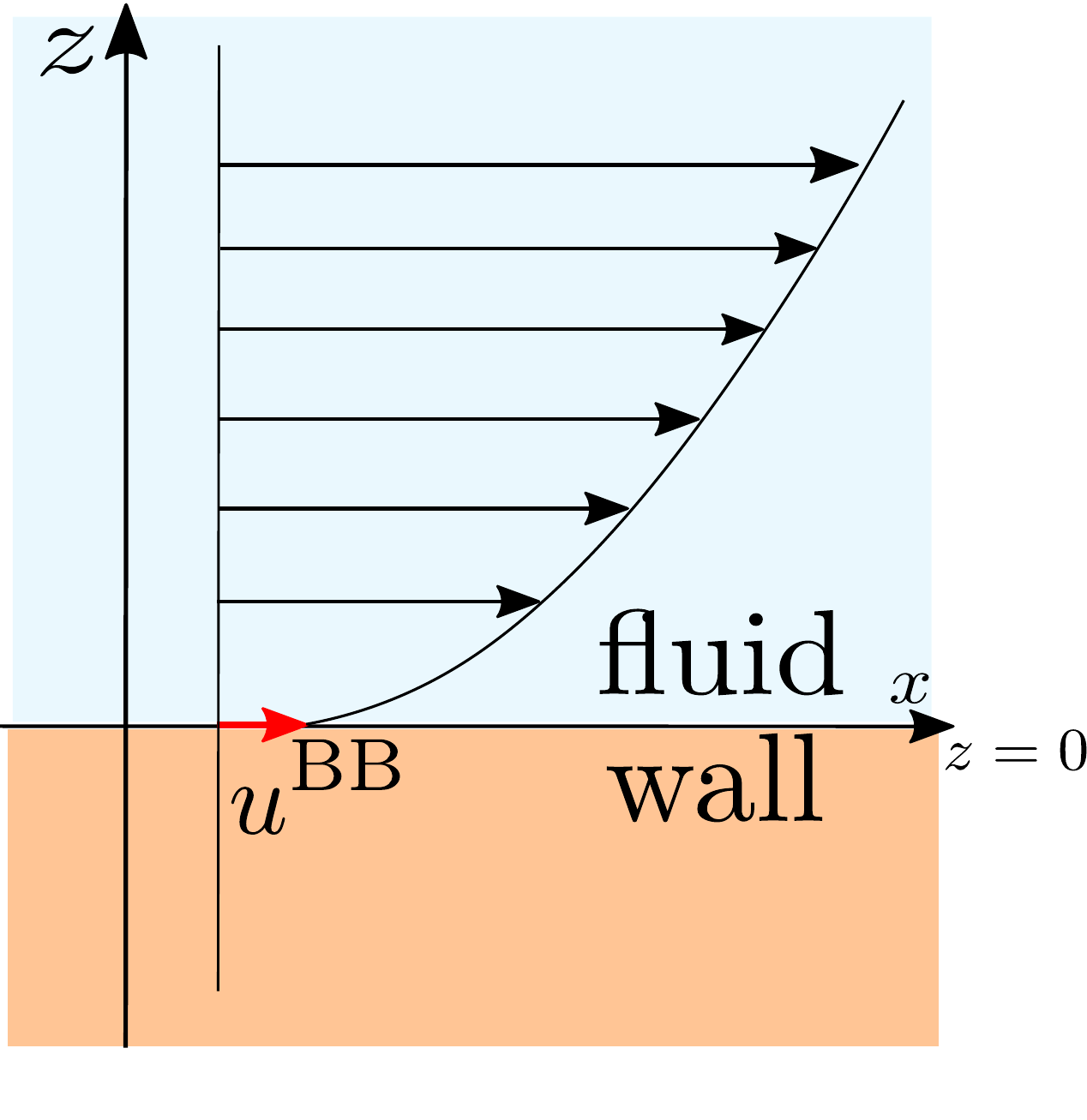}
\caption{model using by Bocquet and Barrat}
\label{fig:intro2}
\end{minipage}
\end{figure}
There is a confusing problem that occurred in the proliferation of the statistical mechanical expressions for the slip length, which was emphasized by Petravic and Harrowell~\cite{petravic2007equilibrium}. Specifically, some expressions with the identical form provide the different value. We below give a detailed explanation on this problem. Solving this problem is our main motivation to explore the complete derivation of the partial slip boundary condition.

The linear response theories concerning the boundary condition may be divided into two groups. One is related to global transport coefficients and another local transport coefficients. First, we give the example of the global linear response theories. We consider the motion of a macroscopic sphere through a fluid (Fig.~\ref{fig:intro1}). When the velocity of the sphere, $\bm{V}^{K}$, is sufficiently small, the total force acting on the sphere, $\bm{F}^K$, is expressed as
\begin{eqnarray}
\bm{F}^{K} = - \gamma_{\rm K} \bm{V}^{\rm K},
\end{eqnarray}
where $\gamma_{\rm K}$ is the friction coefficient of the sphere. From microscopic mechanics, Kirkwood derived a microscopic expression for the friction coefficient, $\gamma_{\rm K}$, in the form~\cite{kirkwood1946statistical}
\begin{eqnarray}
\gamma_{\rm K} = \frac{1}{3k_{\rm B} T}\int_0^{\infty} dt \langle  \hat{\bm{F}}^{\rm K}_t \cdot \hat{\bm{F}}^{\rm K}_0 \rangle_{\rm eq},
\label{eq:Kirkwood's formula}
\end{eqnarray}
where $k_{\rm B}$ denotes the Boltzmann constant, $T$ the temperature of the fluid, $\langle \cdot \rangle_{\rm eq}$ a canonical ensemble average at temperature $T$, and $\hat{\bm{F}}^{\rm K}_t$ the total microscopic force exerted on the sphere by the fluid at time $t$.

When the fluid satisfies the Navier--Stokes equation, Kirkwood's formula (\ref{eq:Kirkwood's formula}) involves the boundary condition. For example, the friction coefficient for a viscous fluid at low Reynolds number is given by
\begin{eqnarray}
\gamma_{\rm K} = \mathcal{C} \eta R,
\label{eq:Stokes' law}
\end{eqnarray}
where $\eta$ is the viscosity of the fluid, and $R$ is the radius of the sphere. The numerical coefficient $\mathcal{C}$ is equal to $4\pi$ for the perfect slip boundary condition and $6\pi$ for the stick boundary condition. Here, the perfect slip boundary condition means that the velocity of the fluid normal to the sphere at the boundary is equal of that of the sphere in this direction and the shear stresses on the sphere are equal to zero. This expression (\ref{eq:Stokes' law}) is directly obtained from (\ref{eq:Kirkwood's formula}) by employing the fluctuating hydrodynamics~\cite{zwanzig1964hydrodynamic,bedeaux1974brownian,itami2015derivation}. By recalling that $\gamma_{K}$ is the friction coefficient defined from the difference between the velocity of the sphere and the fluid at infinity, we notice that Kirkwood's formula (\ref{eq:Kirkwood's formula}) may contain the global information of system. Recently, Itami and Sasa reported such global nature of Kirkwood's formula for some configurations~\cite{itami2015derivation,itami2018singular}.

Second, we explain the local linear response theory for the partial slip boundary condition, which was given by Bocquet and Barrat~\cite{bocquet1994hydrodynamic}. They considered a fluid in a semi-infinite container with a stationary plane wall (Fig.~\ref{fig:intro2}) and focused on the friction force between the fluid and wall, defined by
\begin{eqnarray}
F^{\rm BB} = - \gamma_{\rm BB} u^{\rm BB},
\label{eq:gammaBB definition}
\end{eqnarray}
where $F^{\rm BB}$ is the force acting on the wall and $u^{\rm BB}$ is the difference between the velocity of the wall and the fluid on the wall. Then, they proposed the microscopic expression for the friction coefficient $\gamma_{\rm BB}$ in the form~\cite{bocquet1994hydrodynamic}
\begin{eqnarray}
\gamma_{\rm BB} = \frac{1}{S k_{\rm B} T}\int_0^{\infty} dt \langle  \hat{F}^{\rm BB}_t  \hat{F}^{\rm BB}_0 \rangle_{\rm eq},
\label{eq:gammaBB}
\end{eqnarray}
where $S$ denotes the surface area and $\hat{F}^{\rm BB}_t$ the total force between the wall and fluid at time $t$. Furthermore, they proposed that the slip length is related to $\gamma_{\rm BB}$ as
\begin{eqnarray}
b = \frac{\eta}{\gamma_{\rm BB}}.
\label{eq:slip length and gammaBB}
\end{eqnarray}

The expression (\ref{eq:gammaBB}) provides a useful tool to extract the relationship between the slip length and the microscopic properties of the fluid and wall~\cite{barrat1999influence,priezjev2004molecular,huang2008water,falk2010molecular}. For example, a quasi-universal relationship between the slip length and the static contact angle was derived from this expression~\cite{barrat1999influence,huang2008water}.
We note that (\ref{eq:gammaBB}) contains only the local information of system because it is defined by (\ref{eq:gammaBB definition}). Actually, some theoretical works proposed that the slip length is the local quantity as well as the viscosity of fluid~\cite{priezjev2004molecular,priezjev2010relationship}.

We notice that Kirkwood's formula (\ref{eq:Kirkwood's formula}) and the formula of Bocquet and Barrat (\ref{eq:gammaBB}) (referred to as BB's formula throughout this paper) are same in form. However, they contain the information of the boundary condition in different forms. Specifically, one contains the global effects and another contains only the local effects. Therefore, the problems to be considered are ``What is the difference between these expressions?" and ``How are these expressions distinguished?".

This paper provides the solution of this problem. Our strategy is to introduce the assumption of separation of scales in the explicit form and to carefully employ this assumption at each step of the derivation of the partial slip boundary condition. Indeed, it is accepted that the separation of scales plays the central role in the derivation of hydrodynamics since the pioneer work by Kirkwood~\cite{kirkwood1946statistical}.

As a related study, Bocquet and Barrat suggested a tentative prescription to solve this problem~\cite{bocquet2013green}. They proposed that in order to obtain the correct slip length from (\ref{eq:gammaBB}), the large system size limit $L\to \infty$ should be taken before performing the time integration to infinity:
\begin{eqnarray}
\gamma_{\rm BB} = \lim_{\tau \to \infty} \lim_{L \to \infty} \frac{1}{S k_{\rm B} T}\int_0^{\tau} dt \langle  \hat{F}^{\rm BB}_t \cdot \hat{F}^{\rm BB}_0 \rangle_{\rm eq}.
\label{eq:gammaBB:a}
\end{eqnarray}
In this paper, we demonstrate that their prescription does not provide the final answer to clear up the confusion between Kirkwood's formula (\ref{eq:Kirkwood's formula}) and BB's formula (\ref{eq:gammaBB}), and the separation of scales must be seriously considered when we discuss the microscopic description of the partial slip boundary condition.

\subsection{A quick look of the paper}
\label{sec:A quick look of the paper}
\begin{figure}
\centering
\includegraphics[width=8cm, bb=0 0 363 219]{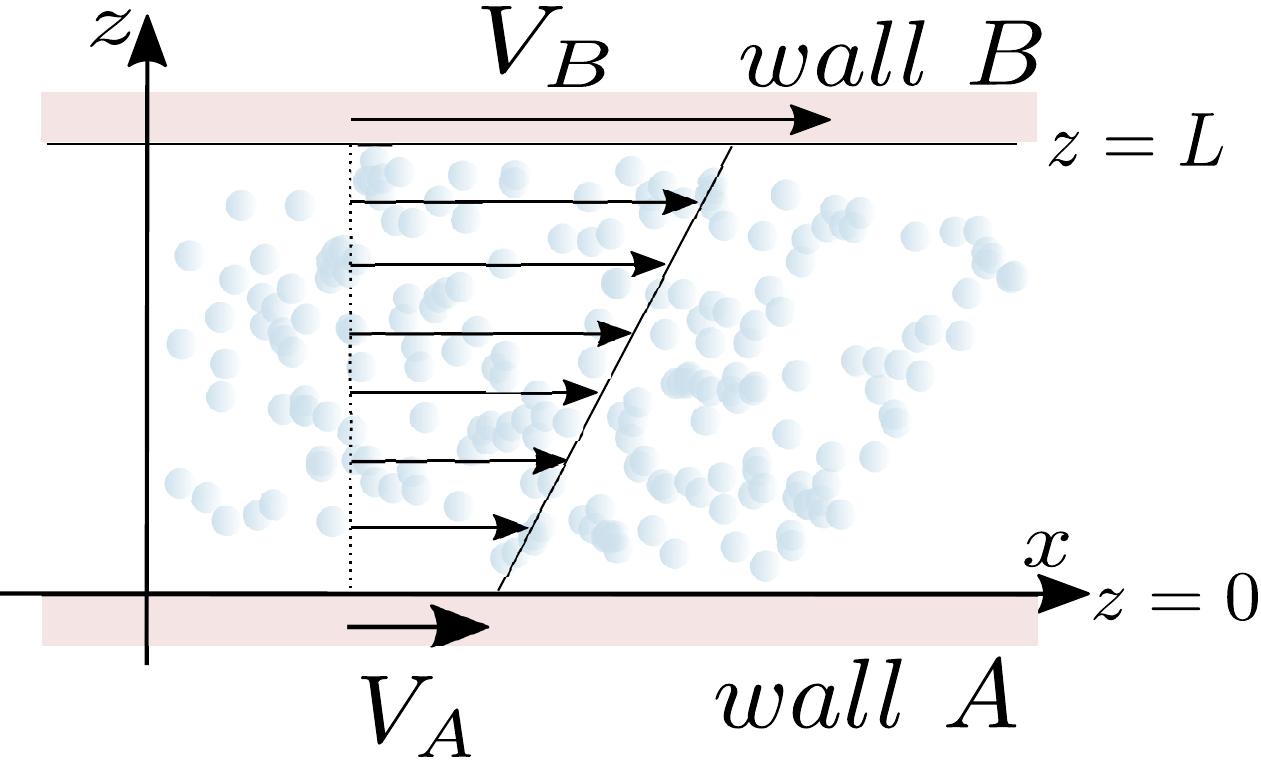}
\caption{Schematic illustration of our model. }
\label{fig:model all}
\end{figure}
To give readers an overview of this long paper, we now briefly summarize our results.

Throughout this paper, we consider the fluid enclosed between two parallel walls separated by a fixed distance. We refer to the lower and upper walls as walls $A$ and $B$, respectively. By moving the walls parallel to each other, uniform shear flow is induced (See Fig.~\ref{fig:model all}). The walls are placed parallel to $xy$-plane and $x$-axis is chosen as the direction in which the walls move. The velocities of walls $A$ and $B$ are represented by $V_A$ and $V_B$, respectively. Then, we focus on the boundary conditions that are imposed on the velocity field at walls $A$ and $B$. This paper is divided into three parts:
\begin{enumerate}
\item Derivation of the partial slip boundary condition from the microscopic particle system (Sects. \ref{sec:Model: Hamiltonian Particle System}, \ref{sec:generalized Green-Kubo formula}, \ref{sec:Separation of length and time scales}, \ref{sec:Linear response theory for the relaxation process of fluid}, \ref{sec:Linear response theory for the  steady state})
\item Verification of our microscopic deviation (and alternative derivation) from the fluctuating hydrodynamics (Sects. \ref{sec:model: linearized fluctuating hydrodynamics}, \ref{sec:Explicit Form of CAA}, \ref{sec:Behaviors of CAA for limiting cases})
\item Relationship with the previous studies and discussion (Sects. \ref{sec:Equilibrium measurement of slip length}, \ref{sec:Concluding remarks})
\end{enumerate}

Throughout this paper, we clarify it inevitable to seriously consider the separation of scales when we discuss the microscopic expressions of the slip length.

\subsubsection{microscopic particle system}
The first part is devoted to the derivation of the partial slip boundary condition from the microscopic particle system. The fluid consists of $N$ particles and walls $A$ and $B$ are represented as collections of material points placed in planes $z=0$ and $z=L$, respectively (Sect.~\ref{sec:Model: Hamiltonian Particle System}). Such walls are often called ``atomically smooth surface". The collection of positions and momenta of all particles is denoted by $\Gamma=(\bm{r}_1,\bm{p}_1,\cdots,\bm{r}_N,\bm{p}_N)$. Similarly, the collections of positions of the material points that comprise walls $\alpha$ are denoted by $\Gamma^{\alpha}=(\bm{q}^{\alpha}_1,\cdots,\bm{q}^{\alpha}_{N_{\alpha}})$, where $N_{\alpha}$ is the number of the material points of wall $\alpha$. The key idea of our derivation is to explicitly employ the assumption of the separation of time and length scales. 

In our system, we may consider two characteristic time scales; one is a time scale of the microscopic motion of molecules, denoted by $\tau_{\rm micro}$, and the other is a time scale of the global equilibrium of fluid, denoted by $\tau_{\rm macro}$. We assume the separation of scales, which is represented by $\tau_{\rm micro} \ll \tau_{\rm macro}$.

More precisely, we may express the separation of scales in terms of the behaviors of time correlation functions (Sect.~\ref{sec:Separation of length and time scales}). For example, we focus on the microscopic momentum density field $\hat{\bm{\pi}}(\bm{r};\Gamma)$. Because this quantity corresponds to a slow variable to describe the macroscopic motion of fluid, the autocorrelation function is expected to behave as
\begin{eqnarray}
\langle \pi^x(\bm{r},\tau) \pi^x(\bm{r},0) \rangle_{\rm eq}
\begin{cases}
>0, \ \ {\rm for} & 0<\tau\simeq \tau_{\rm macro},\\
\simeq0,  \ \ {\rm for} & \tau_{\rm macro} \ll \tau,
\end{cases}
\end{eqnarray}
where $\langle \cdot \rangle_{\rm eq}$ is a canonical ensemble average at temperature $T$.
We next consider the microscopic total force acting on the fluid from wall $A$, $\hat{F}_A(\Gamma;\Gamma^A)$. Because this quantity is not the slow variable, the autocorrelation function decays to $0$ at the microscopic time scale $\tau_{\rm micro}$. However, at the macroscopic time scale $\tau_{\rm marco}$, this quantity is coupled with the macroscopic motion of the fluid. Then, the autocorrelation function becomes non-zero at $\tau_{\rm macro}$, and after a longer time than $\tau_{\rm macro}$, the autocorrelation function completely relaxes. In summary, we conjecture that the autocorrelation function behaves as
\begin{eqnarray}
\langle  F^x_A(\tau) F^x_A(0)\rangle_{\rm eq}
\begin{cases}
>0, \ \ {\rm for} & 0<\tau\simeq \tau_{\rm micro},\\
\simeq0,  \ \ {\rm for} & \tau_{\rm micro}\ll \tau \ll \tau_{\rm macro}, \\
\neq 0,  \ \ {\rm for} & \tau \simeq \tau_{\rm macro}, \\
\simeq0,  \ \ {\rm for} & \tau_{\rm macro} \ll \tau.
\end{cases}
\label{eq:behavior of FAFA}
\end{eqnarray}
We introduce the lower and upper cut-off times $\tau_{\rm meso}$ and $\tau_{\rm ss}$ for the macroscopic motion of fluid. Because $\tau_{\rm macro}$ is the characteristic time related to the macroscopic behavior, $\tau_{\rm meso}$ and $\tau_{\rm ss}$ satisfy the relation
\begin{eqnarray}
\tau_{\rm micro} \ll \tau_{\rm meso} \ll \tau_{\rm macro} \ll \tau_{\rm ss}.
\end{eqnarray}
Based on these behaviors, we consider the following quantity:
\begin{eqnarray}
\gamma_{AA}(\tau) \equiv \frac{1}{k_B T A_{xy}} \int_0^{\tau} dt \langle F^{x}_A(t)F^{x}_A(0) \rangle_{\rm eq},
\label{eq:def gammaAA}
\end{eqnarray}
where $A_{xy}$ is the are of the wall $A$. This quantity resembles to Kirkwood formula or BB's formula, but time integration is performed up to finite time $\tau$.

Combining the fluctuation theorem (Sect.~\ref{sec:generalized Green-Kubo formula}) and assumption of the separation of scales, we can construct two linear response theories; One is related to the lower cut-off scale ($\tau_{\rm meso}$-scale) and another the upper cut-off scale ($\tau_{\rm ss}$-scale). Suppose that the fluid has the velocity field $\bm{u}(\bm{r})=(u^x(z),0,0)$ at time $\tau=0$. We focus on the relaxation process of the fluid. The first linear response theory describes the macroscopic behavior of fluid at the time $\tau=\tau_{\rm meso}$ (Sect.~\ref{sec:Linear response theory for the relaxation process of fluid}). For example, the force acting on wall $A$ at $\tau=\tau_{\rm meso}$, $\langle F^x_A(\tau_{\rm meso}) \rangle$, is expressed as
\begin{eqnarray}
\langle F^x_A(\tau_{\rm meso}) \rangle \simeq -\gamma_{AA} (\tau_{\rm meso}) (u^x(z_A^{\rm 1st})-V_A) ,
\label{eq:linear response theory: Bocquet and Barrat type}
\end{eqnarray}
where $z_A^{\rm 1st}$ is the position of the first epitaxial layer near wall $A$ (see Sect.~\ref{sec:Linear response theory for the relaxation process of fluid} for detail explanation).
This means that the force acting on wall $A$ at $\tau=\tau_{\rm meso}$ is expanded in the difference between the wall velocity and the fluid velocity near wall $A$. The second linear response theory describes the macroscopic behavior of fluid at $\tau= \tau_{\rm ss}$ (Sect.~\ref{sec:Linear response theory for the  steady state}). For example, the force acting on wall $A$ at $\tau=\tau_{\rm ss}$, $\langle F^x_A(\tau_{\rm ss}) \rangle$, is expressed as
\begin{eqnarray}
\langle F^x_A(\tau_{\rm ss}) \rangle \simeq -\gamma_{AA} (\tau_{\rm ss}) (V_B-V_A).
\label{eq:linear response theory: Kirkwood type}
\end{eqnarray}
This means that the force acting on wall $A$ at $\tau=\tau_{\rm ss}$ is expanded in the difference between the wall velocities $V_A$ and $V_B$.

We note that (\ref{eq:linear response theory: Bocquet and Barrat type}) is the local linear response relation and corresponds to the theory given by Bocquet and Barrat~\cite{bocquet1994hydrodynamic}, while (\ref{eq:linear response theory: Kirkwood type}) is global linear response relation and corresponds to the framework of Kirkwood's formula~\cite{kirkwood1946statistical}. By employing these linear response theories, we can derive the partial slip boundary condition and obtain some statistical mechanical expressions of the slip length in the literatures.

\subsubsection{linearized fluctuating hydrodynamics}
In the second part, we exactly calculate the behavior of the time correlation function from $\tau_{\rm meso}$ to $\tau_{\rm ss}$, and discuss importance of the separation of scales and the validity of the expressions of the slip length obtained in the first part. For this purpose, we employ the linearized fluctuating hydrodynamics.

Suppose that the fluid is enclosed between two parallel walls (wall $A$ and $B$), separated by $\mathcal{L}$. We assume that in the bulk the motion of fluid is described by the linearized fluctuating hydrodynamics and on the wall $\alpha=A, B$ by the partial slip boundary condition with the slip length $\mathcal{B}_{\alpha}$ (Sect.~\ref{sec:model: linearized fluctuating hydrodynamics}). Then, we focus on the autocorrelation function, $\langle \mathcal{F}_A(t)\mathcal{F}_A(0)\rangle$, of the force acting on the wall $A$, $\mathcal{F}_A$, in equilibrium.

We can exactly calculate the autocorrelation function $\langle \mathcal{F}_A(t)\mathcal{F}_A(0)\rangle$ owing to the linearity of this problem (Sect.~\ref{sec:Explicit Form of CAA}). The result is given by (\ref{eq:CAA expression: perfect form}). Then, if the linearized fluctuating hydrodynamics accurately describes the fluid motion from $\tau_{\rm meso}$ to $\tau_{\rm ss}$, a part of the assumption (\ref{eq:behavior of FAFA}) imposed in the first part may be replaced with more explicit form (\ref{eq:CAA expression: perfect form}).
 
From the exact expression, we demonstrate that the linearized fluctuating hydrodynamics is consistent with the two linear response theories derived in the first part (Sect.~\ref{sec:Behaviors of CAA for limiting cases}). Specifically, the two relations, (\ref{eq:linear response theory: Bocquet and Barrat type}) and (\ref{eq:linear response theory: Kirkwood type}), are obtained from the linearized fluctuating hydrodynamics. This means that the framework of the linearized fluctuating hydrodynamics provides the alternative derivation of the statistical mechanical expressions for the slip length derived from the underlying microscopic dynamics. We note that these rederivations from the fluctuating hydrodynamics are valid beyond the atomically smooth surfaces in contrast with the derivation from the underlying microscopic theories.

The results mentioned above holds only for finite-size systems. For infinite-size systems with $\mathcal{L} \to \infty$, by carefully dealing with the order of limits, we obtain
\begin{eqnarray}
\lim_{t \to \infty} \lim_{\mathcal{L} \to \infty} \int_0^{t} ds \langle \mathcal{F}_A(s)\mathcal{F}_A(0)\rangle = 0.
\label{eq:slip length plateau no extending: fluctuating hydrodynamics}
\end{eqnarray}
We notice that this equation denies the tentative procedure of Bocquet and Barrat explained in Sect.~\ref{sec:Confusion between two linear response theories: local vs non-local}.
Instead, our calculation proposes that in order to obtain the correct slip length, the integration range of BB's formula must be given by $[0,\tau_{\rm meso}]$ even for the infinite system.

Thus, we confirm that introducing the separation of scales is inevitable when we discuss the statistical mechanical expressions of the slip length.

\subsubsection{relationship with the previous studies}
\label{sec:relationship with the previous studies}
As explained in Sect.~\ref{sec:Confusion between two linear response theories: local vs non-local}, there are several different expressions of the slip length previously suggested. We organize these expressions in terms of two linear response theories belonging to $\tau_{\rm meso}$- and $\tau_{\rm ss}$- scales respectively (Sect.~\ref{sec:Equilibrium measurement of slip length}).

We focus on four different expressions, which were first proposed by Bocquet and Barrat~\cite{bocquet1994hydrodynamic}, Petravic and Harrowell~\cite{petravic2007equilibrium}, Hansen et al.~\cite{hansen2011prediction}, and Huang and Szlufarska~\cite{huang2014green}. In this paper, all of them are derived by employing the fluctuating hydrodynamics, while only a part are derived from the microscopic theory, as shown in Table~\ref{tab:a}. We stress that BB's formula is related to different time scale from otherwise and in order to avoid the confusion between these expressions, existence of two different time scales must be taken into account.

\begin{table}[htbp]
\begin{center}
\begin{tabular}{c|c|c|c} \hline
 & time scale & microscopic theory & fluctuating hydrodynamics \\ \hline \hline
Bocquet and Barrat~\cite{bocquet1994hydrodynamic} & $\tau_{\rm meso}$& Sect.~\ref{sec: expression of slip length: relaxation process} (Eq.~(\ref{eq:slip length: meso})) & Sect.~\ref{sec:Green--Kubo formulae of the slip length} (Eq.~(\ref{eq:160})) \\ \hline
Petravic and Harrowell~\cite{petravic2007equilibrium} & $\tau_{\rm ss}$ & Sect.~\ref{sec:expression of the slip length steady state}  (Eq.~(\ref{eq:slip length steady state modified A})) & Sect.~\ref{sec:Green--Kubo formulae of the slip length} (Eq.~(\ref{eq:bAast: sect6})) \\ \hline
Hansen et al.~\cite{hansen2011prediction} & $\tau_{\rm ss}$ & none & Appendix~\ref{sec:Derivation of bHTD from the linearized fluctuating dynamics} \\ \hline
Huang and Szlufarska~\cite{huang2014green} & $\tau_{\rm ss}$ & Sect.~\ref{sec:expression of the slip length steady state}  (Eq.~(\ref{eq:slip length: steady state expression A})?) & Sect.~\ref{sec:Green--Kubo formulae of the slip length} (Eq.~(\ref{eq:bAast: sect6})) \\ \hline \hline
\end{tabular}
\\
\caption{Comparison with the literatures. Our theories given in this paper cover all four expressions.}
\label{tab:a}
\end{center}
\end{table}

The remainder of this paper is organized as follows. In Sect.~\ref{sec:Model: Hamiltonian Particle System}, the microscopic setup of our model is introduced. In Sect.~\ref{sec:generalized Green-Kubo formula}, we derive the exact formula to describe the time evolution of any ensemble-averaged quantities. This formula is the starting point for developing linear response theories. In Sect.~\ref{sec:Separation of length and time scales}, we introduce the separation of time scales and explain the behavior of correlation functions of the momentum flux. In Sects.~\ref{sec:Linear response theory for the relaxation process of fluid} and \ref{sec:Linear response theory for the  steady state}, we develop the linear response theories describing the fluid motions at $\tau_{\rm meso}$- and $\tau_{\rm ss}$- scales. Then, on the basis of these linear response theories, we derive the partial slip boundary condition and obtain two statistical mechanical expressions of the slip length. In Sect.~\ref{sec:model: linearized fluctuating hydrodynamics}, we introduce the linearized fluctuating hydrodynamics with the partial slip boundary condition. In Sect.~\ref{sec:Explicit Form of CAA}, we calculate the exact form of the autocorrelation function of the force acting on the wall. In Sect.~\ref{sec:Behaviors of CAA for limiting cases}, we obtain more simple forms of the force autocorrelation function for some limiting cases and discuss the relationship with the Green--Kubo formulae obtained from the underlying microscopic theory. In Sect.~\ref{sec:Equilibrium measurement of slip length}, based on our theories, we organize the statistical mechanical expressions of the slip length previously suggested. The final section is devoted to a brief summary and concluding remarks.

Throughout this paper, the superscript $a, b$ represents the indices in Cartesian coordinates $(x,y,z)$.

\section{Model: Hamiltonian Particle System}\label{sec:Model: Hamiltonian Particle System}
\begin{figure}
\centering
\includegraphics[width=8cm,bb=0 0 364 261]{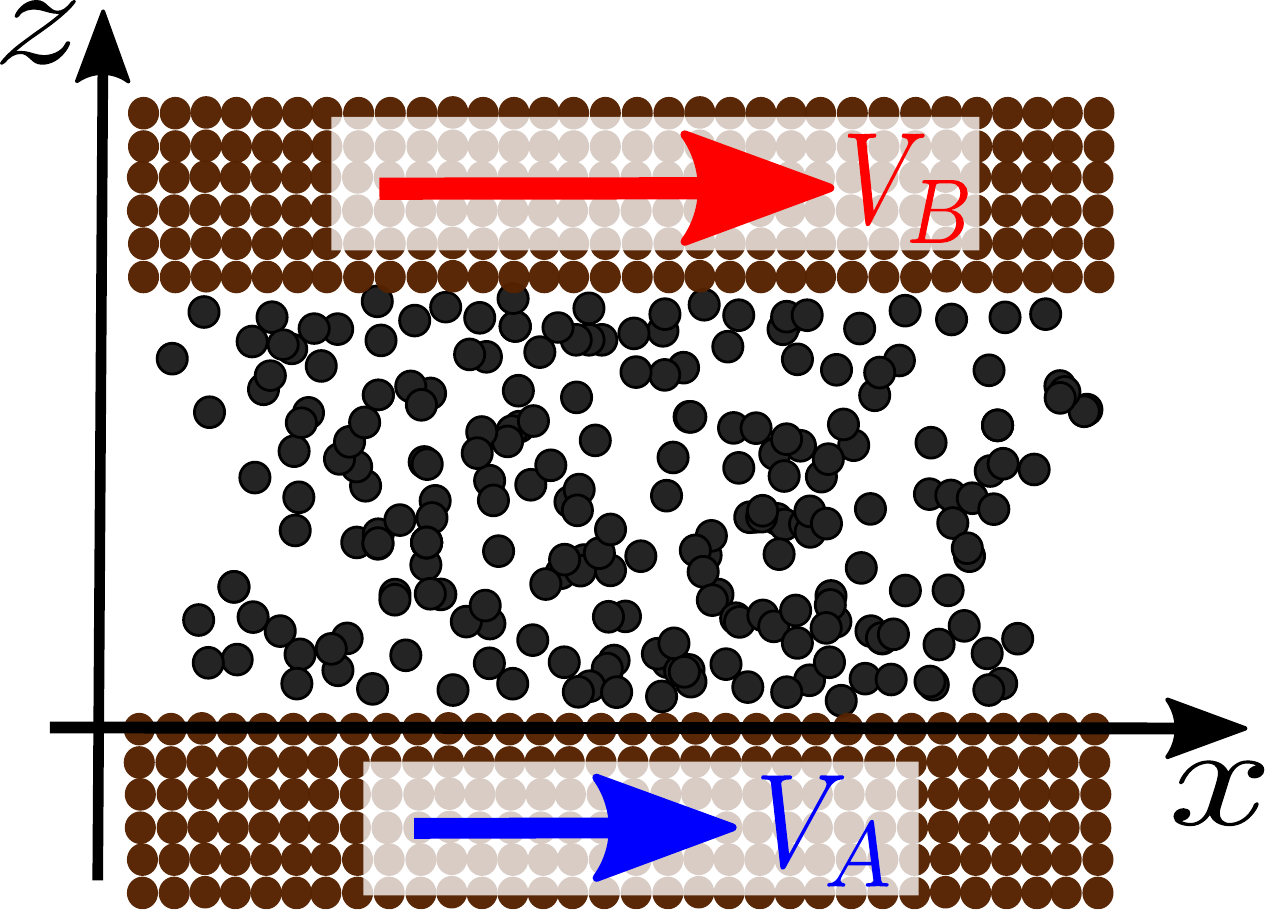}
\caption{Schematic illustration of our microscopic model.}
\label{fig:model microscopic}
\end{figure}
We introduce a model for studying the boundary condition in fluid dynamics. A schematic illustration is shown in Fig.~\ref{fig:model microscopic}. 
The fluid consists of $N$ particles that are confined to an $L_x\times L_y\times L$ cuboid. The position and momentum of the $i$th particle are denoted by $(\bm{r}_i,\bm{p}_i)$ $(i=1,2,\cdots,N)$. We impose periodic boundary conditions along the $x$ and $y$ axes and introduce two parallel walls so as to confine particles in the $z$ direction. We represent the two walls as collections of material points placed in planes $z=0$ and $z=L$, which are referred to as walls $A$ and $B$, respectively. We move the walls in the $x$ direction. Let $\bm{q}^A_i(t)$, $(i=1,2,\cdots,N_A)$, and $\bm{q}^B_i(t)$, $(i=1,2,\cdots,N_B)$, be the position of the material points that consist of walls $A$ and $B$, respectively. The motion of the walls is given by the uniform motion of the material points:
\begin{eqnarray}
\bm{q}^A_i(t) = \bm{q}^A_i(0) + \bm{V}_A t
\end{eqnarray}
for $i=1,2,\cdots,N_A$, and
\begin{eqnarray}
\bm{q}^B_i(t) = \bm{q}^B_i(0) + \bm{V}_B t
\end{eqnarray}
for $i=1,2,\cdots,N_B$, where $\bm{V}_A=(V_A,0,0)$ and $\bm{V}_B = (V_B,0,0)$. 

The collection of positions and momenta of all particles is denoted by $\Gamma=(\bm{r}_1,\bm{p}_1,\cdots,\bm{r}_N,\bm{p}_N)$. Similarly, the collections of positions of the material points that comprise walls $A$ and $B$ are denoted by $\Gamma^A=(\bm{q}^A_1,\cdots,\bm{q}^A_{N_A})$ and $\Gamma^B=(\bm{q}^B_1,\cdots,\bm{q}^B_{N_B})$. The Hamiltonian of the system is given by
\begin{eqnarray}
H(\Gamma;\Gamma^A,\Gamma^B) = \sum_{i=1}^{N} \frac{\bm{p}_i^2}{2m} + U(\{\bm{r}_i\}_{i=1}^{N};\Gamma^A,\Gamma^B)
\end{eqnarray}
with
\begin{eqnarray}
U(\{\bm{r}_i\}_{i=1}^N;\Gamma^A,\Gamma^B) &\equiv& \sum_{i<j} U_{\rm FF}(|\bm{r}_i-\bm{r}_j|) + \sum_{i=1}^N U_{\rm AF}(\bm{r}_i;\Gamma^A) + \sum_{i=1}^N U_{\rm BF}(\bm{r}_i;\Gamma^B) .
\label{eq:potential all}
\end{eqnarray}
$U_{\rm FF}(r)$ describes the interaction potential between particles. $U_{\rm AF}(\bm{r};\Gamma^A)$ and $U_{\rm BF}(\bm{r};\Gamma^B)$ denote the potentials of walls $A$ $B$, respectively. $U_{\rm AF}(\bm{r};\Gamma^A)$ may be expressed as
\begin{eqnarray}
U_{\rm AF}(\bm{r};\Gamma^A) \equiv \sum_{j=1}^{N_A} U_{A}(|\bm{r}-\bm{q}^A_j|),
\label{eq:UAW}
\end{eqnarray}
where we introduce the interaction between the material point and the fluid particle, $U_{A}(r)$. $U_{\rm BF}(\bm{r};V_B)$ has a similar form.

We employ the stochastic thermostat to thermalize the fluid. $\bm{F}_{\rm bath}(\bm{r};T)$ denotes the stochastic force applied to the particle at a point $\bm{r}$ by the thermostat with temperature $T$. Then, the particles obey
\begin{eqnarray}
m\frac{d^2 r^{a}_i}{dt^2} &=& - \sum_{j(\neq i)} \frac{\partial V_{\rm FF}(|\bm{r}_i-\bm{r}_j|)}{\partial r^{a}_i} - \frac{\partial U_{\rm AF}(\bm{r}_i;\Gamma^A)}{\partial r^{a}_i} - \frac{\partial U_{\rm BF}(\bm{r}_i;\Gamma^B)}{\partial r^{a}_i} + F^{a}_{\rm bath}(\bm{r}_i;T).
\label{eq:microscopic dynamics}
\end{eqnarray}
For this study, the configurations of the stochastic thermostat are: i) the dynamics (\ref{eq:microscopic dynamics}) satisfies the local detailed balance condition (Sec.~\ref{sec:fluctuation theorem} gives a detailed account of this condition); ii) the dynamics (\ref{eq:microscopic dynamics}) are Galilean invariant in the $x$ direction, implying that (\ref{eq:microscopic dynamics}) is unchanged under the Galilean transformation $\bm{r}(t) \to \bm{r}'(t)=\bm{r}-\bm{v}t$ with $\bm{v}=(v,0,0)$; and iii) the characteristic strengths of interactions between particles, between a particle and wall, and between a particle and the thermostat, denoted by $|\bm{F}_{\rm int}|$, $|\bm{F}_{\rm wall}|$ and $|\bm{F}_{\rm bath}|$, respectively, obey
\begin{eqnarray}
|\bm{F}_{\rm bath}| \ll |\bm{F}_{\rm int}|,|\bm{F}_{\rm wall}|.
\label{eq:weak bath condition}
\end{eqnarray}

\subsection{Microscopic expression of observed quantities}
Let $\Gamma_t$ denote the collection of positions and momenta of all particles composing the fluid at time $t$. Similarly, let $\Gamma^A_t$ and $\Gamma^B_t$ denote the collection of positions of all material points composing walls $A$ and $B$, respectively. We define the microscopic mass density field $\hat{\rho}(\bm{r};\Gamma)$ and the microscopic momentum density field $\hat{\pi}^a(\bm{r};\Gamma)$ as
\begin{eqnarray}
\hat{\rho}(\bm{r};\Gamma) \equiv \sum_{i=1}^N m \delta\big(\bm{r}-\bm{r}_i\big),
\end{eqnarray}
and
\begin{eqnarray}
\hat{\pi}^a(\bm{r};\Gamma) \equiv \sum_{i=1}^N p^a_i \delta\big(\bm{r}-\bm{r}_i\big)
\label{eq:def of microscopic momentum density}
\end{eqnarray}
for a given microscopic configuration $\Gamma$. These microscopic densities satisfy the microscopic balance equations:
\begin{eqnarray}
\frac{\partial \hat{\rho}(\bm{r};\Gamma_t)}{\partial t} + \frac{\partial \hat{\pi}^a(\bm{r};\Gamma_t)}{\partial r^a} = 0,
\label{eq:equation of continuity for density field}
\end{eqnarray}
and
\begin{eqnarray}
\frac{\partial \hat{\pi}^a(\bm{r};\Gamma_t)}{\partial t} + \frac{\partial \hat{J}^{ab}(\bm{r};\Gamma_t)}{\partial r^b} &=& \sum_{i=1}^N \Big(\hat{F}^a_A(\bm{r}_i(t);\Gamma^A_t) + \hat{F}^a_B(\bm{r}_i(t);\Gamma^B_t) \Big) \delta\big(\bm{r}-\bm{r}_i(t)\big) \nonumber \\[3pt]
&+& \sum_{i=1}^N F^{a}_{\rm bath}(\bm{r}_i(t)) \delta \big(\bm{r}-\bm{r}_i\big).
\label{eq:equation of continuity x,y,z plus heat bath}
\end{eqnarray}
Here, $\hat{F}^a_A(\bm{r};\Gamma^A)$ and $\hat{F}^a_B(\bm{r};\Gamma^B)$ are given by
\begin{eqnarray}
\hat{F}^a_A(\bm{r};\Gamma^A) \equiv - \sum_{j=1}^{N_A} \frac{\partial U_A(\bm{r}-\bm{q}^A_j)}{\partial r^a},
\label{eq:def of FA}
\end{eqnarray}
and
\begin{eqnarray}
\hat{F}^a_B(\bm{r};\Gamma^B) \equiv - \sum_{j=1}^{N_B} \frac{\partial U_B(\bm{r}-\bm{q}^B_j)}{\partial r^a},
\label{eq:def of FB}
\end{eqnarray}
which represent the total force acting on the fluid at a point $\bm{r}$ from walls $A$ and $B$, respectively. The microscopic momentum current $\hat{J}^{ab}(\bm{r};\Gamma)$ is given by
\begin{eqnarray}
\hat{J}^{ab}(\bm{r};\Gamma) \equiv \sum_{i} \frac{p^a_i p^b_i}{m} \delta(\bm{r}-\bm{r}_i) + \sum_{i<j} \hat{F}_{ij}^a(\Gamma)(r^b_i-r^b_j) D(\bm{r};\bm{r}_i,\bm{r}_j)
\end{eqnarray}
with
\begin{eqnarray}
D(\bm{r};\bm{r}_i,\bm{r}_j) \equiv \int_0^1 d \xi \delta(\bm{r}-\bm{r}_i-(\bm{r}_j-\bm{r}_i)\xi),
\label{eq:def of D delta}
\end{eqnarray}
and
\begin{eqnarray}
\hat{F}_{ij}^a(\Gamma) \equiv -\frac{\partial U_{\rm FF}(|\bm{r}_i-\bm{r}_j|)}{\partial r^a_i}.
\end{eqnarray}
Here, we recall the condition (\ref{eq:weak bath condition}), which implies that the interaction between the particle and the thermostat is sufficiently weak. From this condition, the last term on the right-hand side in (\ref{eq:equation of continuity x,y,z plus heat bath}) may be neglected. Hereafter, we employ as the microscopic balance equation for $\hat{\pi}^a(\bm{r};\Gamma)$
\begin{eqnarray}
\frac{\partial \hat{\pi}^a(\bm{r};\Gamma_t)}{\partial t} + \frac{\partial \hat{J}^{ab}(\bm{r};\Gamma_t)}{\partial r^b} &=& \sum_{i=1}^N \Big(\hat{F}^a_A(\bm{r}_i(t);\Gamma^A_t) + \hat{F}^a_B(\bm{r}_i(t);\Gamma^B_t) \Big) \delta\big(\bm{r}-\bm{r}_i(t)\big),
\label{eq:equation of continuity x,y,z}
\end{eqnarray}
instead of (\ref{eq:equation of continuity x,y,z plus heat bath}).

$[\Gamma]=\big(\Gamma_t\big)_{0<t<\tau}$ denotes the history of the system over the time interval $[0,\tau]$. For a given initial state $\Gamma_0^{\rm tot} \equiv (\Gamma_0,\Gamma^A_0,\Gamma^B_0)$, $T\big([\Gamma] \big| \Gamma_0^{\rm tot},V_A,V_B\big)$ represents the probability density that $[\Gamma]$ is realized with wall velocities $V_A$ and $V_B$. We assume that the initial state $\Gamma_0$ is chosen according to a given distribution $P_{\rm ini}(\Gamma_0)$. We now focus on any two microscopic quantities $\hat{f}(\Gamma)$ and $\hat{g}(\Gamma)$ given as functions of microscopic state $\Gamma$. The ensemble average of $\hat{f}(\Gamma_{\tau})$ at time $t=\tau$, $\langle f(\tau) \rangle^{V_A,V_B}$, is expressed in terms of $P_{\rm ini}(\Gamma_0)$ and $T\big([\Gamma] \big| \Gamma_0^{\rm tot},V_A,V_B\big)$ as
\begin{eqnarray}
\langle f(\tau) \rangle^{V_A,V_B} =  \int d [\Gamma] \hat{f}(\Gamma_{\tau}) P_{\rm ini}(\Gamma_0)T\big([\Gamma] \big| \Gamma_0^{\rm tot},V_A,V_B\big).
\label{eq:def of ensemble average quantity}
\end{eqnarray}
Similarly, the ensemble-averaged correlation function between $\hat{f}(\Gamma_{\tau})$ at time $t=\tau$ and $\hat{g}(\Gamma_0)$ at time $t=0$ is expressed as
\begin{eqnarray}
\langle f(\tau)g(0) \rangle^{V_A,V_B} =  \int d[\Gamma] \hat{f}(\Gamma_{\tau}) \hat{g}(\Gamma_0) P_{\rm ini}(\Gamma_0)T\big([\Gamma] \big| \Gamma_0^{\rm tot},V_A,V_B\big).
\label{eq:def of correlation}
\end{eqnarray}

In this paper, the motion of the fluid is characterized by macroscopic mass density field $\rho(\bm{r},t;V_A,V_B)$, macroscopic velocity field $v^a(\bm{r},t;V_A,V_B)$, and stress tensor field $\sigma^{ab}(\bm{r},t;V_A,V_B)$. These quantities are defined as
\begin{eqnarray}
\rho(\bm{r},t;V_A,V_B) \equiv \langle \hat{\rho}(\bm{r},t)\rangle^{V_A,V_B},
\end{eqnarray}
\begin{eqnarray}
\pi^a(\bm{r},t;V_A,V_B) \equiv \langle \hat{\pi}^a(\bm{r},t)\rangle^{V_A,V_B},
\end{eqnarray}
\begin{eqnarray}
v^a(\bm{r},t;V_A,V_B) \equiv \frac{\pi^a(\bm{r},t;V_A,V_B)}{\rho(\bm{r},t;V_A,V_B)},
\end{eqnarray}
and
\begin{eqnarray}
\sigma^{ab}(\bm{r},t;V_A,V_B) \equiv - \Big(J^{ab}(\bm{r},t;V_A,V_B) - \rho(\bm{r},t;V_A,V_B) v^a(\bm{r},t;V_A,V_B) v^b(\bm{r},t;V_A,V_B) \Big)
\label{eq:def of stress tensor}
\end{eqnarray}
with
\begin{eqnarray}
J^{ab}(\bm{r},t;V_A,V_B) = \langle \hat{J}^{ab}(\bm{r},t)\rangle^{V_A,V_B}.
\end{eqnarray}
Furthermore, we focus on the force acting on the fluid from each wall. The microscopic definition is given by
\begin{eqnarray}
\hat{F}^{a}_{\alpha}(\Gamma_t;\Gamma_t^{\alpha}) = \sum_{i=1}^N \hat{F}_{\alpha}^{a}(\bm{r}_i(t);\Gamma_t^{\alpha}),
\label{eq:def of microscopic force acting on the fluid}
\end{eqnarray}
where $\alpha = {\rm A,B}$. The ensemble average is denoted by $F^{a}_{\alpha}(t;V_A,V_B) = \langle \hat{F}^{a}_{\alpha}(\Gamma_t;\Gamma_t^{\alpha}) \rangle^{V_A,V_B}$.

\section{Generalized Green--Kubo formula}\label{sec:generalized Green-Kubo formula}
In this section, we give the exact formula describing the time evolution of any ensemble-averaged quantities.

We prepare initial states as follows. First, we assume that the system is in equilibrium at temperature $T$. Next, we apply an impulse force $m\bm{u}(\bm{r}) \delta(t)$ to this system where we set $\bm{u}=(u^x(z),0,0)$. Then, the probability density just after $t=0$, $P_{\rm leq}(\Gamma_0|\Gamma_0^A,\Gamma_0^B)$, is represented by
\begin{eqnarray}
P_{\rm leq}(\Gamma_0|\Gamma_0^A,\Gamma_0^B) = \frac{1}{Z\big(k_BT,\bm{u}(\bm{r})\big)} \exp\Big[-\frac{1}{k_B T} \Big(H(\Gamma_0;\Gamma_0^A,\Gamma_0^B) - \sum_{i=1}^N u^x(z_i(0))p^x_i(0) \Big)\Big].
\label{eq:initial condition: relaxation process}
\end{eqnarray}
We consider that the initial microscopic state $\Gamma_0$ is chosen according to (\ref{eq:initial condition: relaxation process}). For any physical quantity $\hat{f}(\Gamma)$ and any velocity field $u^x(z)$, we have
\begin{eqnarray}
\langle f(\tau) \rangle^{V_A,V_B}  &=& \langle f(0) \rangle^{V_A,V_B} + \frac{1}{k_B T} \int_0^{\tau} ds \sum_{i=1}^N \left\langle \big(V_A - u^x(z_i(0)) \big)f(s) F^x_A(\bm{r}_i(0),0)\right\rangle^{V_A,V_B} \nonumber \\[3pt]
&+& \frac{1}{k_B T} \int_0^{\tau} ds \sum_{i=1}^N \left\langle \big(V_B - u^x(z_i(0)) \big) f(s) F^x_B(\bm{r}_i(0),0)\right\rangle^{V_A,V_B}  \nonumber \\[3pt]
&-& \frac{1}{k_B T} \int d^3\bm{r} \int_0^{\tau} ds \frac{\partial u^x(z)}{\partial z} \langle f(s) J^{xz}(\bm{r},0) \rangle^{V_A,V_B},
\label{eq:generalized Green-Kubo: local gibbs}
\end{eqnarray}
which is the exact formula to describe the time evolution of $\langle f(\tau) \rangle^{V_A,V_B}$. We refer to (\ref{eq:generalized Green-Kubo: local gibbs}) as the generalized Green--Kubo formula. This formula is the starting point for developing the linear response theories in the subsequent sections.

In this section, we show a proof of the generalized Green--Kubo formula (\ref{eq:generalized Green-Kubo: local gibbs}). The key point in our derivation is an efficient use of a non-equilibrium identity, which is known as the local detailed balance condition.

\subsection{Local detailed balance condition}\label{sec:fluctuation theorem}
By considering the energetics for a given history $[\Gamma]$, we define the heat transferred into the fluid, $Q\big([\Gamma] | \Gamma^{\rm tot}_0,V_A,V_B\big)$, and the work applied to the fluid from the walls, $W\big([\Gamma] | \Gamma^{\rm tot}_0,V_A,V_B\big)$ as
\begin{eqnarray}
Q\big([\Gamma] | \Gamma^{\rm tot}_0,V_A,V_B\big) =\sum_{i=1}^N \int_{\hat{\Gamma}} ds \frac{p^y_i}{m} \Big(\dot{p}^y_i + \frac{\partial H(\Gamma_s;\Gamma^A_s,\Gamma^B_s)}{\partial  r_i^y} \Big),
\end{eqnarray}
\begin{eqnarray}
W\big([\Gamma] | \Gamma^{\rm tot}_0,V_A,V_B\big) &=& \int_{\hat{\Gamma}} ds \Big(V_A \hat{F}^x_A(\Gamma_s;\Gamma_s^A)+ V_B \hat{F}^x_B(\Gamma_s;\Gamma_s^B) \Big).
\label{eq:work definition}
\end{eqnarray}
The first law of the thermodynamics is expressed in the form
\begin{eqnarray}
H(\Gamma_t,t) - H(\Gamma_0,0) = W\big([\Gamma] | \Gamma^{\rm tot}_0,V_A,V_B\big) + Q\big([\Gamma] | \Gamma^{\rm tot}_0,V_A,V_B\big).
\end{eqnarray}
We use the convention that work applied to the fluid is positive and heat dissipated from the fluid is negative.

We write the time reversal of the microscopic state $\Gamma$ as $\Gamma^{\ast} =  (\bm{r}_1,\bm{r}_2,\cdots,\bm{r}_N,-\bm{p}_1,-\bm{p}_2,\cdots,-\bm{p}_N)$, and that of the history of the system $[\Gamma]$ as $[\Gamma^{\dagger}] = \big(\Gamma^{\ast}_{\tau-t} \big)_{0<t<\tau}$. The local detailed balance condition is expressed as
\begin{eqnarray}
\frac{T\big([\Gamma] \big| \Gamma_0^{\rm tot},V_A,V_B\big)}{T\big([\Gamma^{\dagger}] \big| \Gamma_{\tau}^{\ast \rm tot},-V_A,-V_B\big)} = \exp \Big(-\frac{Q\big([\Gamma] | \Gamma^{\rm tot}_0,V_A,V_B\big)}{k_B T} \Big),
\label{eq:fluctuation theorem}
\end{eqnarray}
where $\Gamma^{\ast \rm tot} =(\Gamma^{\ast},\Gamma^A,\Gamma^B)$. As mentioned in Sec.~\ref{sec:Model: Hamiltonian Particle System}, we assume that the dynamics (\ref{eq:microscopic dynamics}) satisfies the condition (\ref{eq:fluctuation theorem}) by choosing the appropriate stochastic thermostat. 

We note that the local detailed balance condition is derived from the microscopic dynamics under some assumptions (e.g., Hamiltonian particle systems~\cite{jarzynski1997nonequilibrium} and Markovian stochastic process~\cite{crooks1999entropy}). In this sense, the local detailed balance condition is often referred to as the local fluctuation theorem (see~\cite{seifert2012stochastic} for review).

\subsection{Derivation of (\ref{eq:generalized Green-Kubo: local gibbs})}
By substituting (\ref{eq:initial condition: relaxation process}) and (\ref{eq:fluctuation theorem}) into (\ref{eq:def of ensemble average quantity}), we obtain
\begin{eqnarray}
\langle f(\tau) \rangle^{V_A,V_B} &=& \int d[\Gamma] f(\Gamma_{\tau}) \exp \Big(\frac{W_{\rm leq}\big([\Gamma] | \Gamma^{\rm tot}_0,V_A,V_B,\bm{u}\big)}{k_B T} \Big) \nonumber \\[3pt]
& & P_{\rm leq}(\Gamma^{\ast}_{\tau}|\Gamma_{\tau}^A,\Gamma_{\tau}^B) T\big([\Gamma^{\dagger}] | \Gamma^{\ast \rm tot}_{\tau},-V_A,-V_B\big)
\label{eq:average value fluctuation theorem 1}
\end{eqnarray}
with
\begin{eqnarray}
W_{\rm leq}\big([\Gamma] | \Gamma^{\rm tot}_0,V_A,V_B,\bm{u}\big) = W\big([\Gamma] | \Gamma^{\rm tot}_0,V_A,V_B\big)- \Big(\sum_{i=1}^N u^x(z_i(t))p^x_i(t) - \sum_{i=1}^N u^x(z_i(0))p^x_i(0) \Big) .
\label{eq: def of Wleq}
\end{eqnarray}
We rewrite the right-hand side of (\ref{eq:average value fluctuation theorem 1}) by using the ensemble average along the time-reversal history. By noting
\begin{eqnarray}
W\big([\Gamma^{\dagger}] | \Gamma^{\ast \rm tot}_{\tau},-V_A,-V_B\big) = - W\big([\Gamma] | \Gamma^{\rm tot}_0,V_A,V_B\big),
\end{eqnarray}
we obtain
\begin{eqnarray}
\langle f(\tau) \rangle^{V_A,V_B} &=& \int d[\Gamma^{\dagger}] f^{\ast}(\Gamma^{\ast}_{\tau}) \exp \Big(-\frac{W_{\rm leq}\big([\Gamma^{\dagger}] | \Gamma^{\ast \rm tot}_{\tau},-V_A,-V_B,-\bm{u}\big)}{k_B T} \Big)  \nonumber \\[3pt]
& & P_{\rm leq}(\Gamma^{\ast}_{\tau}|\Gamma_{\tau}^{A},\Gamma_{\tau}^{B}) T\big([\Gamma^{\dagger}] | \Gamma^{\ast \rm tot}_{\tau},-V_A,-V_B\big) \nonumber \\[3pt]
&=& \Big\langle f^{\ast}(0) \exp \Big(- \frac{W_{\rm leq}\big([\Gamma^{\dagger}] | \Gamma^{\dagger \rm tot}_{0},-V_A,-V_B,-\bm{u}\big)}{k_B T}  \Big)\Big\rangle^{-V_A,-V_B} ,
\label{eq:average value fluctuation theorem}
\end{eqnarray}
where we have defined $f^{\ast}(\Gamma^{\ast}) \equiv f(\Gamma)$. Similarly, we rewrite (\ref{eq:def of correlation}) as
\begin{eqnarray}
\langle f(\tau) g(0) \rangle^{V_A,V_B} = \Big\langle f^{\ast}(0)g^{\ast}(\tau)  \exp \Big(- \frac{W_{\rm leq}\big([\Gamma^{\dagger}] | \Gamma^{\dagger \rm tot}_{0},-V_A,-V_B,-\bm{u}\big)}{k_B T}  \Big)\Big\rangle^{-V_A,-V_B} .
\label{eq:correlation function fluctuation theorem}
\end{eqnarray}

Next, using the microscopic force-balance equation (\ref{eq:equation of continuity x,y,z}), we rewrite $W_{\rm leq}\big([\Gamma] | \Gamma_0,\Gamma^A_0,\Gamma^B_0,V_A,V_B,\bm{u}\big)$. Multiplying (\ref{eq:equation of continuity x,y,z}) by $\bm{u}(\bm{r})$ and integrating over $[0,\tau]$ and all space, we obtain
\begin{eqnarray}
& &\Big(\sum_{i=1}^N u^x(z_i(t))p_i^x(t) - \sum_{i=1}^N u^x(z_i(0))p_i^x(0)\Big) \nonumber \\[3pt]
&=& \int_0^{\tau} ds \sum_{i=1}^N u^x(z_i(s))\Big(\hat{F}^a_A(\bm{r}_i(s);\Gamma^A_s) + \hat{F}^a_B(\bm{r}_i(s);\Gamma^B_s) \Big) + \int_0^{\tau} ds \int d^3\bm{r} \frac{\partial u^x(z)}{\partial z} \hat{J}^{xz}(\bm{r};\Gamma_s).
\label{eq:continuity equation deform 1}
\end{eqnarray}
Substituting (\ref{eq:work definition}) and (\ref{eq:continuity equation deform 1}) into (\ref{eq: def of Wleq}), we obtain
\begin{eqnarray}
& & W_{\rm leq}\big([\Gamma] | \Gamma^{\rm tot}_0,V_A,V_B,\bm{u}\big) \nonumber \\[3pt]
&=&\int_0^{\tau} ds \sum_{i=1}^N \Big(\big(V_A-u^x(z_i(s))\big)\hat{F}^x_A(\bm{r}_i(s);\Gamma^A_s) + \big(V_B-u^x(z_i(s))\big)\hat{F}^x_B(\bm{r}_i(s);\Gamma^B_s) \Big) \nonumber \\[3pt]
&-& \int_0^{\tau} ds \int d^3\bm{r} \frac{\partial u^x(z)}{\partial z} \hat{J}^{xz}(\bm{r};\Gamma_s).
\label{eq:Wleq deform 1}
\end{eqnarray}
Using (\ref{eq:average value fluctuation theorem}), (\ref{eq:correlation function fluctuation theorem}) and (\ref{eq:Wleq deform 1}), we write the derivative of $\langle f(s) \rangle^{V_A,V_B}$ in $s$ as
\begin{eqnarray}
\frac{\partial \langle f(s) \rangle^{V_A,V_B}}{\partial s} &=& \frac{1}{k_B T} \sum_{i=1}^N \left\langle \big(V_A - u^x(z_i(0)) \big)f(s) F^x_A(\bm{r}_i(0),0)\right\rangle^{V_A,V_B} \nonumber \\[3pt] 
&+& \frac{1}{k_B T}\sum_{i=1}^N \left\langle \big(V_B - u^x(z_i(0)) \big) f(s) F^x_B(\bm{r}_i(0),0)\right\rangle^{V_A,V_B} \nonumber \\[3pt]
&-& \int d^3\bm{r} \frac{\partial u^x(z)}{\partial z} \langle f(s) J^{xz}(\bm{r},0) \rangle^{V_A,V_B}.
\label{eq:derivative form 1}
\end{eqnarray}
we obtain (\ref{eq:generalized Green-Kubo: local gibbs}) by integrating (\ref{eq:derivative form 1}) over $[0,\tau]$.

\section{Separation of length and time scales}\label{sec:Separation of length and time scales}
\subsection{Assumption}
In our model, there are two length and time scales. The first scale is characterized by the length and time appearing in the molecular description. The maximum of the microscopic length scales, such as the molecular size, the interaction length, or the mean free path is denoted as $\xi_{\rm micro}$ and the corresponding time scale, which may be the collision time or the mean free time, is denoted as $\tau_{\rm micro}$. The second scale is related to the global equilibration. The characteristic length and time scales are given by the system size $L$ and the time $\tau_{\rm macro}$ necessary for the relaxation of the system to the global equilibrium, respectively. We assume the separation of length and time scales, which is represented by
\begin{eqnarray}
\tau_{\rm micro} \ll \tau_{\rm macro},
\end{eqnarray}
and
\begin{eqnarray}
\xi_{\rm micro} \ll L.
\end{eqnarray}

\subsection{Properties of the correlation functions of the momentum fluxes}\label{sec:4.2}
We now focus on the correlation functions of the momentum fluxes in the equilibrium state. We prepare the initial states according to $P_{\rm leq}(\Gamma_0|\Gamma_0^A,\Gamma_0^B)$ with $u^x(z)=0$ and consider the time evolution with $V_A=0$ and $V_B=0$. $\langle \cdot \rangle_{\rm eq}$ denotes the ensemble average in this system, which corresponds to the average in equilibrium. The correlation functions of the momentum fluxes are, for example, $\langle F^x_A(\tau) F^x_A(0)\rangle_{\rm eq}$, $\langle J^{xz}(\bm{r},\tau) F^x_A(0)\rangle_{\rm eq}$ and $\langle J^{xz}(\bm{r},\tau) J^{xz}(\bm{r}',0) \rangle_{\rm eq}$. First, we focus on the $\tau$-dependence of these correlation functions. We naturally conjecture that these correlation functions behave in the same manner with respect to time $\tau$. Therefore, we choose $\langle  F^x_A(\tau) F^x_A(0)\rangle_{\rm eq}$ as an example. By keeping the separation of time scales in mind, we assume that the $\tau$-dependence of $\langle  F^x_A(\tau) F^x_A(0)\rangle_{\rm eq}$ is divided into four stages. First, when $0 < \tau \simeq \tau_{\rm micro}$, $\langle  F^x_A(\tau) F^x_A(0)\rangle_{\rm eq}$ decays to zero. After that, when $\tau_{\rm micro} \ll \tau \ll \tau_{\rm macro}$, $\langle  F^x_A(\tau) F^x_A(0)\rangle_{\rm eq}$ remains zero. Eventually, when $\tau \simeq \tau_{\rm macro}$, $\langle  F^x_A(\tau) F^x_A(0)\rangle_{\rm eq}$ couples with the slow macroscopic motion and become non-zero. Finally, when $\tau \gg \tau_{\rm macro}$, the relaxation is completed and $\langle  F^x_A(\tau) F^x_A(0)\rangle_{\rm eq}$ again remains zero. In summary, we express this behavior as 
\begin{eqnarray}
\langle  F^x_A(\tau) F^x_A(0)\rangle_{\rm eq}
\begin{cases}
>0, \ \ {\rm for} & 0<\tau\simeq \tau_{\rm micro},\\
\simeq0,  \ \ {\rm for} & \tau_{\rm micro}\ll \tau \ll \tau_{\rm macro}, \\
\neq 0,  \ \ {\rm for} & \tau \simeq \tau_{\rm macro}, \\
\simeq0,  \ \ {\rm for} & \tau_{\rm macro} \ll \tau.
\end{cases}
\label{eq:behavior FF in tau}
\end{eqnarray}
The other correlation functions such as $\langle J^{xz}(\bm{r},\tau) F^x_A(0)\rangle_{\rm eq}$ and $\langle J^{xz}(\bm{r},\tau) J^{xz}(\bm{r}',0) \rangle_{\rm eq}$ behave in the similar manner.

The macroscopic behavior of the fluid is related to the motion at $\tau_{\rm macro}$-scale. By taking this into account, we introduce the lower and upper cut-off times $\tau_{\rm meso}$ and $\tau_{\rm ss}$ for the macroscopic behavior as
\begin{eqnarray}
\tau_{\rm micro} \ll \tau_{\rm meso} \ll \tau_{\rm macro} \ll \tau_{\rm ss}.
\end{eqnarray}
Recalling the separation of length scales, we may choose the length scale corresponding to $\tau_{\rm meso}$, i.e., $\xi_{\rm meso}$, to satisfy $\xi_{\rm micro}\ll \xi_{\rm meso} \ll \xi_{\rm macro}$.
Because the correlation length of the momentum fluxes is reasonably estimated as the order of $\xi_{\rm micro}$ when $\tau=\tau_{\rm meso}$, we assume
\begin{eqnarray}
\langle J^{xz}(\bm{r},\tau_{\rm meso}) J^{xz}(\bm{r}',0) \rangle_{\rm eq} \simeq 0
\label{eq:JxzJxz xi meso 0}
\end{eqnarray}
for $|\bm{r}-\bm{r}'| \geq \xi_{\rm meso}$. Similarly, 
\begin{eqnarray}
\langle J^{xz}(\bm{r},\tau_{\rm meso}) F^x_A(0)\rangle_{\rm eq} \simeq 0
\label{eq:JxzFxA xi meso 0}
\end{eqnarray}
is assumed for $z \geq \xi_{\rm meso}$. Also, because there are no correlations when $\tau=\tau_{\rm ss}$, we have
\begin{eqnarray}
\langle J^{xz}(\bm{r},\tau_{\rm ss}) J^{xz}(\bm{r}',0) \rangle_{\rm eq} \simeq 0,
\end{eqnarray} 
and
\begin{eqnarray}
\langle J^{xz}(\bm{r},\tau_{\rm ss}) F^x_A(0)\rangle_{\rm eq} \simeq 0,
\end{eqnarray}
where $\bm{r}$ is any position.

\subsection{Small parameters}
We introduce two small parameters representing the extent of the separation of scales, 
\begin{eqnarray}
\epsilon \equiv \xi_{\rm micro}/\xi_{\rm macro},
\label{eq:def of epsilon}
\end{eqnarray}
and
\begin{eqnarray}
\zeta \equiv \xi_{\rm meso}/\xi_{\rm macro}.
\label{eq:def of zeta}
\end{eqnarray}
In this study, we are mainly concerned with a macroscopic system of system size $\sim 10^{-2}{\rm m}$ and a shear rate $\sim 1 {\rm sec}^{-1}$. As the molecular size is typically $10^{-10} {\rm m}$, $\epsilon$ is estimated to be of order $10^{-8}$. Recalling that $\xi_{\rm meso}$ satisfies the relation $\xi_{\rm micro}\ll \xi_{\rm meso} \ll \xi_{\rm macro}$, we then assume that $\xi_{\rm meso}$ is of order $10^{-6} {\rm m}$, which leads that $\zeta \sim 10^{-4}$. As a result, the relationship between $\epsilon$ and $\zeta$ is
\begin{eqnarray}
\zeta = \epsilon^{1/2}.
\label{eq:relation between epsilon and delta}
\end{eqnarray}

In the reminder of this paper, we develop two linear response theories to describe the macroscopic motions of the fluid at time $\tau=\tau_{\rm meso}$ and $\tau=\tau_{\rm ss}$.

\section{Linear response theory for the relaxation process of fluid}\label{sec:Linear response theory for the relaxation process of fluid}
We now develop the linear response theory that describes the macroscopic behavior at the time $\tau=\tau_{\rm meso}$. We prepare an initial state chosen according to (\ref{eq:initial condition: relaxation process}), and consider the time evolution during the time interval $[0,\tau_{\rm meso}]$. In this interval, the macroscopic behavior of the fluid is determined by the velocity fields $u^x(z)$ given as the initial condition. We derive the constitutive equations of the fluid with the velocity $u^x(z)$. These constitutive equations lead to the partial slip boundary condition with the expression of the slip length proposed by Bocquet and Barrat~\cite{bocquet1994hydrodynamic}.

In this section, we convert all the quantities to dimensionless forms setting $\xi_{\rm micro}=m=|V_A-V_B|=1$.

\subsection{Scale separation parameter}
Let us assume that $u^{x}(z)$ depends on the small parameter $\epsilon$, 
\begin{eqnarray}
\frac{d u^{x}(z)}{dz} = O(\epsilon),
\label{eq:derivative of ux: epsilon}
\end{eqnarray}
\begin{eqnarray}
\frac{d^2 u^{x}(z)}{dz^2} = O(\epsilon^2),
\label{eq:two derivative of ux: epsilon}
\end{eqnarray}
\begin{eqnarray}
\Big|u^x(0)-V_{A} \Big| = O(\epsilon),
\label{eq:ux0-VA: epsilon}
\end{eqnarray}
and
\begin{eqnarray}
\Big|V_{B}-u^x(L) \Big| = O(\epsilon).
\label{eq:VB-uxL: epsilon}
\end{eqnarray}
For example, $u^x(z)$ is given by
\begin{eqnarray}
u^x(z) = \frac{1}{K}\sum_{k=1}^K \sin\Big(\epsilon \pi k z\Big)+ u^x(0),
\label{eq:velocity field: example}
\end{eqnarray}
where $K>0$ is any integer and $u^x(0)$ is chosen to satisfy (\ref{eq:ux0-VA: epsilon}) and (\ref{eq:VB-uxL: epsilon}).

We note that conditions (\ref{eq:ux0-VA: epsilon}) and (\ref{eq:VB-uxL: epsilon}) confine us to a slip of order $\epsilon$. These conditions are essential to the derivation of the partial slip boundary condition. Previously~\cite{nakano2019microscopic}, we discussed what kind of walls satisfies (\ref{eq:ux0-VA: epsilon}) and (\ref{eq:VB-uxL: epsilon}) when the fluid is in a steady state. In this section, we derive the linear response formula focusing on the first order approximation of (\ref{eq:generalized Green-Kubo: local gibbs}) in $\epsilon$.

\subsection{Constitutive equation in bulk}\label{sec:constitutive equation in bulk}
We focus on the behavior of the shear stress $\sigma^{xz}(\bm{r},t;V_{A},V_{B})$. Our starting point is (\ref{eq:generalized Green-Kubo: local gibbs}) with $\hat{f}(\Gamma)=\hat{J}^{xz}(\bm{r};\Gamma)$ and $\tau=\tau_{\rm meso}$,
\begin{eqnarray}
J^{xz}(\bm{r},\tau_{\rm meso};V_{A},V_{B}) &=& \frac{1}{k_{B} T}\int_0^{\tau_{\rm meso}} \hspace{-0.15cm} ds \sum_{i=1}^N \big\langle \big(V_{A} - u^x(z_{i}(0)) \big)J^{xz}(\bm{r},s) F^x_{A}(\bm{r}_{i}(0),0)\big\rangle^{V_{A},V_{B}} \nonumber \\[3pt]
&+& \frac{1}{k_{B} T} \int_0^{\tau_{\rm meso}} \hspace{-0.15cm} ds \sum_{i=1}^N \big\langle \big(V_{B} - u^x(z_{i}(0)) \big) J^{xz}(\bm{r},s) F^x_{B}(\bm{r}_{i}(0),0)\big\rangle^{V_{A},V_{B}}  \nonumber \\[3pt]
&-& \frac{1}{k_{B} T} \int d^3\bm{r}' \int_0^{\tau_{\rm meso}} \hspace{-0.15cm} ds \frac{d u^x(z')}{d z'} \big\langle J^{xz}(\bm{r},s) J^{xz}(\bm{r}',0) \big\rangle^{V_{A},V_{B}},
\label{eq:generalized Green-Kubo: local gibbs: Jxz}
\end{eqnarray}
where we have used the property $J^{xz}(\bm{r},0;V_{A},V_{B}) = 0$. We now derive a simpler form of (\ref{eq:generalized Green-Kubo: local gibbs: Jxz}) by the separation of scales.

First, we remark on the local approximation. As mentioned in Sect.~\ref{sec:Separation of length and time scales}, during time interval $0<\tau<\tau_{\rm meso}$, the velocity field $v^x(z,\tau;V_{A},V_{B})$ remains $u^x(z)$ given by the initial condition, and the correlation functions of the momentum fluxes decay, e.g., $\langle  F^x_{A}(\tau_{\rm meso}) F^x_{A}(0)\rangle_{\rm eq} \simeq 0$. Therefore, the first term of (\ref{eq:generalized Green-Kubo: local gibbs: Jxz}) may be rewritten as
\begin{eqnarray}
& &\frac{1}{k_B T}\int_0^{\tau_{\rm meso}} \hspace{-0.15cm} ds \sum_{i=1}^N \big\langle \big(V_{A} - u^x(z_{i}(0)) \big)J^{xz}(\bm{r},s) F^x_{A}(\bm{r}_{i}(0),0)\big\rangle^{V_{A},V_{B}} \nonumber \\[3pt]
&\simeq&\frac{1}{k_{B} T} \sum_{i=1}^N \Big(V_{A} - \langle u^x(z_{i})\rangle\Big) \int_0^{\tau_{\rm meso}} \hspace{-0.15cm} ds \big\langle J^{xz}(\bm{r},s) F^x_{A}(\bm{r}_{i}(0),0)\big\rangle^{V_{A},V_{B}} .
\label{eq:ignoring the memory effect}
\end{eqnarray}
With this approximation, we neglect non-local effects. Recalling $\hat{F}^x_{A}(\bm{r},\Gamma^A) \simeq 0$ in $z>\xi_{\rm micro}$, we find that only the particles lying in $0<z<\xi_{\rm micro}$ at time $t=0$ contribute to the sum of (\ref{eq:ignoring the memory effect}). Also, we note that the fluid density is not uniform near the atomically smooth surface and an epitaxial-like layer is observed~\cite{voronov2008review}. Then, we assume that the main contribution to the sum of (\ref{eq:ignoring the memory effect}) stems from the first epitaxial layer normal to the wall. By using this assumption, (\ref{eq:ignoring the memory effect}) is approximated as 
\begin{eqnarray}
& & \frac{1}{k_{B} T} \sum_{i=1}^N \Big(V_{A} - \langle u^x(z_{i})\rangle\Big) \int_0^{\tau_{\rm meso}} \hspace{-0.15cm} ds \big\langle J^{xz}(\bm{r},s) F^x_{A}(\bm{r}_{i}(0),0)\big\rangle^{V_{A},V_{B}} \nonumber \\[3pt]
&\simeq& \frac{1}{k_{B} T} \Big(V_{A} - u^x(z^{\rm 1st}_A)\Big) \int_0^{\tau_{\rm meso}} \hspace{-0.15cm} ds \big\langle J^{xz}(\bm{r},s) F^x_{A}(0)\big\rangle^{V_{A},V_{B}}.
\label{eq:ignoring the memory effect2}
\end{eqnarray}
where $z=z^{\rm 1st}_{\alpha}$ represents the position of the first epitaxial layer near the wall $\alpha=A, B$ and we have used (\ref{eq:def of microscopic force acting on the fluid}).

Next, we consider the linear response approximation. We focus on (\ref{eq:ignoring the memory effect2}). From (\ref{eq:ux0-VA: epsilon}), we note that (\ref{eq:ignoring the memory effect}) is at least of first order in $\epsilon$. Because $\langle J^{xz}(\bm{r},\tau_{\rm meso}) F^x_{A}(0)\rangle^{V_{A},V_{B}}$ is estimated as the value in the steady state with $V_{A}$, $V_{B}$ and $u^x(z)$, we hypothesize that $\langle J^{xz}(\bm{r},(\tau_{\rm meso})) F^x_{A}(0)\rangle^{V_{A},V_{B}}$ may be rewritten as
\begin{eqnarray}
\big\langle J^{xz}(\bm{r},\tau_{\rm meso}) F^x_{A}(0)\big\rangle^{V_{A},V_{B}} \simeq \big\langle J^{xz}(\bm{r},\tau_{\rm meso}) F^x_{A}(0)\big\rangle_{\rm eq} + O\Big(V_{A} - u^x(z_A^{\rm 1st}),\frac{\partial u^x}{\partial z}\Big|_{z_A^{\rm 1st}} \Big),
\label{eq:linear response approximation}
\end{eqnarray}
where the definition of $\langle \cdot \rangle_{\rm eq}$ is given in Sec.~\ref{sec:4.2}. In the linear response theory, we neglect the second term. Collecting the local and linear response approximations, we rewrite (\ref{eq:generalized Green-Kubo: local gibbs: Jxz}) as
\begin{eqnarray}
J^{xz}(\bm{r},\tau_{\rm meso};V_{A},V_{B}) &\simeq& \frac{1}{k_{B} T} \Big(V_{A} - u^x(z^{\rm 1st}_A)\Big) \int_0^{\tau_{\rm meso}} \hspace{-0.15cm} ds  \big\langle J^{xz}(\bm{r},s) F^x_{A}(0)\big\rangle_{\rm eq} \nonumber \\[3pt]
&+& \frac{1}{k_{B} T} \Big(V_{B} - u^x(z^{\rm 1st}_B) \Big) \int_0^{\tau_{\rm meso}}\hspace{-0.15cm} ds  \big\langle J^{xz}(\bm{r},s) F^x_{B}(0)\big\rangle_{\rm eq} \nonumber \\[3pt]
&-& \frac{1}{k_{B} T}  \frac{d u^x(z)}{d z} \int d^3\bm{r}' \int_0^{\tau_{\rm meso}}\hspace{-0.15cm} ds \big\langle J^{xz}(\bm{r},s) J^{xz}(\bm{r}',0) \big\rangle_{\rm eq}.
\label{eq:generalized Green-Kubo: local gibbs: Jxz: linear response theory}
\end{eqnarray}

Here, we define the bulk as the region $\xi_{\rm meso} \leq z \leq L - \xi_{\rm meso}$. From (\ref{eq:JxzFxA xi meso 0}), we find that the first and second terms of (\ref{eq:generalized Green-Kubo: local gibbs: Jxz: linear response theory}) are zero in the bulk. Furthermore, (\ref{eq:JxzJxz xi meso 0}) implies that the integral of the third term is independent of the walls when $\bm{r}$ is chosen in the bulk. Hence, we obtain
\begin{eqnarray}
J^{xz}(\bm{r},\tau_{\rm meso};V_{A},V_{B})  = - \eta \frac{d u^x(z)}{d z} 
\label{eq:generalized Green-Kubo: local gibbs: Jxz: linear response regime in bulk}
\end{eqnarray}
with
\begin{eqnarray}
\eta =  \frac{1}{k_{B} T}\int d^3\bm{r}' \int_0^{\tau_{\rm meso}} \hspace{-0.15cm} ds \big\langle J^{xz}(\bm{r},s) J^{xz}(\bm{r}',0) \big\rangle_{\rm eq},
\end{eqnarray}
where $\eta$ is independent of the walls. By substituting (\ref{eq:generalized Green-Kubo: local gibbs: Jxz: linear response regime in bulk}) into (\ref{eq:def of stress tensor}), we obtain
\begin{eqnarray}
\sigma^{xz}(\bm{r},\tau_{\rm meso};V_{A},V_{B}) = \eta \frac{d u^x(z)}{d z} ,
\label{eq:sigmaxz: linear response regime in bulk}
\end{eqnarray}
where we have used $v^z(z,\tau_{\rm meso};V_{A},V_{B}) =0$.

\subsection{Constitutive equation at the wall}
We next focus on the behavior of the force per unit area acting on the fluid from wall $A$, which is written as $F_{A}^x(t;V_{A},V_{B})/A_{xy}$. Here, $A_{xy}=L_xL_y$. By substituting $\hat{f}(\Gamma)=\hat{F}^x_{A}(\Gamma;\Gamma^A)/A_{xy}$ and $\tau=\tau_{\rm meso}$ into (\ref{eq:generalized Green-Kubo: local gibbs}), we obtain
\begin{eqnarray}
\frac{F_{A}^x(\tau_{\rm meso};V_{A},V_{B})}{A_{xy}} &=& \frac{1}{k_{B} T A_{xy}} \int_0^{\tau_{\rm meso}} \hspace{-0.15cm} ds \sum_{i=1}^N \big\langle \big(V_{A} - u^x(z_{i}(0)) \big)F^x_{A}(s) F^x_{A}(\bm{r}_{i}(0),0)\big\rangle^{V_{A},V_{B}} \nonumber \\[3pt]
&+& \frac{1}{k_{B} T A_{xy}} \int_0^{\tau_{\rm meso}} \hspace{-0.15cm} ds \sum_{i=1}^N \big\langle \big(V_{B} - u^x(z_{i}(0)) \big) F^x_{A}(s) F^x_{B}(\bm{r}_{i}(0),0)\big\rangle^{V_{A},V_{B}}  \nonumber \\[3pt]
&-& \frac{1}{k_{B} T A_{xy}} \int d^3\bm{r}' \int_0^{\tau_{\rm meso}} \hspace{-0.15cm} ds \frac{d u^x(z')}{d z'} \big\langle F^x_{A}(s) J^{xz}(\bm{r}',0) \big\rangle^{V_{A},V_{B}}.
\label{eq:generalized Green-Kubo: local gibbs: FxA: linear response regime}
\end{eqnarray}
We apply again the procedure given in Sect.~\ref{sec:constitutive equation in bulk}. For example, the first term is rewritten as
\begin{eqnarray}
& & \frac{1}{k_{B} T A_{xy}} \int_0^{\tau_{\rm meso}} \hspace{-0.15cm} ds \sum_{i=1}^N \big\langle \big(V_{A} - u^x(z_{i}(0)) \big)F^x_{A}(s) F^x_{A}(\bm{r}_{i}(0),0)\big\rangle^{V_{A},V_{B}} \nonumber \\[3pt]
&\simeq& \Big(V_{A} - u^x(z_A^{\rm 1st}) \Big) \frac{1}{k_{B} T A_{xy}} \int_0^{\tau_{\rm meso}} \hspace{-0.15cm} ds \big\langle F^x_{A}(s) F^x_{A}(0) \big\rangle^{V_{A},V_{B}} .
\end{eqnarray}
In the same manner, we rewrite (\ref{eq:generalized Green-Kubo: local gibbs: FxA: linear response regime}) in the approximate form
\begin{eqnarray}
\frac{F_{A}^x(\tau_{\rm meso};V_{A},V_{B})}{A_{xy}} &=& \gamma_{AA}(\tau_{\rm meso}) \Big(V_{A} - u^x(z_A^{\rm 1st}) \Big)  + \gamma_{AB}(\tau_{\rm meso}) \Big(V_{B} - u^x(z_B^{\rm 1st}) \Big) \nonumber \\[3pt]
&-& \gamma_{AJ}(\tau_{\rm meso}) \frac{d u^x}{d z} \Big|_{z=0}
\label{eq:generalized Green-Kubo: local gibbs: FxA: linear response regime at surface A}
\end{eqnarray}
with
\begin{eqnarray}
\gamma_{\alpha \beta}(\tau) =\frac{1}{k_{B} T A_{xy}} \int_0^{\tau} ds \big\langle F^x_{\alpha}(s) F^x_{\beta}(0) \big\rangle_{\rm eq},
\label{eq:def of gamma_ab}
\end{eqnarray}
\begin{eqnarray}
\gamma_{\alpha J}(\tau) = \frac{1}{k_{B}  T A_{xy} }\int d^3\bm{r}' \int_0^{\tau} ds \big\langle F^x_{\alpha}(s) J^{xz}(\bm{r}',0) \big\rangle_{\rm eq},
\end{eqnarray}
where $\alpha,\beta=A,B$. As the distance between walls $A$ and $B$ is of order $\xi_{\rm macro}$, we have
\begin{eqnarray}
 \gamma_{AB}(\tau_{\rm meso}) \simeq 0.
\end{eqnarray}
Then, (\ref{eq:generalized Green-Kubo: local gibbs: FxA: linear response regime at surface A}) is rewritten as
\begin{eqnarray}
\frac{F_{A}^x(\tau_{\rm meso};V_{A},V_{B})}{A_{xy}}  \simeq - \gamma_{AA}(\tau_{\rm meso}) \Big(u^x(z_{A}^{\rm 1st})-V_{A} \Big) - \gamma_{AJ}(\tau_{\rm meso}) \frac{d u^x}{d z} \Big|_{z=0} .
\label{eq:generalized Green-Kubo: local gibbs: FxA: linear response regime at surface A final form}
\end{eqnarray}
Similarly, by focusing on the force acting on the fluid from wall $B$, we have 
\begin{eqnarray}
\frac{F_{B}^x(\tau_{\rm meso};V_{A},V_{B})}{A_{xy}}  \simeq \gamma_{BB}(\tau_{\rm meso}) \Big(V_{B} -u^x(z_{B}^{\rm 1st}) \Big) - \gamma_{BJ}(\tau_{\rm meso}) \frac{d u^x}{d z} \Big|_{z=L} .
\label{eq:generalized Green-Kubo: local gibbs: FxB: linear response regime at surface B final form}
\end{eqnarray}

\subsection{Expression of the slip length}
\label{sec: expression of slip length: relaxation process}
To connect the constitutive equation in the bulk (\ref{eq:sigmaxz: linear response regime in bulk}) and at wall $A$ (\ref{eq:generalized Green-Kubo: local gibbs: FxA: linear response regime at surface A final form}), we assume the force-balance at time $\tau_{\rm meso}$ near the wall,
\begin{eqnarray}
-\frac{1}{A_{xy}}\int_{z=\xi_{\rm meso}} \hspace{-0.3cm} dxdy \sigma^{xz}(\bm{r},(\tau_{\rm meso});V_{A},V_{B}) \simeq \frac{F^x_{A}(\tau_{\rm meso};V_{A},V_{B})}{A_{xy}}.
\label{eq:the force-balance at time meso near the wall}
\end{eqnarray}
Substituting (\ref{eq:sigmaxz: linear response regime in bulk}) and (\ref{eq:generalized Green-Kubo: local gibbs: FxA: linear response regime at surface A final form}) into (\ref{eq:the force-balance at time meso near the wall}), we rewrite (\ref{eq:the force-balance at time meso near the wall}) in terms of $u^x(z)$ as
\begin{eqnarray}
- \eta \frac{d u^x(z)}{d z} \Big|_{z=\xi_{\rm meso}} \simeq - \gamma_{AA}(\tau_{\rm meso}) \Big(u^x(z_A^{\rm 1st})-V_{A} \Big) - \gamma_{AJ}(\tau_{\rm meso}) \frac{d u^x}{d z} \Big|_{z=0} .
\end{eqnarray}
Furthermore, by expanding the left-hand side around $z=z_A^{\rm 1st}$, we obtain
\begin{eqnarray}
u^x(z_A^{\rm 1st})-V_{A} \simeq (d_{A}-z_{A}) \frac{d u^x}{d z} \Big|_{z=z_A^{\rm 1st}} + d_{A} (\xi_{\rm meso} -z_A^{\rm 1st}) \frac{d^2 u^x}{d z^2} \Big|_{z=z_A^{\rm 1st}} 
\label{eq:partial slip boundary condition: wall A: linear response theory mid}
\end{eqnarray}
with
\begin{eqnarray}
d_{A} = \frac{\eta}{\gamma_{AA}(\tau_{\rm meso})},
\label{eq:slip length A meso}
\end{eqnarray}
and
\begin{eqnarray}
z_{A} =  \frac{\gamma_{AJ}(\tau_{\rm meso})}{\gamma_{AA}(\tau_{\rm meso})} .
\label{eq:wall position A meso}
\end{eqnarray}
We now consider the order of each term of (\ref{eq:partial slip boundary condition: wall A: linear response theory mid}). From (\ref{eq:relation between epsilon and delta}) and (\ref{eq:two derivative of ux: epsilon}), the second term on the right-hand side is estimated as $O(\epsilon^{3/2})$. Because we are interested in the first order approximation in $\epsilon$, we may neglect this term. Then, (\ref{eq:partial slip boundary condition: wall A: linear response theory mid}) is rewritten as 
\begin{eqnarray}
u^x(z_A^{\rm 1st})-V_{A} = b^{\rm meso}_{A} \frac{d u^x}{d z} \Big|_{z=z_A^{\rm 1st}}
\label{eq:partial slip boundary condition: wall A: linear response theory 1}
\end{eqnarray}
with 
\begin{eqnarray}
b^{\rm meso}_{A} = d_{A} - z_{A},
\label{eq:slip length: meso}
\end{eqnarray}
which represents the partial slip boundary condition with slip length $b^{\rm meso}_A$. $d_{A}$ and $z_{A}$ are independent of a specific value of $\tau_{\rm meso}$ because of the separation of scales, e.g., (\ref{eq:behavior FF in tau}). This expression is the same as that proposed by Bocquet and Barrat~\cite{bocquet1994hydrodynamic}. We note that the assumption (\ref{eq:ux0-VA: epsilon}) leads to $b^{\rm meso}_{A} \simeq O(\epsilon^0)$, which implies that $d_A$ and $z_A$ remain microscopic,
\begin{eqnarray}
b_A^{\rm meso} \simeq d_A\simeq z_A \simeq \xi_{\rm micro}.
\label{eq:slip length is microscopic order}
\end{eqnarray}
Because the typical value of the slip length reported in recent experiments is about $10 {\rm nm}$~\cite{lauga2007microfluidics,cao2009molecular,bocquet2010nanofluidics}, the result (\ref{eq:slip length is microscopic order}) is reasonable.

We may rewrite (\ref{eq:partial slip boundary condition: wall A: linear response theory 1}) in the form
\begin{eqnarray}
u^x(z_{A}+z_A^{\rm 1st}) - V_{A} = d_{A}\frac{d u^x}{d z} \Big|_{z=z_A+z_A^{\rm 1st}}.
\label{eq:partial slip boundary condition: wall A: linear response theory 2}
\end{eqnarray}
In Ref.~\cite{bocquet1994hydrodynamic}, taking the expression (\ref{eq:partial slip boundary condition: wall A: linear response theory 2}) into account, Bocquet and Barrat called $z_A+z_A^{\rm 1st}$ the ``hydrodynamic position of the wall" and defined $d_A$ as the slip length. However, we shall define the slip length as the length measured from the first epitaxial layer near wall $A$, $z=z_{A}^{\rm 1st}$. That is, we refer to $b^{\rm meso}_A$ as the slip length.

\section{Linear response theory for the  steady state}\label{sec:Linear response theory for the  steady state}
We next develop the linear response theory that describes the macroscopic behaviors at time $\tau=\tau_{\rm ss}$. The fluid at time $\tau=\tau_{\rm ss}$ is assumed to be in the non-equilibrium steady state. We note that the steady state is independent of the velocity $u^x(z)$ given as the initial condition. Instead, the steady state is characterized only by the wall velocities $V_A$ and $V_B$. Then, we focus on the linear response regime for $V_A$ and $V_B$.

In this section, calculations are performed with quantities retaining their dimensional units.

\subsection{Linear response regime}
First, we consider the momentum density field $\pi^x(\bm{r},\tau_{\rm ss};V_{A},V_{B})$ in the steady state. Substituting $\hat{f}(\Gamma) =\hat{\pi}^x(\bm{r};\Gamma)$ and $\tau=\tau_{\rm ss}$ into (\ref{eq:generalized Green-Kubo: local gibbs}) yields
\begin{eqnarray}
\pi^x(\bm{r},\tau_{\rm ss};V_{A},V_{B}) &=& \frac{1}{k_{B} T} \int_0^{\tau_{\rm ss}} \hspace{-0.15cm} ds \sum_{i=1}^N \big\langle \big(V_{A} - u^x(z_i(0)) \big) \pi^x(\bm{r},s) F^x_{A}(\bm{r}_{i}(0),0)\big\rangle^{V_{A},V_{B}} \nonumber \\[3pt]
&+& \frac{1}{k_{B} T} \int_0^{\tau_{\rm ss}} ds \sum_{i=1}^N \big\langle \big(V_{B} - u^x(z_{i}(0)) \big) \pi^x(\bm{r},s) F^x_{B}(\bm{r}_{i}(0),0)\big\rangle^{V_{A},V_{B}}  \nonumber \\[3pt]
&-& \frac{1}{k_{B} T} \int d^3\bm{r}' \int_0^{\tau_{\rm ss}} ds \frac{d u^x(z')}{d z'} \big\langle \pi^x(\bm{r},s) J^{xz}(\bm{r}',0) \big\rangle^{V_{A},V_{B}}.
\label{eq:generalized Green-Kubo: local gibbs: pix}
\end{eqnarray}
As the steady state is independent of the initial condition, the right-hand side of (\ref{eq:generalized Green-Kubo: local gibbs: pix}) is independent of $u^x(z)$. Then, by using the initial condition with $u^x(z)=0$, we simplify (\ref{eq:generalized Green-Kubo: local gibbs: pix}) as 
\begin{eqnarray}
\pi^x(\bm{r},\tau_{\rm ss};V_{A},V_{B}) = \frac{V_{A} }{k_{B} T} \int_0^{\tau_{\rm ss}} \hspace{-0.15cm} ds \big\langle \pi^x(\bm{r},s) F^x_{A}(0)\big\rangle^{V_{A},V_{B}} + \frac{V_ {B}}{k_{B} T} \int_0^{\tau_{\rm ss}} \hspace{-0.15cm} ds \big\langle \pi^x(\bm{r},s) F^x_{B}(0)\big\rangle^{V_{A},V_{B}}  .
\label{eq:generalized Green-Kubo: local gibbs: pix: simplify}
\end{eqnarray}
We apply a linear approximation to (\ref{eq:generalized Green-Kubo: local gibbs: pix: simplify}) in $V_{A}$ and $V_{B}$. We assume that $\langle \hat{\pi}^x(\bm{r},\tau_{\rm ss}) \rangle^{V_{A},V_{B}}$ may be expanded in $V_{A}$ and $V_B$, which leads to the expansion of the right-hand side of (\ref{eq:generalized Green-Kubo: local gibbs: pix: simplify}) in $V_A$ and $V_B$. Within the linear response regime, (\ref{eq:generalized Green-Kubo: local gibbs: pix: simplify}) may be rewritten as
\begin{eqnarray}
\pi^x(\bm{r},\tau_{\rm ss};V_A,V_B)  &=& \gamma_{PA}(\bm{r},\tau_{\rm ss}) V_A + \gamma_{PB}(\bm{r},\tau_{\rm ss}) V_B
\label{eq:generalized Green-Kubo: local gibbs: pix: linear response regime}
\end{eqnarray}
with
\begin{eqnarray}
\gamma_{P\alpha}(\bm{r},\tau_{\rm ss}) =  \frac{1}{k_B T} \int_0^{\tau_{\rm ss}} \hspace{-0.15cm} ds \big\langle \pi^x(\bm{r},s) F^x_{\alpha}(0)\big\rangle_{\rm eq},
\label{eq:gamma Palpha} 
\end{eqnarray}
where $\alpha=A,B$.

We can obtain linear approximations of other quantities by a similar procedure. In particular, we employ in the reminder of this section the following approximate expressions:
\begin{eqnarray}
\frac{F_A^x(\tau_{\rm ss};V_A,V_B)}{A_{xy}}  &=& \gamma_{AA}(\tau_{\rm ss}) V_A + \gamma_{AB}(\tau_{\rm ss})V_B,
\label{eq:generalized Green-Kubo: local gibbs: FxA: linear response regime: ss}
\end{eqnarray}
and
\begin{eqnarray}
\frac{F_B^x(\tau_{\rm ss};V_A,V_B)}{A_{xy}}  &=& \gamma_{BA}(\tau_{\rm ss}) V_A + \gamma_{BB}(\tau_{\rm ss}) V_B,
\label{eq:generalized Green-Kubo: local gibbs: FxB: linear response regime: ss}
\end{eqnarray}
where $\gamma_{\alpha \beta}(\tau)$ $(\alpha={\rm A,B})$ is defined by (\ref{eq:def of gamma_ab}).

\subsection{Force balance in steady state}
Because of the nature of the steady state, the microscopic balance equation for the momentum density, (\ref{eq:equation of continuity x,y,z}), leads to
\begin{eqnarray}
\frac{1}{A_{xy}}\int_{z} dxdy \sigma^{xz}(\bm{r},\tau_{\rm ss};V_A,V_B) = -\frac{F_A^x(\tau_{\rm ss};V_A,V_B)}{A_{xy}} = \frac{F_B^x(\tau_{\rm ss};V_A,V_B)}{A_{xy}}
\label{eq:force-balance equation: ss}
\end{eqnarray}
for $\xi_{\rm micro} \ll z \ll L-\xi_{\rm micro}$. This is physically reasonable, but in this subsection, we derive (\ref{eq:force-balance equation: ss}) from the microscopic description.

First, we extend the microscopic momentum current $\hat{J}^{ab}(\bm{r};\Gamma)$ to the system with walls. Using (\ref{eq:def of FA}) and (\ref{eq:def of D delta}), we rewrite the first term on the right-hand side of (\ref{eq:equation of continuity x,y,z}) as 
\begin{eqnarray}
\sum_{i=1}^N \hat{F}^a_A(\bm{r}_i;\Gamma^A) \delta(\bm{r}-\bm{r}_i) 
= -\frac{\partial \hat{J}^{ab}_A(\bm{r};\Gamma,\Gamma^A)}{\partial r^a} - \sum_{i=1}^N \sum_{j=1}^{N_A} \frac{\partial U_A(\bm{r}-\bm{q}^A_j)}{\partial r^a} \Big|_{\bm{r}=\bm{r}_i}\delta(\bm{r}-\bm{q}^A_j) 
\label{eq:rewrite microscopic current wall A}
\end{eqnarray}
with
\begin{eqnarray}
\hat{J}^{ab}_A(\bm{r};\Gamma,\Gamma^A) = - \sum_{i=1}^N \sum_{j=1}^{N_A} \frac{\partial U_A(\bm{r}-\bm{q}^A_j)}{\partial r^a} \Big|_{\bm{r}=\bm{r}_i}(r^b_i-q^{Ab}_j) D(\bm{r};\bm{r}_i,\bm{q}^A_j )
\label{eq:def of microscopic current wall A}.
\end{eqnarray}
Similarly, we introduce $\hat{J}^{ab}_B(\bm{r};\Gamma,\Gamma^B)$ and rewrite the second term on the right-hand side of (\ref{eq:equation of continuity x,y,z}). Substituting these equations into (\ref{eq:equation of continuity x,y,z}), we obtain
\begin{eqnarray}
& & \frac{\partial \hat{\pi}^x(\bm{r};\Gamma_t)}{\partial t} + \frac{\partial \hat{J}^{xb}(\bm{r};\Gamma_t)}{\partial r^b} + \frac{\partial \hat{J}_A^{xb}(\bm{r};\Gamma_t)}{\partial r^b} + \frac{\partial \hat{J}_B^{xb}(\bm{r};\Gamma_t)}{\partial r^b} \nonumber \\[3pt]
&=& - \sum_{i=1}^N \sum_{j=1}^{N_A} \frac{\partial U_A(\bm{r}-\bm{q}^A_j)}{\partial r^a} \Big|_{\bm{r}=\bm{r}_i(t)}\delta(\bm{r}-\bm{q}^A_j) - \sum_{i=1}^N \sum_{j=1}^{N_B} \frac{\partial U_B(\bm{r}-\bm{q}^B_j)}{\partial r^a} \Big|_{\bm{r}=\bm{r}_i(t)}\delta(\bm{r}-\bm{q}^B_j) .
\label{eq:microscopic force-balance equation: deform}
\end{eqnarray}
As our concern is the fluid in the region $0<z<L$, we may neglect the right-hand side of (\ref{eq:microscopic force-balance equation: deform}). Equation (\ref{eq:microscopic force-balance equation: deform}) implies that $\hat{J}^{ab}_A(\bm{r};\Gamma,\Gamma^A)$ and $\hat{J}^{ab}_B(\bm{r};\Gamma,\Gamma^B)$ correspond to the microscopic momentum current induced by walls A and B, respectively. 

We now focus on the force balance equation in the steady state. Because $\langle \hat{\pi}^x(\bm{r},t) \rangle^{V_A,V_B}$ is independent of time $t$ in the steady state, (\ref{eq:microscopic force-balance equation: deform}) is rewritten as
\begin{eqnarray}
\frac{\partial}{\partial r^b} \Big( J^{xb}(\bm{r},\tau_{\rm ss};V_A,V_B) +  J_A^{xb}(\bm{r},\tau_{\rm ss};V_A,V_B) +  J_B^{xb}(\bm{r},\tau_{\rm ss};V_A,V_B) \Big) = 0
\label{eq:ensemble average of the microscopic force-balance equation}
\end{eqnarray}
for $0<z<L$, where $J^{ab}_{\alpha}(\bm{r},t;V_A,V_B)$ ($\alpha=A,B$) is defined by
\begin{eqnarray}
J^{ab}_{\alpha}(\bm{r},t;V_A,V_B) = \langle \hat{J}^{ab}_{\alpha}(\bm{r},t)\rangle^{V_A,V_B}.
\label{eq:def of JAab}
\end{eqnarray}
Spatially averaging (\ref{eq:ensemble average of the microscopic force-balance equation}) in the $xy$ plane at a position $z$ gives
\begin{eqnarray}
\frac{1}{A_{xy}}\int_{z} dxdy \Big( J^{xz}(\bm{r},\tau_{\rm ss};V_A,V_B) +  J_A^{xz}(\bm{r},\tau_{\rm ss};V_A,V_B) +  J_B^{xz}(\bm{r},\tau_{\rm ss};V_A,V_B) \Big) = {\rm const}
\label{eq:macroscopic force-balance equation}
\end{eqnarray}
in $z$, where we have used  the periodic boundary conditions along the $x$ and $y$ axes. 

By definition (\ref{eq:def of microscopic current wall A}), $\hat{J}^{ab}_A(\bm{r};\Gamma,\Gamma^A)$ is zero in the region $z\gg \xi_{\rm micro}$ where the potential force of wall $A$ cannot reach. Similarly, $\hat{J}^{ab}_B(\bm{r};\Gamma,\Gamma^A)$ is zero in the region $z\ll L-\xi_{\rm micro}$. Using these properties, (\ref{eq:macroscopic force-balance equation}) is rewritten as
\begin{eqnarray}
\frac{1}{A_{xy}}\int_{z} dxdy J^{xz}(\bm{r},\tau_{\rm ss};V_A,V_B)  = {\rm const}
\label{eq:macroscopic force-balance equation:bulk}
\end{eqnarray}
in $\xi_{\rm micro} \ll z \ll L -\xi_{\rm micro}$.
Next, we introduce $\xi_{A}$ so that $0 < \xi_{A} \ll \xi_{\rm micro}$ holds. Because fluid particles are seldom present in the thin layer $0<z<\xi_{A}$ due to the repulsive force of wall $A$, it follows that
\begin{eqnarray}
\hat{J}^{ab}(\bm{r};\Gamma) \simeq 0
\label{eq:101}
\end{eqnarray}
in this layer $0<z<\xi_{A}$. Using Gauss's divergence theorem and relation (\ref{eq:def of microscopic current wall A}), we have
\begin{eqnarray}
& & \frac{1}{A_{xy}}\int_{z=\xi_{A}} \hspace{-0.3cm} dxdy \hat{J}^{xz}_A(\bm{r};\Gamma,\Gamma^A) \nonumber \\[3pt]
&=& \frac{1}{A_{xy}}\int_{z\leq \xi_{A}} \hspace{-0.3cm} dxdydz \frac{\partial}{\partial z} \hat{J}^{xz}_A(\bm{r};\Gamma,\Gamma^A) \nonumber \\[3pt]
&=& - \frac{1}{A_{xy}}\int_{z\leq \xi_{A}} \hspace{-0.3cm} dxdydz \Big(\sum_{i=1}^N \hat{F}^a_A(\bm{r}_i;\Gamma,\Gamma^A) \delta(\bm{r}-\bm{r}_i) + \sum_{i=1}^N \sum_{j=1}^{N_A} \frac{\partial U_A(\bm{r}-\bm{q}^A_j)}{\partial r^a} \Big|_{\bm{r}=\bm{r}_i}\delta(\bm{r}-\bm{q}^A_j) \Big).
\label{eq:surface integral JA to FA mid}
\end{eqnarray}
Similarly, as particles are seldom present in the region $z<\xi_{A}$, the first term of (\ref{eq:surface integral JA to FA mid}) is zero. Therefore, we obtain
\begin{eqnarray}
\frac{1}{A_{xy}}\int_{z=\xi_{A}} \hspace{-0.3cm} dxdy \hat{J}^{xz}_A(\bm{r};\Gamma,\Gamma^A) \simeq \frac{\hat{F}_A^x(\Gamma_t;\Gamma^A_t)}{A_{xy}},
\label{eq:surface integral JA to FA}
\end{eqnarray}
where we have used (\ref{eq:def of microscopic force acting on the fluid}). From (\ref{eq:macroscopic force-balance equation}), (\ref{eq:macroscopic force-balance equation:bulk}), (\ref{eq:101}), and (\ref{eq:surface integral JA to FA}), we find
\begin{eqnarray}
J^{xz}(z,\tau_{\rm ss};V_A,V_B) \simeq \frac{F_A^x(\tau_{\rm ss};V_A,V_B)}{A_{xy}} 
\end{eqnarray}
in $\xi_{\rm micro} \ll z \ll L -\xi_{\rm micro}$. Finally, taking into account that  $v^z(z,\tau_{\rm ss};V_A,V_B)=0$ in the steady state, we obtain the first equality of (\ref{eq:force-balance equation: ss}). By a similar procedure for wall $B$, we obtain the second equality of (\ref{eq:force-balance equation: ss}).

\subsection{Galilean invariance of steady state}
The transport coefficients in (\ref{eq:generalized Green-Kubo: local gibbs: pix: linear response regime}), (\ref{eq:generalized Green-Kubo: local gibbs: FxA: linear response regime: ss}) and (\ref{eq:generalized Green-Kubo: local gibbs: FxB: linear response regime: ss}) are dependent on each other because of the Galilean invariance of the steady state. In this subsection, we derive relations between these transport coefficients. 

Let the probability density in the non-equilibrium steady state be $P_{\rm ss}\big((\bm{r}_i)_{i=1}^N,(\bm{p}_i)_{i=1}^N;V_A,V_B\big)$. The Galilean invariance in the steady state is expressed in terms of $P_{\rm ss}\big((\bm{r}_i)_{i=1}^N,(\bm{p}_i)_{i=1}^N;V_A,V_B\big)$ as
\begin{eqnarray}
P_{\rm ss}\big((\bm{r}_i)_{i=1}^N,(\bm{p}'_i)_{i=1}^N;V_A+V,V_B+V\big) = P_{\rm ss}\big((\bm{r}_i)_{i=1}^N,(\bm{p}_i)_{i=1}^N;V_A,V_B\big),
\label{eq:Galilean invariance in steady state}
\end{eqnarray} 
where $\bm{p}'_i=(p^x_i+V, p^y_i, p^z_i)$ for any $V$. As the microscopic force acting on the fluid from wall $A$, (\ref{eq:def of microscopic force acting on the fluid}), is independent of the momentum of the particles composing the fluid, $F^x_A(\tau_{\rm ss};V_A+V,V_B+V)$ is independent of $V$. In the linear response regime, this implies that $F^x_A(\tau_{\rm ss};V_A,V_B)$ is proportional to the relative velocity between the walls, $V_A-V_B$. Using this property and (\ref{eq:generalized Green-Kubo: local gibbs: FxA: linear response regime: ss}), we find 
\begin{eqnarray}
\gamma_{AA}(\tau_{\rm ss}) + \gamma_{AB}(\tau_{\rm ss}) = 0.
\label{eq:relation AA and AB}
\end{eqnarray}
In the same manner, we obtain
\begin{eqnarray}
\gamma_{BA}(\tau_{\rm ss}) + \gamma_{BB}(\tau_{\rm ss}) = 0.
\label{eq:relation BA and BB}
\end{eqnarray}
Next, because the microscopic momentum density $\hat{\pi}^a(\bm{r};\Gamma)$ depends on the the momentum of the particles composing the fluid, $\langle \hat{\pi}^x(\bm{r},\tau_{\rm ss}) \rangle^{V_A+V,V_B+V}$ depends on $V$. Using (\ref{eq:def of microscopic momentum density}) and (\ref{eq:Galilean invariance in steady state}), we obtain
\begin{eqnarray}
\langle \hat{\pi}^x(\bm{r},\tau_{\rm ss}) \rangle^{V_A,V_B} &=& \int d \Gamma \sum_{i=1}^N p^x_i \delta\big(\bm{r}-\bm{r}_i\big) P_{\rm ss}\big((\bm{r}_i)_{i=1}^N,(\bm{p}_i)_{i=1}^N;V_A,V_B\big) \nonumber \\[3pt]
&=& \int d \Gamma' \sum_{i=1}^N \big( p'^{x}_i - V \big) \delta\big(\bm{r}-\bm{r}_i\big) P_{\rm ss}\big((\bm{r}_i)_{i=1}^N,(\bm{p}'_i)_{i=1}^N;V_A+V,V_B+V\big) \nonumber \\[3pt]
&=& \langle \hat{\pi}^x(\bm{r},\tau_{\rm ss}) \rangle^{V_A+V,V_B+V} - V \rho(\bm{r},\tau_{\rm ss};V_A+V,V_B+V) .
\label{eq:Galilean invariance of the momentum density}
\end{eqnarray}
In particular, by focusing on the linear response regime, (\ref{eq:Galilean invariance of the momentum density}) is rewritten as
\begin{eqnarray}
\langle \hat{\pi}^x(\bm{r},\tau_{\rm ss}) \rangle^{V_A+V,V_B+V} = \langle \hat{\pi}^x(\bm{r},\tau_{\rm ss}) \rangle^{V_A,V_B} + V \rho(\bm{r},\tau_{\rm ss};V_A,V_B) .
\label{eq:Galilean invariance of the momentum density: linear response regime}
\end{eqnarray}
From (\ref{eq:generalized Green-Kubo: local gibbs: pix: linear response regime}) and (\ref{eq:Galilean invariance of the momentum density: linear response regime}), we have
\begin{eqnarray}
\gamma_{PA}(\bm{r},\tau_{\rm ss}) + \gamma_{PB}(\bm{r},\tau_{\rm ss}) =  \rho(\bm{r},\tau_{\rm ss};0,0).
\label{eq:relation PA and PB}
\end{eqnarray}
Finally, by applying the Onsager reciprocity theorem to (\ref{eq:generalized Green-Kubo: local gibbs: FxA: linear response regime: ss}) and (\ref{eq:generalized Green-Kubo: local gibbs: FxB: linear response regime: ss}), we obtain
\begin{eqnarray}
\gamma_{AB}(\tau_{\rm ss}) = \gamma_{BA}(\tau_{\rm ss}).
\label{eq:relation Onsager reciprocity theorem}
\end{eqnarray}

In summary, we find that the linear response relations (\ref{eq:generalized Green-Kubo: local gibbs: pix: linear response regime}), (\ref{eq:generalized Green-Kubo: local gibbs: FxA: linear response regime: ss}) and (\ref{eq:generalized Green-Kubo: local gibbs: FxB: linear response regime: ss}) may be expressed using only two dependent transport coefficients.


\subsection{Expression of the slip length}
\label{sec:expression of the slip length steady state}
The linear response theory derived in this section is consistent with the standard hydrodynamics. Then, we assume that the macroscopic behavior of the fluid in the bulk is described as an incompressible Newtonian liquid. That is, we have
\begin{eqnarray}
\rho(\bm{r},t;V_A,V_B) = \rho,
\label{eq:incompressible condition in bulk: assumption}
\end{eqnarray}
and
\begin{eqnarray}
\sigma^{xz}(\bm{r},t;V_A,V_B) = \eta \frac{\partial v^x(\bm{r},t;V_A,V_B)}{\partial z}
\label{eq:constitutive equation in bulk: assumption}
\end{eqnarray}
in the bulk, where $\rho$ is a constant value and $\eta$ is known. In this subsection, we derive the most general boundary condition under this assumption.

From (\ref{eq:force-balance equation: ss}) and (\ref{eq:constitutive equation in bulk: assumption}), the velocity field in the steady state is expressed as
\begin{eqnarray}
\frac{1}{A_{xy}}\int_{z} dxdy v^x(\bm{r},\tau_{\rm ss};V_A,V_B) = - \frac{F_A^x(\tau_{\rm ss};V_A,V_B)}{\eta A_{xy}}(z-z_0) + \frac{1}{A_{xy}}\int_{z_0} dxdy v^x(\bm{r},\tau_{\rm ss};V_A,V_B),
\label{eq:velocity fields in the steady state}
\end{eqnarray}
in the bulk, where $z_0$ is any position in the bulk. Here, it is plausible that the velocity fields in the steady state, $v^x(\bm{r},\tau_{\rm ss};V_A,V_B)$, are independent of $x,y$. Then, we focus on the $z$-dependence of $v^x(\bm{r},\tau_{\rm ss};V_A,V_B)$ and omit the $x,y$-dependence. By substituting the linear response relations (\ref{eq:generalized Green-Kubo: local gibbs: pix: linear response regime}), (\ref{eq:generalized Green-Kubo: local gibbs: FxA: linear response regime: ss}) and (\ref{eq:generalized Green-Kubo: local gibbs: FxB: linear response regime: ss}) into (\ref{eq:velocity fields in the steady state}), we obtain
\begin{eqnarray}
v^x(z,\tau_{\rm ss};V_A,V_B) = - \frac{\gamma_{AA}(\tau_{\rm ss}) V_A + \gamma_{AB}(\tau_{\rm ss})V_B}{\eta}(z-z_0) + \frac{\gamma_{PA}(z_0,\tau_{\rm ss}) V_A + \gamma_{PB}(z_0,\tau_{\rm ss}) V_B}{\rho}.
\label{eq:vx steady state intermediate expression}
\end{eqnarray}
Using (\ref{eq:relation AA and AB}), (\ref{eq:relation BA and BB}), (\ref{eq:relation PA and PB}) and (\ref{eq:relation Onsager reciprocity theorem}), (\ref{eq:vx steady state intermediate expression}) is expressed in terms of two transport coefficients as
\begin{eqnarray}
\hspace{-0.15cm} v^x(z,\tau_{\rm ss};V_A,V_B) = - \frac{\gamma_{AA}(\tau_{\rm ss})}{\eta}(V_A-V_B) z + \Big( - \frac{\gamma_{PB}(z_0,\tau_{\rm ss}) }{\rho} +\frac{\gamma_{AA}(\tau_{\rm ss})}{\eta} z_0 \Big) (V_A - V_B) + V_A.
\label{eq:vx steady state expression}
\end{eqnarray}
Here, we extrapolate the velocity field in the bulk, (\ref{eq:vx steady state expression}), to the whole region. Then, we obtain the boundary condition for the extrapolated velocity field:
\begin{eqnarray}
v_x(z=0,\tau_{\rm ss};V_A,V_B) - V_A = b^{\rm ss}_A \frac{\partial v_x(z,\tau_{\rm ss};V_A,V_B)}{\partial z} \Big|_{z=0}
\label{eq:partial slip boundary condition: steady state expression A}
\end{eqnarray}
with
\begin{eqnarray}
b^{\rm ss}_A =  \frac{\eta \gamma_{PB}(z_0,\tau_{\rm ss}) }{\rho \gamma_{AA}(\tau_{\rm ss})} - z_0 .
\label{eq:slip length: steady state expression A}
\end{eqnarray}
As long as we are interested in the velocity field in the bulk, we may use the boundary condition for the extrapolated velocity field (\ref{eq:partial slip boundary condition: steady state expression A}). Because (\ref{eq:slip length: steady state expression A}) is independent of $v_x(z=0,\tau_{\rm ss};V_A,V_B)$, $V_A$, and $V_B$, we identify (\ref{eq:partial slip boundary condition: steady state expression A}) as the partial slip boundary condition with constant slip length $b^{\rm ss}_A$. Therefore, we confirm that the partial slip boundary condition is the most general boundary condition consistent with the linear response theory in this section.

We here show that the slip length $b^{\rm ss}_A$ is independent of $z_0$. Note that $z_0$ in (\ref{eq:slip length: steady state expression A}) comes from the reference point for the velocity field $v^x(z,\tau_{\rm ss};V_A,V_B)$, (\ref{eq:velocity fields in the steady state}). Because $v^x(z,\tau_{\rm ss};V_A,V_B)$ is independent of the reference point, we find that the slip length $b^{\rm ss}_A$ is also independent of $z_0$:
\begin{eqnarray}
\frac{db^{\rm ss}_A}{dz_0} = 0.
\end{eqnarray}

Furthermore, by adding the assumption that walls $A$ and $B$ have the same microscopic structure, we can obtain more simple expression of the slip length. Substituting $z=L/2$ into (\ref{eq:relation PA and PB}) leads to
\begin{eqnarray}
\gamma_{PA}(L/2,\tau_{\rm ss}) = \gamma_{PB}(L/2,\tau_{\rm ss})  = \frac{\rho}{2}.
\end{eqnarray}
Then, by using this relation, the expression of the slip length (\ref{eq:slip length: steady state expression A}) is rewritten as
\begin{eqnarray}
b_A^{\rm ss} = \frac{\eta}{2\gamma_{AA}(\tau_{\rm ss})} - \frac{L}{2}.
\label{eq:slip length steady state modified A}
\end{eqnarray}

Finally, we stress that $b_A^{\rm ss}$ is defined as the slip length measured from the true wall position. By recalling that $b_A^{\rm meso}$ is measured from the first epitaxial layer $z=z_A^{\rm 1st}$, we find that these expressions are related to the relation
\begin{eqnarray}
b_A^{\rm meso} = b_A^{\rm ss} + z_A^{\rm 1st}.
\end{eqnarray}

\section{Model: Linearized Fluctuating Hydrodynamics}\label{sec:model: linearized fluctuating hydrodynamics}
In the previous sections, we employed the microscopic Hamiltonian particle system and derived the expressions of the slip length. Our key assumptions are: i) the correlation functions of the momentum fluxes behave as (\ref{eq:behavior FF in tau}); and ii) the length and time scales are separated. In the reminder of this paper, we employ the linearized fluctuating hydrodynamics instead of the microscopic particle system. Assuming that the macroscopic behavior of the fluid is described by the linearized fluctuating hydrodynamics, we demonstrate the validity of results obtained in the previous sections. In this section, we explain the details of our model.

\subsection{Linearized fluctuating hydrodynamics}
We study a linearized fluctuating hydrodynamics with the partial slip boundary condition. The geometry of the system is the same as that studied in Sect.~\ref{sec:Model: Hamiltonian Particle System}. The system is defined in the region $[0,\mathcal{L}_x]\times[0,\mathcal{L}_y]\times[0,\mathcal{L}]$.  Let $\tilde{v}^a(\bm{r},t)$ be a fluctuating velocity field in an incompressible fluid. The time evolution of $\tilde{v}^a(\bm{r},t)$ is described by the incompressible Navier--Stokes equation with stochastic fluxes,
\begin{eqnarray}
\rho \frac{\partial \tilde{v}^a}{\partial t} + \frac{\partial \tilde{J}^{ab}}{\partial r^b} = 0,
\label{eq:Navier-Stokes equation subjected to a Gaussian random stress tensor}
\end{eqnarray}
where twice repeated indices are assumed to be summed over. The momentum flux tensor $\tilde{J}^{ab}(\bm{r},t)$ is given by
\begin{eqnarray}
\tilde{J}^{ab} =  \tilde{p} \delta_{ab} - \eta \Big(\frac{\partial \tilde{v}^a}{\partial r^b}  + \frac{\partial \tilde{v}^b}{\partial r^a}\Big) + \tilde{s}^{ab},
\label{eq:momentum flux tensor}
\end{eqnarray}
where $\tilde{s}^{ab}(\bm{r},t)$ is the Gaussian random stress tensor satisfying
\begin{eqnarray}
\big\langle \tilde{s}^{ab}(\bm{r},t) \tilde{s}^{cd}(\bm{r}',t') \big\rangle = 2 k_B T \eta \Big[\delta_{ac}\delta_{bd} + \delta_{ad}\delta_{bc} - \frac{2}{3} \delta_{ab} \delta_{cd} \Big] \delta^3(\bm{r}-\bm{r}') \delta(t-t').
\label{eq:random stress tensor}
\end{eqnarray}
The pressure term in (\ref{eq:momentum flux tensor}) is used to enforce the incompressibility condition,
\begin{eqnarray}
\frac{\partial \tilde{v}^a(\bm{r},t)}{\partial r^a} = 0.
\label{eq:divergence free condition}
\end{eqnarray}
In this study, we ignore the nonlinear effect induced by the advection term. We impose periodic boundary conditions along the $x$ and $y$ axes and the partial slip boundary condition with slip length $\mathcal{B}_A$ and $\mathcal{B}_B$ at $z=0$ and $z=\mathcal{L}$ respectively, specifically, 
\begin{eqnarray}
\tilde{v}^x(\bm{r}) \Big|_{z=0} - V_A = \mathcal{B}_A \frac{\partial \tilde{v}^x(\bm{r})}{\partial z} \Big|_{z=0},
\label{eq:partial slip boundary condition z=0}
\end{eqnarray}
and 
\begin{eqnarray}
V_B - \tilde{v}^x(\bm{r}) \Big|_{z=\mathcal{L}} = \mathcal{B}_B \frac{\partial \tilde{v}^x(\bm{r})}{\partial z} \Big|_{z=\mathcal{L}},
\label{eq:partial slip boundary condition z=L}
\end{eqnarray}
where $V_A$ and $V_B$ are the wall velocities. In this section, we focus on stationary walls. That is, we set to $V_A=V_B=0$.

In the calculations using the fluctuating hydrodynamics, we convert all the quantities to the dimensionless form by setting $\mathcal{L}=\tau_{\rm marco}=k_B T = 1$ in order to measure all the quantities on the macroscopic scale. Here, $\tau_{\rm macro}$ is given by
\begin{eqnarray}
\tau_{\rm macro} = \frac{\mathcal{L}^2 \rho}{\eta},
\end{eqnarray}
which represents the relaxation time of the incompressible fluid. We note that the units in this section are different from those in Sec.~\ref{sec:Linear response theory for the relaxation process of fluid}, where all quantities are measured on the microscopic scale.

We remark on the ultraviolet cutoff of the fluctuating hydrodynamics. The ultraviolet cutoff is given by $(\xi_{\rm uv},\tau_{\rm uv})$. In particular, $\tau_{\rm uv}$ in the hydrodynamic description is assumed to be
\begin{eqnarray}
\tau_{\rm uv} = \frac{\xi_{\rm uv}^2 \rho}{\eta}.
\end{eqnarray}
Using an ultraviolet cutoff, we rewrite the properties of the stochastic flux, (\ref{eq:random stress tensor}), in a more precise form
\begin{eqnarray}
\big\langle \tilde{s}^{ab}(\bm{r},t) \tilde{s}^{cd}(\bm{r}',t') \big\rangle = 2 k_B T \eta \Big[\delta_{ac}\delta_{bd} + \delta_{ad}\delta_{bc} - \frac{2}{3} \delta_{ab} \delta_{cd} \Big] \mathcal{D}_3(|\bm{r}-\bm{r}'|) \mathcal{T}(|t-t'|).
\label{eq:stochastic properties of flux: full dimensionless form}
\end{eqnarray}
where $\mathcal{D}_3(d)$ and $\mathcal{T}(s)$ are given by
\begin{eqnarray}
\mathcal{D}_3(d)
\begin{cases}
= (4 \pi \xi_{\rm uv}^3/3)^{-1}, & {\rm for}  \ \ d < \xi_{\rm uv}, \\
= 0, & {\rm for} \ \ d \geq \xi_{\rm uv},
\end{cases}
\label{eq:def of D}
\end{eqnarray}
and
\begin{eqnarray}
\mathcal{T}(s)
\begin{cases}
= 1/2\tau_{\rm uv}, & {\rm for} \ \ s < \tau_{\rm uv}, \\
= 0, & {\rm for} \ \ s \geq \tau_{\rm uv}.
\end{cases}
\label{eq:def of T}
\end{eqnarray}
The functions $\mathcal{D}_3(d)$ and $\mathcal{T}(s)$ correspond to regularized functions of $\delta^3(\bm{r})$ and $\delta(t)$. 

We assume that the cutoff length $\xi_{\rm uv}$ satisfies the condition $\xi_{\rm micro} \ll \xi_{\rm uv} \ll \xi_{\rm macro}$. Furthermore, we assume that the slip lengths $\mathcal{B}_A$ and $\mathcal{B}_B$ are $O(\xi_{\rm micro})$. Through later discussions, this assumption is found to be inevitable when we compare the results obtained from the linearized fluctuating hydrodynamics with those obtained from the underlying microscopic theory.

\subsection{Calculated quantities}
\label{sec:Calculated quantities}
We focus on the time correlation functions of the forces that the fluid exerts on walls $A$ and $B$. For this purpose, we define $\tilde{\mathcal{F}}_A(t)$ as
\begin{eqnarray}
\tilde{\mathcal{F}}_A(t) &=& \int_{z=\xi_{\rm uv}} \hspace{-0.15cm} dx dy~\tilde{J}^{xz}(\bm{r},t)  \nonumber \\[3pt]
&=& - \eta \partial_{z} \tilde{\mathcal{V}}^{x}(\xi_{\rm uv},t) + \tilde{\mathcal{S}}^{xz}(\xi_{\rm uv},t) ,
\label{eq:explicit form of FA}
\end{eqnarray} 
where 
\begin{eqnarray}
\tilde{\mathcal{V}}^{x}(z,t) = \int_{z} dx dy~\tilde{v}^{x}(\bm{r},t),
\label{eq:def of bar(v)}
\end{eqnarray}
and
\begin{eqnarray}
\tilde{\mathcal{S}}^{ab}(z,t) = \int_{z} dx dy~\tilde{s}^{ab}(\bm{r},t).
\label{eq:def of bar(s)}
\end{eqnarray}
$\tilde{\mathcal{F}}_B(t)$ is also written as the similar form. Then, we define the time correlation functions:
\begin{eqnarray}
C_{\alpha \beta}(t,t') \equiv \Big\langle \tilde{\mathcal{F}}_{\alpha}(t)\tilde{\mathcal{F}}_{\beta}(t') \Big\rangle,
\label{eq:def of Caa}
\end{eqnarray}
where $\alpha,\beta = A,B$. We note that $\tilde{\mathcal{F}}_A(t)$ is defined by the surface integral at $z=\xi_{\rm uv}$, not $z=0$. Therefore, $\tilde{\mathcal{F}}_A(t)$ is not necessarily the force acting on wall $A$. However, when we focus on $C_{\alpha \beta}(t,0)$ in the time region $t\gg \tau_{\rm uv}$, we conjecture that  $C_{\alpha \beta}(t,0)$ is nearly equal to the time correlation function of the forces acting on walls by smoothing out the motion of the fluid occurring in the region of $\xi_{\rm uv}$ scale. Then, we introduce a time $\tau^f_{\rm meso}$ as the minimum time that $C_{\alpha \beta}(t,0)$ can be identified with the correlation function of the forces acting on walls. We assume that $\tau^f_{\rm meso}$ satisfies the condition:
\begin{eqnarray}
\tau_{\rm uv} \ll \tau^f_{\rm meso} \ll \tau_{\rm macro}.
\end{eqnarray}
Hereafter, we consider only the region $t\gtrsim \tau^f_{\rm meso}$. Also, in order to avoid a proliferation of symbols, we below set $\xi_{\rm uv}=+0$. For example, $\tilde{\mathcal{F}}_A(t)$ is written as
\begin{eqnarray}
\tilde{\mathcal{F}}_A(t) = \Big[- \eta \partial_{z} \tilde{\mathcal{V}}^{x}(+0,t) + \tilde{\mathcal{S}}^{xz}(+0,t) \Big].
\label{eq:explicit form of FA: dam}
\end{eqnarray} 

By the definition (\ref{eq:random stress tensor}), $\tilde{\mathcal{S}}^{xz}$-$\tilde{\mathcal{S}}^{xz}$ correlation function is calculated as
\begin{eqnarray}
\Big\langle \tilde{\mathcal{S}}^{xz}(z,t)\tilde{\mathcal{S}}^{xz}(z',t') \Big\rangle = 2 k_B T \eta A_{xy} \delta(z-z') \delta(t-t'),
\label{eq:random stress: bar}
\end{eqnarray}
where $\delta(z-z')$ is interpreted as the same form as (\ref{eq:def of D}).
Using the explicit form of $\tilde{\mathcal{F}}_A(t)$ given in (\ref{eq:explicit form of FA}), (\ref{eq:def of Caa}) is rewritten as the sum of four terms,
\begin{eqnarray}
C_{AA}(t,t') &=& \eta^2 \Big\langle \partial_{z} \tilde{\mathcal{V}}^{x}(+0,t) \partial_{z} \tilde{\mathcal{V}}^{x}(+0,t') \Big\rangle + \Big\langle \tilde{\mathcal{S}}^{xz}(+0,t) \tilde{\mathcal{S}}^{xz}(+0,t') \Big \rangle \nonumber \\[3pt]
&-& \eta \Big\langle \partial_{z} \tilde{\mathcal{V}}^{x}(+0,t)\tilde{\mathcal{S}}^{xz}(+0,t') \Big\rangle - \eta \Big\langle \partial_{z}\tilde{\mathcal{V}}^{x}(+0,t') \tilde{\mathcal{S}}^{xz}(+0,t) \Big\rangle .
\label{eq:CAA: real time domain}
\end{eqnarray}
For a function $f(t)$, we define its Fourier transform as
\begin{eqnarray}
f(\omega) = \int_{-\infty}^{\infty} dt e^{i\omega t} f(t),
\end{eqnarray}
where we use the notation $f(\omega)$ for the Fourier transform of $f(t)$. Due to the time translational invariance in equilibrium, $C_{AA}(t,t')$ is expressed in the form
\begin{eqnarray}
 C_{AA}(t,t') = \int \frac{d\omega}{2\pi} C_{AA}(\omega) e^{-i\omega (t-t')}
\end{eqnarray}
with
\begin{eqnarray}
C_{AA}(\omega) &=& \eta^2 \Big\langle \partial_{z} \tilde{\mathcal{V}}^{x}(+0,\omega) \partial_{z} \tilde{\mathcal{V}}^{x}(+0,-\omega) \Big\rangle + \Big\langle \tilde{\mathcal{S}}^{xz}(+0,\omega) \tilde{\mathcal{S}}^{xz}(+0,-\omega)  \Big\rangle \nonumber \\[3pt]
&-& \eta \Big\langle \partial_{z}\tilde{\mathcal{V}}^{x}(+0,\omega) \tilde{\mathcal{S}}^{xz}(+0,-\omega) \Big\rangle - \eta \Big\langle \partial_{z} \tilde{\mathcal{V}}^{x}(+0,-\omega)\tilde{\mathcal{S}}^{xz}(+0,\omega) \Big\rangle .
\label{eq:CAA: frequency domain}
\end{eqnarray}

\section{Explicit Form of $C_{AA}(\omega)$}\label{sec:Explicit Form of CAA}
The linearized fluctuating hydrodynamics describes the macroscopic motion of the fluid. Therefore, by calculating $C_{AA}(t,t')$ in the linearized fluctuating hydrodynamics, a part of (\ref{eq:behavior FF in tau}) can be replaced with more appropriate form. The explicit form of $C_{AA}(\omega)$ in $\omega \lesssim 2\pi/\tau^f_{\rm meso}$
%
%
is given by
\begin{eqnarray}
\frac{C_{AA}(\omega)}{2\eta k_B T A_{xy}} &=& \Big|\frac{q}{\Delta} \Big|^2 \Big[\big(\frac{1}{q_R} + \frac{|q|^2}{q_R} \mathcal{B}_B^2 + 4q_R\mathcal{B}_A\mathcal{B}_B \big)\sinh(2q_R \mathcal{L})  + 2\big(\mathcal{B}_A + \mathcal{B}_B + |q|^2\mathcal{B}_A \mathcal{B}_B^2\big) \cosh(2q_R\mathcal{L}) \nonumber \\[3pt]
&+& \big(\frac{1}{q_R} - \frac{|q|^2}{q_R} \mathcal{B}_B^2 - 4q_R\mathcal{B}_A\mathcal{B}_B \big) \sin(2q_R\mathcal{L}) + 2\big(\mathcal{B}_A + \mathcal{B}_B - |q|^2\mathcal{B}_A \mathcal{B}_B^2\big) \cos(2q_R\mathcal{L})
 \Big] \nonumber \\[3pt]
 &+& \delta(0) \Big|\frac{q}{\Delta} \Big|^2 \Big[4 q_R \mathcal{B}_A^2\mathcal{B}_B \sinh(2q_R \mathcal{L}) - 4 q_R \mathcal{B}_A^2\mathcal{B}_B \sin(2q_R \mathcal{L}) \nonumber \\[3pt]
 &+& 2 \mathcal{B}_A^2\big(1+|q|^2 \mathcal{B}_B^2 \big) \cosh(2q_R\mathcal{L}) + 2 \mathcal{B}_A^2\big(1-|q|^2 \mathcal{B}_B^2 \big) \cos(2q_R\mathcal{L}) + 4\mathcal{B}_B^2 \Big] 
\label{eq:CAA expression: perfect form}
\end{eqnarray}
with
\begin{eqnarray}
\Delta = (1+q\mathcal{B}_A)(1+q\mathcal{B}_B) e^{q\mathcal{L}} - (1-q\mathcal{B}_A)(1-q\mathcal{B}_B) e^{-q\mathcal{L}} ,
\end{eqnarray}
where $q$ is given by
\begin{eqnarray}
q = q_R - i q_R
\end{eqnarray}
with
\begin{eqnarray}
q_R = \sqrt{\frac{\omega \rho}{2 \eta}}.
\end{eqnarray}
In this section, we show the derivation of (\ref{eq:CAA expression: perfect form}). The method of calculation is essentially the same as that in Monahan et al.~\cite{monahan2016hydrodynamic}. They claimed that the cross correlations between the random stress tensor and other fluctuating fields vanish. However, by directly calculating these correlations, we shall show that these correlations do not vanish and the previous results are modified.

\subsection{Green function}
Integrating (\ref{eq:Navier-Stokes equation subjected to a Gaussian random stress tensor}) and (\ref{eq:momentum flux tensor}) over the $xy$ plane yields the equation of motion for $\tilde{\mathcal{V}}^{x}(z,t)$,
\begin{eqnarray}
\rho \frac{\partial \tilde{\mathcal{V}}^{x}}{\partial t} = \eta \partial_{z}^2 \tilde{\mathcal{V}}^{x}+ \partial_{z} \tilde{\mathcal{S}}^{xz}.
\label{eq:starting equation: barvx, real time}
\end{eqnarray}
In the frequency domain, (\ref{eq:starting equation: barvx, real time}) is expressed as
\begin{eqnarray}
\big(\partial_{z}^2 - q^2 \big) \tilde{\mathcal{V}}^{x} = - \frac{1}{\eta} \partial_{z} \tilde{\mathcal{S}}^{xz}
\label{eq:starting equation: barvx, frequency}
\end{eqnarray}
with
\begin{eqnarray}
q^2 = - \frac{i\omega \rho}{\eta}.
\end{eqnarray}
The stochastic properties of $\tilde{\mathcal{S}}^{xz}$ in the frequency domain are given by
\begin{eqnarray}
\Big\langle \tilde{\mathcal{S}}^{xz}(z,\omega) \tilde{\mathcal{S}}^{xz}(z',\omega') \Big\rangle = 2 \eta k_B T A_{xy} \delta(z-z') \delta(\omega+\omega').
\label{eq:random stress: bar: frequency}
\end{eqnarray}
The Green function of (\ref{eq:starting equation: barvx, frequency}) is defined as the solution of the equation
\begin{eqnarray}
\big(\partial_{z}^2 - q^2 \big) G(z,z_2;\omega) = - \frac{1}{\eta} \delta(z-z_2)
\label{eq:def of Green function}
\end{eqnarray}
with boundary conditions
\begin{eqnarray}
G(z=0,z_2;\omega) = \mathcal{B}_A \frac{\partial G(z,z_2;\omega)}{\partial z} \Big|_{z=0},
\label{eq:partial slip boundary condition z=0}
\end{eqnarray}
and
\begin{eqnarray}
G(z=\mathcal{L},z_2;\omega) = - \mathcal{B}_B \frac{\partial G(z,z_2;\omega)}{\partial z} \Big|_{z=\mathcal{L}}.
\label{eq:partial slip boundary condition z=L}
\end{eqnarray}
As discussed in Refs.~\cite{kim2013microhydrodynamics,schwinger1998classical}, the general solutions of (\ref{eq:def of Green function}) are given by 
\begin{eqnarray}
G(z,z_2;\omega) = g_1 e^{-q z} + g_2 e^{q(z-\mathcal{L})} + \frac{1}{2\eta q}e^{-q|z-z_2|},
\label{eq:Green function: explicit form}
\end{eqnarray}
where $g_1$ and $g_2$ are arbitrary constants. Using the boundary conditions, $g_1$ and $g_2$ are determined to be
\begin{eqnarray}
g_1 = - \frac{1}{2\eta q \Delta} \Big((1-q\mathcal{B}_A)(1+q\mathcal{B}_B) e^{q(\mathcal{L}-z_2)} - (1-q\mathcal{B}_A)(1-q\mathcal{B}_B) e^{-q(\mathcal{L}-z_2)} \Big),
\label{eq:g1}
\end{eqnarray}
and
\begin{eqnarray}
g_2 = - \frac{1}{2\eta q \Delta} \Big((1+q\mathcal{B}_A)(1-q\mathcal{B}_B) e^{qz_2} - (1-q\mathcal{B}_A)(1-q\mathcal{B}_B) e^{-qz_2} \Big)
\label{eq:g2}
\end{eqnarray}
with
\begin{eqnarray}
\Delta = (1+q\mathcal{B}_A)(1+q\mathcal{B}_B) e^{q\mathcal{L}} - (1-q\mathcal{B}_A)(1-q\mathcal{B}_B) e^{-q\mathcal{L}} .
\end{eqnarray}
Therefore, the Green function is expressed as (\ref{eq:Green function: explicit form}) with (\ref{eq:g1}) and (\ref{eq:g2}).

\subsection{Expression of $C_{AA}(\omega)$ in terms of Green function} 
The solution of (\ref{eq:starting equation: barvx, frequency}) with the partial slip boundary condition is expressed in terms of the Green function as
\begin{eqnarray}
\tilde{\mathcal{V}}^{x}(z,\omega) = \int_{0}^{\mathcal{L}} dz_2 G(z,z_2,\omega) \partial_{z_2} \tilde{\mathcal{S}}^{xz}(z_2,\omega).
\label{eq:vx in terms of Green function}
\end{eqnarray}
We note that the integration limits on the right-hand side are $z_2=0$ and $z_2=\mathcal{L}$, not $z_2=\xi_{\rm uv}$ and $z_2=\mathcal{L}-\xi_{\rm uv}$.
Using this expression for the fluctuating velocity fields (\ref{eq:vx in terms of Green function}), we may rewrite (\ref{eq:CAA: frequency domain}) in terms of the Green function. We sketch this calculation below; see Appendix~\ref{sec:Technical details of the calculation of CAA(omega)} for technical aspects of the calculation.

We consider the velocity-velocity correlation function. By using (\ref{eq:vx in terms of Green function}), we obtain
\begin{eqnarray}
\hspace{-0.5cm} \Big\langle \tilde{\mathcal{V}}^{x}(z,\omega) \tilde{\mathcal{V}}^{x}(z',-\omega) \Big\rangle = \int_{0}^{\mathcal{L}} dz_2 \int_{0}^{\mathcal{L}} dz'_2 G(z,z_2,\omega) G(z',z'_2,-\omega) \Big\langle  \partial_{z_2}\tilde{\mathcal{S}}^{xz}(z_2,\omega) \partial_{z'_2} \tilde{\mathcal{S}}^{xz}(z'_2,-\omega)\Big\rangle.
\end{eqnarray}
Using the properties of the random stress (\ref{eq:random stress: bar: frequency}) and performing an integration by parts yield
\begin{eqnarray}
\Big\langle \tilde{\mathcal{V}}^{x}(z,\omega) \tilde{\mathcal{V}}^{x}(z',-\omega) \Big\rangle &=& 2 A_{xy} \eta k_B T \int_{0}^{\mathcal{L}} dz_2 \int_{0}^{\mathcal{L}} dz'_2 G(z,z_2,\omega) G(z',z'_2,\omega)  \partial_{z_2} \partial_{z'_2} \delta(z_2-z'_2) \nonumber \\[3pt]
&=& 2 A_{xy} \eta k_B T \int_{0}^{\mathcal{L}} dz_2  \partial_{z_2} G(z,z_2;-\omega) \partial_{z_2}G(z',z_2;-\omega)  \nonumber \\[3pt]
&+& 2 A_{xy} \eta k_B T \Big(G(z,0;\omega) \partial_{z_2} G(z',0;-\omega) - G(z,\mathcal{L};\omega) \partial_{z_2} G(z',\mathcal{L};-\omega) \Big) \nonumber \\[3pt]
&+& 2 A_{xy} \eta k_B T \delta(0)  \Big(G(z,\mathcal{L};\omega) G(z',\mathcal{L};\omega) + G(z,0;\omega) G(z',0;-\omega) \Big) .
\label{eq:velocity-velocity correlation function}
\end{eqnarray}
Here, by recalling the regularization of the delta function, we find that $\delta(0)$ in the last line equals $2/\xi_{\rm uv}$. Then, the first term of (\ref{eq:CAA: frequency domain}) is expressed as
\begin{eqnarray}
& & \Big\langle \partial_{z} \tilde{\mathcal{V}}^{x}(+0,\omega) \partial_{z}\tilde{\mathcal{V}}^{x}(+0,-\omega) \Big\rangle \nonumber \\[3pt]
&=& 2 A_{xy} \eta k_B T \int_{0}^{\mathcal{L}} dz_2  \partial_{z} \partial_{z_2} G(+0,z_2;\omega) \partial_{z} \partial_{z_2}G(+0,z_2;-\omega)  \nonumber \\[3pt]
&+& 2 A_{xy} \eta k_B T \Big( \partial_{z} G(+0,0;\omega) \partial_{z}\partial_{z_2} G(+0,0;-\omega) - \partial_z G(+0,\mathcal{L};\omega) \partial_{z}\partial_{z_2} G(+0,\mathcal{L};-\omega) \Big) \nonumber \\[3pt]
&+& 2 A_{xy} \eta k_B T \delta(0) \Big( \partial_{z} G(+0,\mathcal{L};\omega) \partial_{z} G(+0,\mathcal{L};-\omega) +  \partial_{z} G(+0,0;\omega) \partial_{z}G(+0,0;-\omega) \Big) .
\label{eq:first term: explicit expression}
\end{eqnarray}
Similar calculations apply to the remaining terms of (\ref{eq:CAA: frequency domain}) and lead to
\begin{eqnarray}
& & \Big\langle \partial_{z} \tilde{\mathcal{V}}^{x}(+0,\omega) \tilde{\mathcal{S}}^{xz}(+0,-\omega) \Big\rangle + \Big\langle\partial_{z} \tilde{\mathcal{V}}^{x}(+0,-\omega) \tilde{\mathcal{S}}^{xz}(+0,\omega) \Big\rangle \nonumber \\[3pt]
&=& - 2 A_{xy} \eta k_B T\Big(\partial_{z} \partial_{z_2} G(+0,+0;\omega) + \partial_{z} \partial_{z_2} G(+0,+0;-\omega) \Big).
\end{eqnarray}

Finally, by recalling (\ref{eq:random stress: bar: frequency}), we obtain the full expression of $C_{AA}(\omega)$ in terms of the Green function,
\begin{eqnarray}
\frac{C_{AA}(\omega)}{2\eta k_B T A_{xy}} &=& \delta(0) - \eta \Big(\partial_{z} \partial_{z_2} G(+0,+0;\omega) + \partial_{z} \partial_{z_2} G(+0,+0;-\omega) \Big) \nonumber \\[3pt]
&+& \eta^2  \int_{0}^{\mathcal{L}} dz_2  \partial_{z} \partial_{z_2} G(+0,z_2;\omega) \partial_{z} \partial_{z_2}G(+0,z_2;-\omega)  \nonumber \\[3pt]
&+& \eta^2 \Big( \partial_{z} G(+0,0;\omega) \partial_{z}\partial_{z_2} G(+0,0;-\omega) -  \partial_{z} G(+0,\mathcal{L};\omega) \partial_{z}\partial_{z_2} G(+0,\mathcal{L};-\omega) \Big) \nonumber \\[3pt]
&+& \eta^2 \delta(0) \Big( \partial_{z} G(+0,\mathcal{L};\omega) \partial_{z} G(+0,\mathcal{L};-\omega) +  \partial_{z} G(+0,0;\omega) \partial_{z}G(+0,0;-\omega) \Big) ,
\label{eq:CAA green function expression}
\end{eqnarray}
where $\delta(0) = 2/\xi_{\rm uv}$. We note that the third and fourth lines of (\ref{eq:CAA green function expression}) stem from the surface terms of the integration by parts. When we consider the infinite system or the stick boundary condition, these surface terms are zero. Having imposed the partial slip boundary condition on the walls, we cannot neglect these terms.

\subsection{explicit form of $C_{AA}(\omega)$}
Now, substituting the explicit form of the Green function (\ref{eq:Green function: explicit form}) into (\ref{eq:CAA green function expression}), we obtain the explicit form of $C_{AA}(\omega)$, (\ref{eq:CAA expression: perfect form}).

By straightforward calculation, the derivatives of the Green function contained in (\ref{eq:CAA green function expression}) are given by
\begin{eqnarray}
\partial_z G(+0,0;\omega) = - \frac{q}{\eta\Delta} \mathcal{B}_A (1 + q\mathcal{B}_B)^{q\mathcal{L}} - \frac{q}{\eta\Delta} \mathcal{B}_A (1 - q\mathcal{B}_B)^{-q\mathcal{L}},
\end{eqnarray}
\begin{eqnarray}
\partial_z G(+0,\mathcal{L};\omega) = \frac{2q}{\eta\Delta} \mathcal{B}_B,
\end{eqnarray}
\begin{eqnarray}
\partial_z \partial_{z_2} G(+0,z_2;\omega) = - \frac{q}{\eta \Delta} (1+q\mathcal{B}_B) e^{q (\mathcal{L}-z_2)}  - \frac{q}{\eta \Delta} (1-q\mathcal{B}_B) e^{-q (\mathcal{L} - z_2)} + \frac{1}{\eta}\delta(+0-z_2)
\end{eqnarray}
for $\omega \lesssim 2\pi/\tau^f_{\rm meso}$. Here, the condition $\omega \lesssim 2\pi/\tau^f_{\rm meso}$ is used to yield the relation
\begin{eqnarray}
e^{q \xi_{\rm uv}} \simeq 1.
\end{eqnarray}
Substituting these derivatives into (\ref{eq:CAA green function expression}) leads to the explicit form of $C_{AA}(\omega)$, (\ref{eq:CAA expression: perfect form}).

\section{Behaviors of $C_{AA}(\omega)$ for limiting cases}
\label{sec:Behaviors of CAA for limiting cases}
In the previous section, we obtained the explicit form of $C_{AA}(\omega)$, (\ref{eq:CAA expression: perfect form}). This equation is so complicated and inconvenient to argue the behavior of $C_{AA}(t,t')$. In this section, we obtain more simple forms of $C_{AA}(\omega)$ for some limiting cases and discuss the relationship with the Green--Kubo formulae obtained from the underlying microscopic theory.

\subsection{Behavior of $C_{AA}(\omega)$ in low- and high- frequency regions}
\label{sec:Behavior of CAA in low- and high- frequency regions}
 We focus on $C_{AA}(\omega)$ in the low- and high- frequency regions, which are defined as
\begin{eqnarray}
\omega \ll \frac{2\pi}{\tau_{\rm macro}},
\end{eqnarray}
and
\begin{eqnarray}
 \frac{2\pi}{\tau_{\rm macro}} \ll \omega \simeq \frac{2\pi}{\tau^f_{\rm meso}},
\label{eq:omega:ast meso}
\end{eqnarray}
respectively. $C_{AA}(\omega)$ in these regions can be calculated by straightforwardly taking the limits $\omega \to 0$ and $\omega \to \infty$ of (\ref{eq:CAA expression: perfect form}). Then, we simply express these regions as $\omega \to 0$ and $\omega \to \infty$. Here, we again stress that all the quantities are converted to the dimensionless form by setting $\mathcal{L}=\tau_{\rm marco}=k_B T = 1$.

We first consider $\omega \to \infty$. By noting
\begin{eqnarray}
\lim_{\omega \to \infty} \Big|\frac{q}{\Delta} \Big|^2  = \frac{e^{2q_R L}}{|q|^2 \mathcal{B}_A^2\mathcal{B}_B^2},
\label{eq:omega to infty: finite contribution}
\end{eqnarray}
we find that the finite contribution of (\ref{eq:CAA expression: perfect form}) is calculated as 
\begin{eqnarray}
\lim_{\omega \to \infty} \frac{C_{AA}(\omega)}{2 \eta k_B TA_{xy}}  &=& \lim_{\omega \to \infty} \Big|\frac{q}{\Delta} \Big|^2 \Big[2|q|^2 \mathcal{B}_A \mathcal{B}_B^2 \cosh(2q_R L) + \delta(0) 2|q|^2 \mathcal{B}^2_A \mathcal{B}_B^2 \cosh(2q_R L) \Big] \nonumber \\[3pt]
&=& \frac{1}{\mathcal{B}_A} + \delta(0).
\end{eqnarray}

Recalling that $\delta(0) = O(\xi^{-1}_{\rm uv})$, the assumption $\mathcal{B}_A=\mathcal{B}_B=O(\xi_{\rm micro})$ leads to
\begin{eqnarray}
\lim_{\omega \to \infty} \frac{C_{AA}(\omega)}{2k_B T A_{xy}} = \frac{\eta}{\mathcal{B}_A}\Big(1 + O(\frac{\xi_{\rm micro}}{\xi_{\rm uv}}) \Big),
\label{eq:CAA: omega infty}
\end{eqnarray}
where we have used the relation (\ref{eq:relation between epsilon and delta}). By recalling the assumption $\xi_{\rm micro} \ll \xi_{\rm uv} \ll \xi_{\rm macro}$, we find that the second term may be neglected.

We next consider $\omega \to 0$. By using the expansion of $\Delta$ as
\begin{eqnarray}
\Delta \sim 2 q (\mathcal{B}_A+\mathcal{B}_B + \mathcal{L})
\end{eqnarray}
in $\omega \to 0$, we obtain
\begin{eqnarray}
\lim_{\omega \to 0} \frac{C_{AA}(\omega)}{2\eta k_B T A_{xy}}  &=& \frac{1}{\mathcal{B}_A + \mathcal{B}_B + \mathcal{L}} + \delta(0) \frac{\mathcal{B}_A^2 + \mathcal{B}_B^2}{(\mathcal{B}_A + \mathcal{B}_B + \mathcal{L})^2}.
\label{eq:CAA: omega 0 transient}
\end{eqnarray}
By empoying the assumptions $\mathcal{B}_A=\mathcal{B}_B=O(\xi_{\rm micro})$, $\delta(0)=O(\xi^{-1}_{\rm uv})$ and $\mathcal{L} = O(\xi_{\rm macro})$, we expand (\ref{eq:CAA: omega 0 transient}) in $\epsilon = \xi_{\rm micro}/\xi_{\rm macro}$ as
\begin{eqnarray}
\lim_{\omega \to 0} \frac{C_{AA}(\omega)}{2k_B T A_{xy}} = \frac{\eta}{\mathcal{L}} \big\{1 - (\frac{\mathcal{B}_A}{\mathcal{L}} + \frac{\mathcal{B}_B}{\mathcal{L}}) + o(\epsilon) \big\} .
\label{eq:CAA: omega 0}
\end{eqnarray}
Then, we find that the final term may be neglected.

Equations (\ref{eq:CAA: omega infty}) and (\ref{eq:CAA: omega 0}) lead to the fact that the terms proportional to $\delta(0)$ are the higher terms in $\epsilon$. Furthermore, we can show that this fact is satisfied for all the equations in the reminder of this paper. Therefore, in the reminder of this paper, we neglect the terms proportional to $\delta(0)$.

\subsection{Green--Kubo formulae of the slip length}
\label{sec:Green--Kubo formulae of the slip length}
The behaviors of $C_{AA}(\omega)$ in the high- and low- frequency regions are given by relations (\ref{eq:CAA: omega infty}) and (\ref{eq:CAA: omega 0}), respectively, which are then written as integrals over time correlation functions and correspond to the Green--Kubo formulae. 

We start with the expression (\ref{eq:CAA: omega infty}) to obtain
\begin{eqnarray}
\frac{\eta}{\mathcal{B}_A} &\simeq& \lim_{\omega \to \infty} \frac{C_{AA}(\omega)}{2 k_B T A_{xy}} \nonumber \\[3pt]
&=& \frac{1}{k_B T A_{xy}} \lim_{\omega \to \infty} \int_{0}^{\infty} dt C_{AA}(t,0) e^{i \omega t} \nonumber \\[3pt]
&=& \frac{1}{k_B T A_{xy}} \int_{0}^{\infty} dt C_{AA}(t,0) e^{i t/\tau^f_{\rm meso}},
\label{159}
\end{eqnarray}
where we have used the property $C_{AA}(t,0) = C_{AA}(-t,0)$. Here, we recall that we ignore the time region $t<\tau^f_{\rm meso}$ because $C_{AA}(t,0)$ in this region dose not correspond to the autocorrelation function of the force acting on wall $A$. Then, we impose as the additional assumption that the behavior of $C_{AA}(t,0)$ in $t<\tau^f_{\rm meso}$ is coarse-grained into the generation of the fluctuation at $t=0$. Because $C_{AA}(t,0)$ is expressed as a smooth function in $t>\tau_{\rm meso}$, the additional assumption yields that $C_{AA}(t,0)$ is written in the form
\begin{eqnarray}
C_{AA}(t,0) =  \delta(t) \int_0^{\tau^f_{\rm meso}} \hspace{-0.2cm} ds C_{AA}(s,0) + f(t),
\label{2000}
\end{eqnarray}
where $f(t)$ is the smooth function part of $C_{AA}(t,0)$. By substituting (\ref{2000}) into (\ref{159}) and using that the factor $e^{i t/\tau^f_{\rm meso}}$ describes a rapid oscillation, (\ref{159}) may be rewritten in the Green--Kubo form
\begin{eqnarray}
\frac{\eta}{\mathcal{B}_A} \simeq \tilde{\gamma}_{AA}(\tau^f_{\rm meso})
\label{eq:160}
\end{eqnarray}
with
\begin{eqnarray}
\tilde{\gamma}_{\alpha \beta}(\tau) = \frac{1}{k_B T A_{xy}} \int_0^{\tau} ds \big\langle \tilde{\mathcal{F}}_{\alpha}(s)\tilde{\mathcal{F}}_{\beta}(0) \big\rangle.
\label{eq:def of gamma_ab: tilde}
\end{eqnarray}

$\tilde{\gamma}_{\alpha \beta}(\tau)$ defined in the fluctuating hydrodynamics corresponds to (\ref{eq:def of gamma_ab}) in the linear response theory. We now recall that the statistical mechanical expression of the slip length derived from the underlying microscopic dynamics is given by (\ref{eq:slip length: meso}), which contains two parameters $d_A$ and $z_A$. The Green--Kubo formula (\ref{eq:160}) corresponds to the statistical mechanical expression of $d_A$ given by (\ref{eq:slip length A meso}). From this correspondence, we conjecture that $\tau_{\rm meso}^f$ is contained in the crossover region of $\langle  F^x_A(\tau) F^x_A(0)\rangle_{\rm eq}$ from the microscopic behavior to the macroscopic behavior. In summary, we find that the expression of the slip length in the fluctuating hydrodynamic description is given by (\ref{eq:slip length: meso}) with $z_A=0$.

Next, from the expression (\ref{eq:CAA: omega 0}), we obtain
\begin{eqnarray}
\frac{\eta}{\mathcal{L}}\big\{1 - (\frac{\mathcal{B}_A}{\mathcal{L}} + \frac{\mathcal{B}_B}{\mathcal{L}})\big\} &\simeq& \lim_{\omega \to 0} \frac{C_{AA}(\omega)}{2 k_B T A_{xy}} \nonumber \\[3pt]
&=& \frac{1}{k_B T A_{xy}} \lim_{\omega \to 0} \int_{0}^{\infty} dt C_{AA}(t,0) e^{i \omega t} \nonumber \\[3pt]
&=& \frac{1}{k_B T A_{xy}}  \int_{0}^{\infty} dt \big \langle \tilde{\mathcal{F}}_{A}(t)\tilde{\mathcal{F}}_{A}(0)  \big\rangle.
\label{eq:161}
\end{eqnarray}
Here, we focus on the deterministic motion of the fluid with the wall velocities $V_A$ and $V_B$. By solving the deterministic Navier--Stokes equation, we have
\begin{eqnarray}
\frac{ \big\langle \tilde{\mathcal{F}}_A(\infty)  \big\rangle}{A_{xy}} &=& \frac{\eta}{\mathcal{L}+\mathcal{B}_A+\mathcal{B}_B} (V_A-V_B).
\end{eqnarray}
Assuming $\mathcal{B}_A=\mathcal{B}_B=O(\xi_{\rm micro})$, we obtain
\begin{eqnarray}
\frac{\big\langle \tilde{\mathcal{F}}_A(\infty) \big\rangle}{A_{xy}} = \tilde{\gamma}_{AA}(\infty) (V_A-V_B) + o(\epsilon).
\label{eq:162}
\end{eqnarray}
Comparing (\ref{eq:162}) with (\ref{eq:generalized Green-Kubo: local gibbs: FxA: linear response regime: ss}) and (\ref{eq:relation AA and AB}), we find that the linear response relation for the force acting on wall $A$ derived in Sec.~\ref{sec:Linear response theory for the  steady state} holds to the first order in $\epsilon$.

We comment on the other linear response relations derived in Sect.~\ref{sec:Linear response theory for the  steady state}. Because the calculation is performed in a similar manner to $C_{AA}(\omega)$, we present only the results. See Appendix~\ref{sec:Sketch of the derivation of (153) and (154)} for several technical details of this calculation. First, we have the relation
\begin{eqnarray}
\lim_{\omega \to 0}C_{AA}(\omega) = - \lim_{\omega \to 0}C_{AB}(\omega) = - \lim_{\omega \to 0}C_{BA}(\omega) = \lim_{\omega \to 0}C_{BB}(\omega) ,
\end{eqnarray}
which corresponds to (\ref{eq:relation AA and AB}), (\ref{eq:relation BA and BB}) and (\ref{eq:relation Onsager reciprocity theorem}). Second, we obtain
\begin{eqnarray}
\frac{1}{k_B T A_{xy}}  \int_{0}^{\infty} dt \big\langle \tilde{\mathcal{V}}^{x}(z,t) \tilde{\mathcal{F}}_{A}(0) \big\rangle = (1-\frac{z}{\mathcal{L}} +\frac{\mathcal{B}_B}{\mathcal{L}}) + (1-\frac{z}{\mathcal{L}})(\frac{\mathcal{B}_A}{\mathcal{L}}+\frac{\mathcal{B}_B}{\mathcal{L}}) + o(\epsilon),
\label{eq:gamma PA: fluctuating hydro}
\end{eqnarray}
and
\begin{eqnarray}
\frac{1}{k_B T A_{xy}} \int_{0}^{\infty} dt \big\langle  \tilde{\mathcal{V}}^{x}(z,t) \tilde{\mathcal{F}}_{B}(0) \big\rangle = (\frac{z}{\mathcal{L}}+\frac{\mathcal{B}_A}{\mathcal{L}}) - \frac{z}{\mathcal{L}}(\frac{\mathcal{B}_A}{\mathcal{L}}+ \frac{\mathcal{B}_B}{\mathcal{L}}) + o(\epsilon).
\label{eq:gamma PB: fluctuating hydro}
\end{eqnarray}
Moreover, by focusing on the deterministic motion of the fluid between the walls moving with velocities $V_A$ and $V_B$, we have
\begin{eqnarray}
\frac{1}{A_{xy}}\big\langle \tilde{\mathcal{V}}^{x}(z,\infty) \big\rangle = \frac{\mathcal{L}+\mathcal{B}_B-z}{\mathcal{L}+\mathcal{B}_A+\mathcal{B}_B} V_A + \frac{\mathcal{B}_A+z}{\mathcal{L}+\mathcal{B}_A+\mathcal{B}_B} V_B.
\label{eq:deterministic motion velocity}
\end{eqnarray}
By combining (\ref{eq:gamma PA: fluctuating hydro}), (\ref{eq:gamma PB: fluctuating hydro}) and (\ref{eq:deterministic motion velocity}), we obtain
\begin{eqnarray}
\frac{1}{A_{xy}}\big\langle \tilde{\mathcal{V}}^{x}(z,\infty) \big\rangle = \tilde{\gamma}_{vA}(z,\infty) V_A + \tilde{\gamma}_{vB}(z,\infty) V_B + o(\epsilon),
\label{eq:linear response relation v: fluctuating hydrodynamics}
\end{eqnarray}
where $\tilde{\gamma}_{v\alpha}(z,\tau)$ is defined as
\begin{eqnarray}
\tilde{\gamma}_{v\alpha}(z,\tau) =  \frac{1}{k_B T A_{xy}} \int_0^{\tau} ds \big\langle \tilde{\mathcal{V}}^{x}(z,s) \tilde{\mathcal{F}}_{A}(0)\big\rangle,
\label{eq:gamma Valpha: tilde} 
\end{eqnarray}
where $\alpha=A,B$. $\tilde{\gamma}_{v\alpha}(z,\tau)$ defined in the fluctuating hydrodynamics corresponds to (\ref{eq:gamma Palpha}) in the linear response theory. Equation (\ref{eq:linear response relation v: fluctuating hydrodynamics}) is the linear response relation associated with (\ref{eq:generalized Green-Kubo: local gibbs: pix: linear response regime}).

Combining the linear response relations (\ref{eq:162}) and (\ref{eq:linear response relation v: fluctuating hydrodynamics}), we obtain an expression of the slip length,
\begin{eqnarray}
\mathcal{B}_A \big(1 + o(\epsilon)\big) =  \frac{\eta \tilde{\gamma}_{vB}(z,\infty) }{\tilde{\gamma}_{AA}(\infty)} - z
\label{eq:bAast: sect6}
\end{eqnarray}
for any position $z$. This corresponds to the expression (\ref{eq:slip length: steady state expression A}) derived in Sect.~\ref{sec:Linear response theory for the  steady state}. Using the exact expressions of $\tilde{\gamma}_{vB}(z,\infty)$, (\ref{eq:gamma PB: fluctuating hydro}), and $\tilde{\gamma}_{AA}(\infty)$, (\ref{eq:161}), we find that the right-hand side of (\ref{eq:bAast: sect6}) is independent of $z$.

\subsection{Behavior of $C_{AA}(\omega)$ for the limit $\mathcal{L} \to \infty$}
\label{sec:Behavior of CAA for infinite system}
The results in Sects.~\ref{sec:Behavior of CAA in low- and high- frequency regions} and \ref{sec:Green--Kubo formulae of the slip length} are obtained for the finite size system. We now consider the behavior of $C_{AA}(\omega)$ for the limit $\mathcal{L} \to \infty$.

By noting 
\begin{eqnarray}
\lim_{\mathcal{L} \to \infty} \Big|\frac{q}{\Delta} \Big|^2 = \frac{|q|^2}{|1+q\mathcal{B}_A|^2|1+q\mathcal{B}_B|^2e^{2q_R\mathcal{L}}},
\end{eqnarray}
we calculate (\ref{eq:CAA expression: perfect form}) in $\mathcal{L} \to \infty$ as
\begin{eqnarray}
\hspace{-1cm} \lim_{\mathcal{L} \to \infty} \frac{C_{AA}(\omega)}{2\eta k_B T A_{xy}} = \frac{q_R (1+2q_R \mathcal{B}_A)}{1+2q_R \mathcal{B}_A + 2 q_R^2 \mathcal{B}_A^2},
\label{eq:CAA: Linfty: transient}
\end{eqnarray}
where we have ignored the terms proportional to $\delta(0)$. The expression (\ref{eq:CAA: Linfty: transient}) is independent of $\mathcal{B}_B$. Then, by considering (\ref{eq:CAA: Linfty: transient}) in the low- and high- frequency regions, we obtain
\begin{eqnarray}
\lim_{\omega \to \infty} \lim_{\mathcal{L} \to \infty} \frac{C_{AA}(\omega)}{2\eta k_B T A_{xy}} = \lim_{\omega \to \infty} \Big[\frac{1}{\mathcal{B}_A} - \frac{1}{q_R} \frac{1}{2\mathcal{B}_A^2} \Big] = \frac{1}{\mathcal{B}_A},
\label{eq:CAA Linfty: omegainfty}
\end{eqnarray}
\begin{eqnarray}
\lim_{\omega \to 0} \lim_{\mathcal{L} \to \infty} \frac{C_{AA}(\omega)}{2\eta k_B T A_{xy}} = \lim_{\omega \to 0}\sqrt{\frac{\rho}{2 \eta}}\omega^{\frac{1}{2}} = 0.
\label{eq:CAA Linfty: omega0}
\end{eqnarray}
By comparing (\ref{eq:CAA Linfty: omegainfty}) and (\ref{eq:CAA Linfty: omega0}) to (\ref{eq:CAA: omega infty}) and (\ref{eq:CAA: omega 0}), we find that two limits $\omega \to \infty$ (or $\omega \to 0$) and $\mathcal{L}\to \infty$ can be interchanged. Furthermore, we note that these properties (\ref{eq:CAA Linfty: omegainfty}) and (\ref{eq:CAA Linfty: omega0}) may be expressed in terms of $\tilde{\gamma}_{AA}(\tau)$ defined by (\ref{eq:def of gamma_ab: tilde}):
\begin{eqnarray}
\lim_{\tau \to \tau^f_{\rm meso}}  \lim_{\mathcal{L} \to \infty} \tilde{\gamma}_{AA}(\tau) = \frac{\eta}{\mathcal{B}_A},
\end{eqnarray}
and
\begin{eqnarray}
\lim_{\tau \to \infty} \lim_{\mathcal{L} \to \infty} \tilde{\gamma}_{AA}(\tau) = 0.
\label{eq:gammaAA Linfty: t infty}
\end{eqnarray}

This nature is in contrast to that of the correlation function of momentum current $\tilde{J}^{ab}(\bm{r},t)$. The correlation function of momentum current $\tilde{J}^{ab}(\bm{r},t)$ is calculated as~\cite{mazenko2008nonequilibrium,das2011statistical}
\begin{eqnarray}
\frac{1}{2k_B T \eta} \big\langle \tilde{J}^{xz}(\bm{k},\omega)\tilde{J}^{xz}(-\bm{k},-\omega) \big\rangle = \frac{(\rho \omega)^2}{\eta^2k^4 +(\rho \omega)^2}
\label{eq:JxzJxz: intermediate}
\end{eqnarray}
with
\begin{eqnarray}
\tilde{J}^{ab}(\bm{k},\omega) = \int d^3\bm{r} \int dt e^{i \bm{k} \cdot \bm{r}} e^{i\omega t} \tilde{J}^{ab}(\bm{r},t),
\end{eqnarray}
where we have imposed the periodic boundary conditions in all directions. We focus on $\tilde{\eta}(t;\mathcal{L})$ defined by
\begin{eqnarray}
\tilde{\eta}(\tau,\mathcal{L}) =  \frac{1}{k_B T} \int_0^{\tau} dt\big\langle \tilde{J}^{xz}(\bm{k},t)\tilde{J}^{xz}(-\bm{k},0) \big\rangle \Big|_{\bm{k}=(\frac{2\pi}{\mathcal{L}},\frac{2\pi}{\mathcal{L}},\frac{2\pi}{\mathcal{L}})}. 
\end{eqnarray}
By using the exact equation (\ref{eq:JxzJxz: intermediate}) and repeating the similar calculation in Sect.~\ref{sec:Green--Kubo formulae of the slip length}, we obtain the properties of $\tilde{\eta}(t;\mathcal{L})$:
\begin{eqnarray}
\tilde{\eta}(\tau_{\rm uv},\mathcal{L}) = \eta,
\label{eq:JJ omega infty}
\end{eqnarray}
\begin{eqnarray}
\lim_{\tau \to \infty}\tilde{\eta}(\tau,\mathcal{L}) = 0,
\label{eq:JJ omega 0 finite system}
\end{eqnarray}
and
\begin{eqnarray}
\lim_{\tau \to \infty}\lim_{\mathcal{L} \to \infty}\tilde{\eta}(\tau,\mathcal{L}) = \eta.
\label{eq:JJ omega 0}
\end{eqnarray}
Equation (\ref{eq:JJ omega infty}) is equivalent to the properties of the stochastic flux, (\ref{eq:random stress tensor}), and is just the definition of the viscosity $\eta$. On the other hand, (\ref{eq:JJ omega 0}) is the Green--Kubo-like formula that holds only in the linearized fluctuating hydrodynamics. From (\ref{eq:JJ omega 0 finite system}) and (\ref{eq:JJ omega 0}), we find that two limits $\tau \to \infty$ and $\mathcal{L} \to \infty$ cannot commute for $\tilde{\eta}(t;\mathcal{L})$.

Finally, we explain the relationship between the result obtained in the linearized fluctuating hydrodynamics and the tentative prescription suggested by Bocquet and Barrat in \cite{bocquet1994hydrodynamic,bocquet2013green}. In these papers, they proposed that in order to obtain the correct slip length from (\ref{eq:160}) (or Eq.~(\ref{eq:slip length: meso})), the system size limit $\mathcal{L} \to \infty$ should be taken before performing the time integration to infinity. However, from (\ref{eq:gammaAA Linfty: t infty}), we find that this prescription does not yield the correct slip length. We note that such prescription is useful to obtain the correct viscosity from $\tilde{\eta}(t;\mathcal{L})$ in the linearized fluctuating hydrodynamics, as shown in (\ref{eq:JJ omega 0 finite system}) and (\ref{eq:JJ omega 0}).

\section{Equilibrium measurement of slip length}
\label{sec:Equilibrium measurement of slip length}
\subsection{Previous studies}
Previously, several different methods were suggested to calculate the slip length using equilibrium molecular dynamics (EMD). Because each method is based on different models and assumptions, the relationship among the methods remains unclear and the research group who proposed one method often misunderstood the other methods. In this subsection, we organize representative methods based on our theory. 

Here, we recall the existence of two characteristic time $\tau_{\rm meso}$ and $\tau^f_{\rm meso}$. $\tau_{\rm meso}$ is defined as the time to characterize the crossover region from the microscopic behavior to the macroscopic behavior of $\langle  F^x_A(\tau) F^x_A(0)\rangle_{\rm eq}$ (See Sect.~\ref{sec:4.2}). $\tau^f_{\rm meso}$ is defined as the minimum time that $C_{AA}(t,0)$ calculated in the fluctuating hydrodynamics can be identified with the autocorrelation function of the force acting on wall $A$ (See Sect.~\ref{sec:Calculated quantities}). As shown in Sect.~\ref{sec:Green--Kubo formulae of the slip length}, the assumption $\tau^f_{\rm meso} = \tau_{\rm meso}$ leads to the consistency between the results obtained in the underlying microscopic theory and in the linearized fluctuating hydrodynamics. Therefore, in this section, we assume $\tau^f_{\rm meso} = \tau_{\rm meso}$.

The first method to predict the slip length using EMD was presented by Bocquet and Barrat~\cite{bocquet1994hydrodynamic}. They evaluated the slip length from the Green--Kubo formula
\begin{eqnarray}
\frac{\eta}{b^{\rm BB}_A} = \frac{1}{A_{xy}k_B T} \int_0^{\infty} ds \big\langle F_A^x(s) F_A^x(0) \big\rangle_{\rm eq}.
\end{eqnarray}
They treated the integration to infinity by introducing a cutoff upper time $\tau^{\rm BB}$, which is defined from the first zero of $\big\langle F_A^x(s) F_A^x(0) \big\rangle_{\rm eq}$. Because $\tau^{\rm BB}$ is related to the relaxation time of $\big\langle F_A^x(s) F_A^x(0) \big\rangle_{\rm eq}$~\cite{lagar1978molecular,espanol1993force}, we interpret that $\tau^{\rm BB}$ is a practical estimation of $\tau_{\rm meso}$. Then, we conclude that $b^{\rm BB}_A=b^{\rm meso}_A$ (See Sect.~\ref{sec: expression of slip length: relaxation process}).

Next, Petravic and Harrowell~\cite{petravic2007equilibrium} numerically investigated the behavior of $\big\langle F_A^x(s) F_A^x(0) \big\rangle_{\rm eq}$ and proposed the following Green--Kubo formula
\begin{eqnarray}
\frac{\eta}{L+2b^{\rm PH}_A} = \frac{1}{A_{xy}k_B T} \int_0^{\infty} ds \big\langle F_A^x(s) F_A^x(0) \big\rangle_{\rm eq},
\end{eqnarray}
where walls $A$ and $B$ are assumed to have microscopically same structure. They evaluated the integration to infinity as the value of the time integral after it had reached a constant value. This time corresponds to $\tau_{\rm ss}$ in our theory. Then, we find that $b^{\rm PH}_A = b^{\rm ss}_A$ (See Sect.~\ref{sec:expression of the slip length steady state}). Here, we note that $b^{\rm PH}_A$ is dependent on the choice of the system size $L$. They chose $L$ as the true system size, which coincides with our choice in Sect.~\ref{sec:expression of the slip length steady state}.

Recently, Huang and Szlufarska~\cite{huang2014green} proposed a new method, which employs the following Green--Kubo formula
\begin{eqnarray}
\frac{\eta}{b^{HS}_A} =  \frac{1}{A_{xy}k_B T(1-I^{HS})} \int_0^{\infty} ds \sum_{i=1}^N \big\langle F_A^x(\bm{r}_i(s),s) F_A^x(\bm{r}_i(0),0) \big\rangle_{\rm eq}
\label{eq:slip length Huand and Szlufarska}
\end{eqnarray}
with
\begin{eqnarray}
I^{HS} = \frac{1}{k_B T} \int_0^{\infty} ds \big\langle u^x_{slab}(s) F_A^x(\bm{r}_i(0),0) \big\rangle_{\rm eq},
\end{eqnarray}
where $\hat{u}^x_{slab}(\Gamma)$ is defined as the average velocity of the particles in the slab $[0,L_x] \times[0,L_y]\times[0,\Delta]$:
\begin{eqnarray}
\hat{u}^x_{slab}(\Gamma) = \frac{1}{N_{slab}} \sum_{z_i \in [0,\Delta]} \frac{\hat{p}^x_i}{m}.
\end{eqnarray}
Here, $N_{slab}$ is the particle number in the slab and $\Delta$ is chosen as sufficiently small value. The integration to infinity was evaluated by the same method as Petravic and Harrowell. This expression might appear to be related to the expression of $b^{\rm ss}_A$, (\ref{eq:slip length: steady state expression A}). Actually, by dividing the correlation function of the total force $\hat{F}^x_A(\Gamma)$ into the correlations between the same particle and the ones between the different particles:
\begin{eqnarray}
\frac{1}{A_{xy}k_B T} \int_0^{\infty} ds \big\langle F_A^x(s) F_A^x(0) \big\rangle_{\rm eq} &=& \frac{1}{A_{xy}k_B T} \int_0^{\infty} ds \sum_{i=1}^N \big\langle F_A^x(\bm{r}_i(s),s) F_A^x(\bm{r}_i(0),0) \big\rangle_{\rm eq} \nonumber \\[3pt]
&+&  \frac{1}{A_{xy}k_B T} \int_0^{\infty} ds \sum_{i\neq j} \big\langle F_A^x(\bm{r}_i(s),s) F_A^x(\bm{r}_j(0),0) \big\rangle_{\rm eq} ,
\end{eqnarray}
and ignoring correlation between the different particles (second term), (\ref{eq:slip length: steady state expression A}) is rewritten as
\begin{eqnarray}
\frac{\eta}{b_A^{\rm ss}+z_0} = \frac{1}{A_{xy} k_B T (1-\gamma_{PA}(z_0,\tau_{\rm ss})/\rho)} \int_0^{\tau_{\rm ss}} ds \sum_{i=1}^N \big\langle F_A^x(\bm{r}_i(s),s) F_A^x(\bm{r}_i(0),0) \big\rangle_{\rm eq},
\label{eq:slip length Huand and Szlufarska intermediate}
\end{eqnarray}
where we have used (\ref{eq:relation PA and PB}). If
\begin{eqnarray}
\frac{1}{k_B T} \int_0^{\infty} ds \big\langle u^x_{slab}(s) F_A^x(\bm{r}_i(0),0) \big\rangle_{\rm eq} = \frac{1}{k_B T \rho} \int_0^{\tau_{\rm ss}} \hspace{-0.15cm} ds \big\langle \pi^x(z,s) F^x_{\alpha}(0)\big\rangle_{\rm eq}
\end{eqnarray}
holds where $z$ is an appropriate point near wall $A$, by replacing $\gamma_{PA}(z_0,\tau_{\rm ss})/\rho$ in (\ref{eq:slip length Huand and Szlufarska intermediate}) with $I^{HS}$, we can obtain (\ref{eq:slip length Huand and Szlufarska}). However, the expression of $b^{\rm ss}_A$, (\ref{eq:slip length: steady state expression A}), holds only when $z_0$ is chosen in the bulk and we cannot substitute $z_0$ near wall $A$ into (\ref{eq:slip length Huand and Szlufarska intermediate}). Therefore, because this replacement is beyond our theory, our microscopic theory does not validate the expression (\ref{eq:slip length Huand and Szlufarska}). On the other hand, the linearized fluctuating hydrodynamics validates the expression (\ref{eq:slip length Huand and Szlufarska}) because the expression (\ref{eq:bAast: sect6}) does not contain this problem.

Finally, we explain the work by Hansen et al.~\cite{hansen2011prediction}. They predicted the slip length by the following formula
\begin{eqnarray}
\frac{\eta}{b^{HTD}_A} = \frac{\int_0^{\infty} ds \langle u^x_{slab}(0) F_A^x(s) \rangle_{\rm eq}}{ \int_0^{\infty} ds \langle u^x_{slab}(0) u^x_{slab}(s) \rangle_{\rm eq}},
\label{eq:expression of slip length by Hansen et al}
\end{eqnarray}
where the integration to infinity was evaluated by the same method as Petravic and Harrowell, specifically, taking the value of the time integral after it had reached a constant value. Therefore, their expression is related to the linear response theory at $\tau_{\rm ss}$-scale. Indeed, this expression is derived from the linearized fluctuating hydrodynamics, as briefly explained in Appendix~\ref{sec:Derivation of bHTD from the linearized fluctuating dynamics}.

These results are summarized in Table~\ref{tab:a} (See Sect.~\ref{sec:A quick look of the paper}).

\subsection{Proposal of a new method}
As shown in Table~\ref{tab:a}, almost all methods for calculating the slip length from EMD employ the data of the correlation functions at $t=\tau_{\rm ss}$. Because $\tau_{\rm ss}$ is longer than the relaxation time of the fluid, the calculation of the correlation functions to $t=\tau_{\rm ss}$ demands a high computational cost. Therefore, we conjecture that the methods related to $\tau_{\rm meso}$-scale are more practical.

The method for employing the linear response theory at $\tau_{\rm meso}$-scale was proposed only by Bocquet and Barrat. The difficulty of their method stems from the uncertain nature of $\tau_{\rm meso}$. Actually, in the EMD, the behavior of $\tilde{\gamma}_{AA}(\tau)$ at $\tau_{\rm meso}$-scale is observed as the crossover from the microscopic behavior to the macroscopic behavior~\cite{lagar1978molecular,espanol1993force}. Then, the determination of $\tau_{\rm meso}$ from the crossover region is rather difficult.

We propose a new method that determines the slip length without any determination of $\tau_{\rm meso}$. This method employs the explicit form of the time correlation function obtained from the linearized fluctuating hydrodynamics, (\ref{eq:CAA expression: perfect form}). First, if we can numerically obtain $\big\langle F_A^x(t) F_A^x(0) \big\rangle_{\rm eq}$ from $t=\tau_{\rm micro}$ to $t=\tau_{\rm ss}$, then $\mathcal{B}_A$, $\mathcal{B}_B$, and $\mathcal{L}$ can be obtained by fitting (\ref{eq:CAA expression: perfect form}) to this data. In order to obtain $\mathcal{B}_A$ from the data up to smaller $t$, we expand (\ref{eq:CAA expression: perfect form}) at $\omega \to \infty$. By using the expansion
\begin{eqnarray}
\Big|\frac{q^2}{\Delta} \Big|^2 = \frac{e^{-2q_R L} }{\mathcal{B}_A^2 \mathcal{B}_B^2} - \frac{2q_Re^{-2q_R L}}{|q|^2} \frac{\mathcal{B}_A+\mathcal{B}_B}{\mathcal{B}_A^3 \mathcal{B}_B^3} + O(\frac{e^{-\sqrt{\omega}}}{\omega}),
\end{eqnarray}
(\ref{eq:CAA expression: perfect form}) is expanded at $\omega \to \infty$ as
\begin{eqnarray}
C_{AA}(\omega) = \frac{1}{\mathcal{B}_A} - \sqrt{\frac{\eta}{2 \rho}} \frac{1}{\mathcal{B}_A^2}  \omega^{-\frac{1}{2}} + O(\omega^{-1}) ,
\label{eq:CAA omega infty expansion}
\end{eqnarray}
where we have ignored the terms proportional to $\delta(0)$. As shown in Sect.~\ref{sec:Behaviors of CAA for limiting cases}, the first term of (\ref{eq:CAA omega infty expansion}) is related to BB's formula. The second term provides how $\tilde{\gamma}_{AA}(\tau)$ goes out from the crossover region at $\tau_{\rm meso}$-scale. We conjecture that by numerically observing the $\omega^{-1/2}$ diffusion $\mathcal{B}_A$ can be determined regardless of the uncertain nature of $\tau_{\rm meso}$. This method provides a straightforward extension of the method by Bocquet and Barrat.

Conversely, we can confirm the validity of the assumptions employing throughout this paper (see Sect.~\ref{sec:Separation of length and time scales}) by comparing between the theoretical result (\ref{eq:CAA omega infty expansion}) and the result of EMD. We consider that numerical verification is important to more deeply understand our theory.

\section{Concluding remarks}
\label{sec:Concluding remarks}
\begin{figure}
\centering
\includegraphics[angle=90,width=16cm,bb=0 0 595 842]{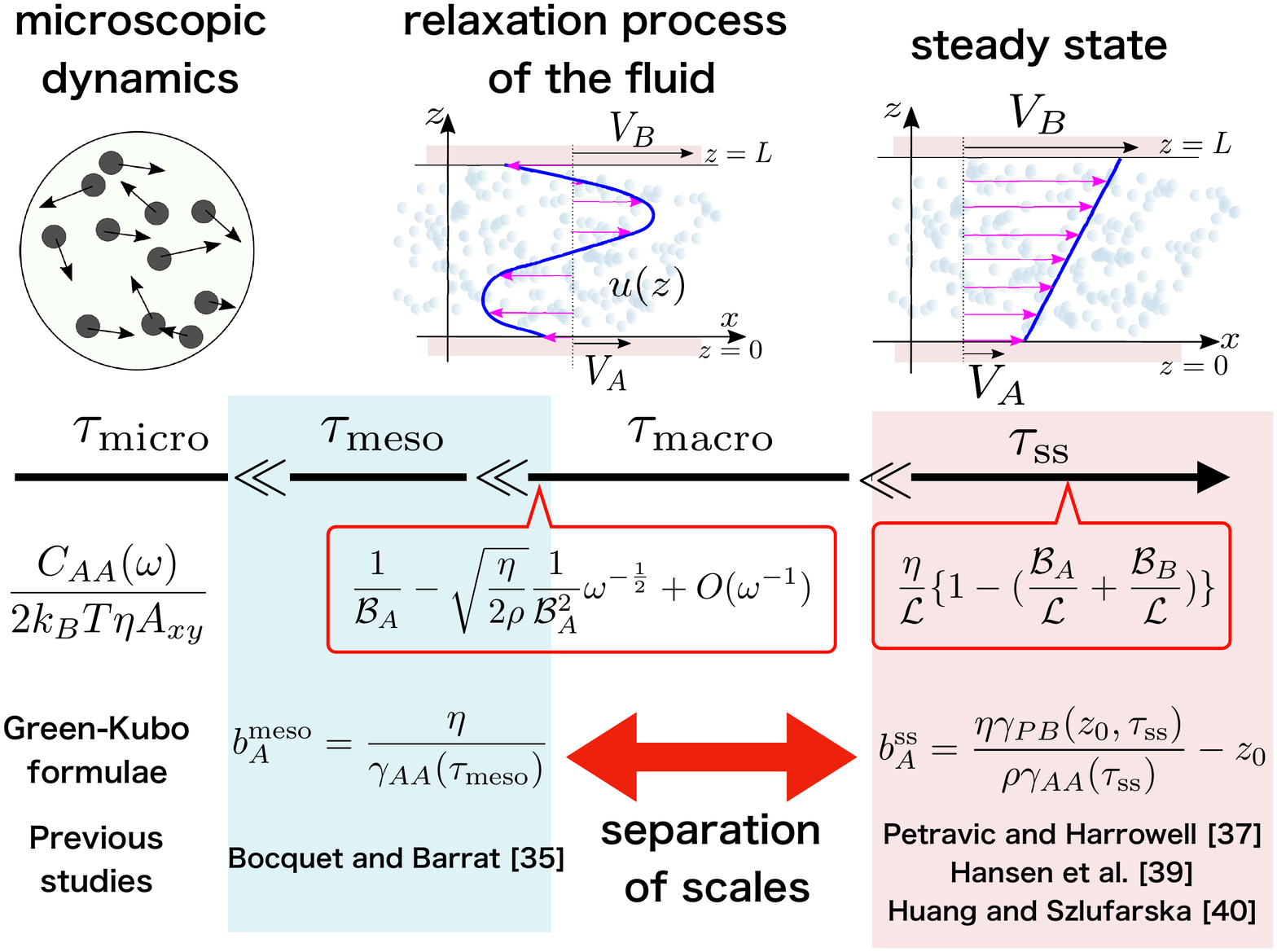}
\caption{Diagram illustrating the relationship between the derivations of two statistical mechanical expressions of the slip length.}
\label{fig2}
\end{figure}

In this paper, we derived several statistical mechanical expressions of the slip length on the basis of the microscopic dynamics and the linearized fluctuating hydrodynamics. These expressions were suggested in the literatures, but our derivation is more general than the previous ones in that all expressions are derived from the single model. Our derivation clarifies that the existence of two different time scales must be carefully taken in account in order to avoid the confusion between these expressions. More precise summary of this paper is given in Sect.~\ref{sec:A quick look of the paper}. In closing the paper, we make some remarks.

In Sect.~\ref{sec:model: linearized fluctuating hydrodynamics}, the nonlinear fluctuation effects were neglected. It has been known that mixing of the fluctuations of the macroscopic density fields by the nonlinear streaming effect leads to a long-time tail in the correlation function of the momentum flux and breaks the separation of scales~\cite{forster1977large}. As a result, the bare transport coefficients that appear in the nonlinear fluctuating hydrodynamics are different from the transport coefficients in the linearized fluctuating hydrodynamics. In Sect.~\ref{sec:Separation of length and time scales}, we assumed the separation of scales and neglected the long-time tail in the time correlation functions of the momentum fluxes. Then, the nonlinear fluctuating effects were neglected in the phenomenological argument for the formulae obtained from the microscopic dynamics. Therefore, because we were not concerned with the long-time tail effects in the discussion employing the fluctuating hydrodynamics, we neglected the nonlinear advection term that leads to the long time tail.
As a related study, Wolynes developed the mode-mode coupling theory of the compressible fluid in the presence of a plane wall and reported that the correction to the bare slip length is of order $10^{-9} {\rm m}$~\cite{wolynes1976hydrodynamic,nieuwoudt1984long}. Recalling that in our derivation the slip length is $O(\xi_{\rm micro})$, we conjecture that the contribution of the nonlinear effect is non-negligible. Therefore, developing a more careful discussion taking the long-time tail into account is an open problem for subsequent study.

Our derivations suggested that the partial slip boundary condition holds when the slip length is $O(\xi_{\rm micro})$. Here, we briefly discuss what happens when the slip length is $O(\xi_{\rm meso})$. We conjecture that there is no partial slip boundary condition with the mesoscopic slip length. For example, consider (\ref{eq:partial slip boundary condition: wall A: linear response theory mid}). If $d_A$ and $z_A$ were $O(\xi_{\rm meso})$, there would be no reason why the second term on the right-hand side of (\ref{eq:partial slip boundary condition: wall A: linear response theory mid}) is neglected in the first order approximation in $\epsilon$. Therefore, we consider that a boundary condition containing the higher-order derivative of the velocity field is plausible. In Ref.~\cite{nakano2019microscopic}, we discussed deriving such boundary conditions based on several phenomenological assumptions.

Furthermore, we comment the relation between our derivation and the experiments concerning the boundary condition. The typical value of the slip length reported from recent experiments is about $10 {\rm nm}$~\cite{lauga2007microfluidics,bocquet2010nanofluidics}, which is consistent with our derivation. However, some pioneer experiments for special situations reported that a slip length is of order $100{\rm nm}$~\cite{zhu2001rate,schmatko2005friction,ulmanella2008molecular,joseph2006slippage,lee2008structured}. In such situations, the slip lengths were dependent on the fluid velocity at walls~\cite{thompson1997general,priezjev2004molecular,martini2008slip}. Then, it remains to be elucidated what boundary conditions must be imposed in such situations. We expect that our careful considerations regarding the slip length provide the first step concerning this aspect.

In the present work, the microscopic approach is limited to atomically smooth surfaces, which are represented as a collection of material points placed in a plane. Actually, most equilibrium measurements of the slip length were performed only for such surfaces. However, our analysis of the fluctuating hydrodynamics proposes that the statistical mechanical expressions derived from underlying microscopic dynamics may hold beyond the atomically smooth surfaces because the model in Sect.~\ref{sec:model: linearized fluctuating hydrodynamics} does not mention the type of surfaces. Atomically smooth surfaces are very rare in nature, and most realistic surfaces have finite roughness. It has not been elucidated how such roughness affects the value of the slip length. We expect that the Green--Kubo formulae provide some solutions to this problem.


\begin{acknowledgements}
The authors would like to thank A.~Yoshimori, M. Itami and Y.~Minami for helpful comments. The present study was supported by KAKENHI (Nos. 17H01148).
\end{acknowledgements}

\appendix\normalsize
\renewcommand{\theequation}{\Alph{section}.\arabic{equation}}
\setcounter{equation}{0}
\makeatletter
  \def\@seccntformat#1{%
    \@nameuse{@seccnt@prefix@#1}%
    \@nameuse{the#1}%
    \@nameuse{@seccnt@postfix@#1}%
    \@nameuse{@seccnt@afterskip@#1}}
  \def\@seccnt@prefix@section{Appendix }
  \def\@seccnt@postfix@section{:}
  \def\@seccnt@afterskip@section{\ }
  \def\@seccnt@afterskip@subsection{\ }
\makeatother

\section{Technical details of the calculation of $C_{AA}(\omega)$}\label{sec:Technical details of the calculation of CAA(omega)}
In the calculation of (\ref{eq:velocity-velocity correlation function}), we detail how to proceed from the first to the second line. We perform an integration by parts
\begin{eqnarray}
& & \int_{0}^{\mathcal{L}} dz_2 \int_{0}^{\mathcal{L}} dz'_2 G(z,z_2,\omega) G(z',z'_2,-\omega)  \partial_{z_2} \partial_{z'_2} \delta(z_2-z'_2) \nonumber \\[3pt]
&=& \Big[\int_{0}^{\mathcal{L}} dz_2 G(z,z_2,\omega) G(z',z'_2,-\omega)  \partial_{z_2} \delta(z_2-z'_2) \Big]_{z'_2=0}^{z'_2=\mathcal{L}}\nonumber \\[3pt]
&-& \int_{0}^{\mathcal{L}} dz_2 \int_{0}^{\mathcal{L}} dz'_2 G(z,z_2,\omega) \partial_{z'_2} G(z',z'_2,-\omega)  \partial_{z_2} \delta(z_2-z'_2) \nonumber \\[3pt]
&=& \Big[G(z,z_2,\omega) G(z',\mathcal{L},-\omega) \delta(z_2-\mathcal{L}) \Big]_{z_2=0}^{z_2=\mathcal{L}} - \int_{0}^{\mathcal{L}} dz_2 \partial_{z_2}G(z,z_2,\omega) G(z',\mathcal{L},-\omega) \delta(z_2-\mathcal{L}) \nonumber \\[3pt]
&-& \Big[G(z,z_2,\omega) G(z',0,-\omega)  \delta(z_2-0) \Big]_{z_2=0}^{z_2=\mathcal{L}} + \int_{0}^{\mathcal{L}} dz_2  \partial_{z_2}G(z,z_2,\omega) G(z',0,-\omega) \delta(z_2-0) \nonumber \\[3pt]
&-&  \int_{0}^{\mathcal{L}} dz'_2 G(z,\mathcal{L},\omega) \partial_{z'_2} G(z',z'_2,-\omega) \delta(\mathcal{L}-z'_2) +  \int_{0}^{\mathcal{L}} dz'_2 G(z,0,\omega) \partial_{z'_2} G(z',z'_2,-\omega) \delta(0-z'_2) \nonumber \\[3pt]
&+& \int_{0}^{\mathcal{L}} dz_2 \int_{0}^{\mathcal{L}} dz'_2 \partial_{z_2} G(z,z_2,\omega) \partial_{z'_2} G(z',z'_2,-\omega) \delta(z_2-z'_2).
\end{eqnarray}
The last expression contains integrals of the delta function weighted only at the edge of the integration range. These integrals are calculated using the property:
\begin{eqnarray}
\int_0^{\mathcal{L}} dz \delta(z) = \frac{1}{2}.
\label{eq:delta function at the edge}
\end{eqnarray}
The Integration of the delta functions in (\ref{eq:delta function at the edge}) yields the second line of (\ref{eq:velocity-velocity correlation function}).

\section{Sketch of the derivation of (\ref{eq:gamma PA: fluctuating hydro}) and (\ref{eq:gamma PB: fluctuating hydro})}\label{sec:Sketch of the derivation of (153) and (154)}
We sketch the derivation of (\ref{eq:gamma PA: fluctuating hydro}) and (\ref{eq:gamma PB: fluctuating hydro}). We introduce the time correlation functions $C_{v\alpha}(z,t,t')$ as
\begin{eqnarray}
C_{v\alpha}(z,t,t') \equiv \langle \tilde{\mathcal{V}}^{x}(z,t) \tilde{\mathcal{F}}_{\alpha}(t') \rangle,
\end{eqnarray}
where $\alpha=A,B$. Equations (\ref{eq:gamma PA: fluctuating hydro}) and (\ref{eq:gamma PB: fluctuating hydro}) are written in terms of $C_{v\alpha}(z,t,t')$ as
\begin{eqnarray}
\frac{1}{k_B T A_{xy}} \int_{0}^{\infty} dt C_{vA}(z,t,0) =  (1-\frac{z}{\mathcal{L}} +\frac{\mathcal{B}_B}{\mathcal{L}}) + (1-\frac{z}{\mathcal{L}})(\frac{\mathcal{B}_A}{\mathcal{L}}+\frac{\mathcal{B}_B}{\mathcal{L}}) + o(\epsilon),
\label{eq:gamma PA: fluctuating hydro: deform}
\end{eqnarray}
and
\begin{eqnarray}
\frac{1}{k_B T A_{xy}}  \int_{0}^{\infty} dt C_{vB}(z,t,0)  = (\frac{z}{\mathcal{L}}+\frac{\mathcal{B}_A}{\mathcal{L}}) - \frac{z}{\mathcal{L}}(\frac{\mathcal{B}_A}{\mathcal{L}} + \frac{\mathcal{B}_B}{\mathcal{L}}) + o(\epsilon).
\label{eq:gamma PB: fluctuating hydro: deform}
\end{eqnarray}
Since $C_{v\alpha}(z,t,0)$ is the correlation function between the odd and even quantities under time reversal, the equality
\begin{eqnarray}
C_{v\alpha}(z,t,0) = -C_{v\alpha}(z,-t,0)
\end{eqnarray}
holds. This is in contrast to $C_{\alpha \beta}(t,0) \ (\alpha,\beta=A, B)$, which satisfies
\begin{eqnarray}
C_{\alpha \beta}(t,0) = C_{\alpha \beta}(-t,0).
\end{eqnarray}
Because of this property, we have
\begin{eqnarray}
\lim_{\omega \to 0} C_{v\alpha}(z,\omega) = \lim_{\omega \to 0} \int_{-\infty}^{\infty} dt C_{v\alpha}(z,t,0)e^{i\omega t} = 0.
\end{eqnarray}
Therefore, we cannot repeat the calculation of (\ref{eq:161}) in deriving (\ref{eq:gamma PA: fluctuating hydro: deform}) and (\ref{eq:gamma PB: fluctuating hydro: deform}). We calculate (\ref{eq:gamma PA: fluctuating hydro: deform}) and (\ref{eq:gamma PB: fluctuating hydro: deform}) using the following steps. First, we calculate the Fourier transform $C_{v\alpha}(z,\omega)$ and obtain the real-time correlation function in the form
\begin{eqnarray}
C_{v\alpha}(z,t,0) = \int \frac{d\omega}{2\pi} C_{v \alpha}(z,\omega)e^{-i\omega t}.
\label{eq:C23}
\end{eqnarray}
Next, we perform the integral in (\ref{eq:C23}). Here, we assume that main contribution of $C_{v\alpha}(z,t,0)$ in (\ref{eq:gamma PA: fluctuating hydro: deform}) and (\ref{eq:gamma PB: fluctuating hydro: deform}) comes from the region $\omega \simeq 0$. Finally, using this result, we calculate the time integral on the left-hand side of (\ref{eq:gamma PA: fluctuating hydro: deform}) and (\ref{eq:gamma PB: fluctuating hydro: deform}).

We focus on (\ref{eq:gamma PA: fluctuating hydro: deform}). $C_{vA}(z,\omega)$ is written in terms of the Green function as
\begin{eqnarray}
\frac{C_{vA}(z;\omega)}{2k_B TA_{xy}} &=& \eta \partial_{z_2} G(z,+0;\omega) -  \eta^2  \int_0^{\mathcal{L}} dz_2 \partial_{z_2} G(z,z_2;\omega) \partial_{z} \partial_{z_2} G(+0,z_2;-\omega) \nonumber \\[3pt]
&-& \eta^2 G(z,0;\omega)  \partial_{z}\partial_{z_2} G(+0,0;-\omega) + \eta^2 G(z,\mathcal{L};\omega)  \partial_{z} \partial_{z_2} G(+0,\mathcal{L};-\omega) ,
\label{eq:DVA in terms of green function}
\end{eqnarray}
where we ignore terms proportional to $\delta(0)$, such as the fourth line of (\ref{eq:CAA green function expression}). The leading contribution of the Green function in $\omega$ is expressed as
\begin{eqnarray}
G(z,z_2;\omega) \simeq \frac{(\mathcal{L}+\mathcal{B}_B-z)(\mathcal{B}_A+z_2)}{\eta(\mathcal{B}_A+\mathcal{B}_B+\mathcal{L})-i\omega \rho c}
\label{eq:the green function omega 01}
\end{eqnarray}
for $z>z_2$, and
\begin{eqnarray}
G(z,z_2;\omega) \simeq \frac{(1+\mathcal{B}_B-z_2)(\mathcal{B}_A+z)}{\eta(\mathcal{B}_A+\mathcal{B}_B+\mathcal{L})-i\omega \rho c}
\label{eq:the green function omega 02}
\end{eqnarray}
for $z<z_2$, with
\begin{eqnarray}
c \equiv \mathcal{B}_A \mathcal{B}_B \mathcal{L} + \frac{1}{2} (\mathcal{B}_A+\mathcal{B}_B) \mathcal{L}^2 + \frac{1}{6} \mathcal{L}^3.
\end{eqnarray}
We consider the first term of (\ref{eq:DVA in terms of green function}). Substituting (\ref{eq:the green function omega 01}) and (\ref{eq:the green function omega 02}) into the first term of (\ref{eq:DVA in terms of green function}) produces
\begin{eqnarray}
\eta \partial_{z_2} G(z,+0;\omega) \simeq - \frac{\eta}{i\rho c} \frac{\mathcal{L}+\mathcal{B}_B-z}{\omega-\frac{\eta(\mathcal{B}_A+\mathcal{B}_B+\mathcal{L})}{i\rho c}}.
\end{eqnarray}
Next, after performing the integral in $\omega$, we obtain
\begin{eqnarray}
\eta \int\frac{d\omega}{2\pi} \partial_{z_2} G(z,+0;\omega)  e^{-i\omega t} = \frac{\eta(\mathcal{L}+\mathcal{B}_B-z)}{\rho c} e^{-t\frac{\eta(\mathcal{L}+\mathcal{B}_A+\mathcal{B}_B)}{\rho c}},
\label{eq:first term DVA intermediate}
\end{eqnarray}
which is a part of $C_{vA}(z,t,0)/2k_B TA_{xy}$ obtained from the first term of (\ref{eq:DVA in terms of green function}). Finally, integrating the right-hand side of (\ref{eq:first term DVA intermediate}) in $t$ gives
\begin{eqnarray}
\int_0^{\infty} dt \frac{\eta(\mathcal{L}+\mathcal{B}_B-z)}{\rho c} e^{-t\frac{\eta(\mathcal{L}+\mathcal{B}_A+\mathcal{B}_B)}{\rho c}} = \frac{\mathcal{L}+\mathcal{B}_B-z}{\mathcal{L}+\mathcal{B}_A+\mathcal{B}_B}.
\end{eqnarray}
We calculate the remaining terms of (\ref{eq:DVA in terms of green function}) in the same manner. By collecting these results and extracting the first order terms in $\epsilon$, we obtain (\ref{eq:gamma PA: fluctuating hydro: deform}). Finally, we note that (\ref{eq:gamma PB: fluctuating hydro: deform}) is obtained by a similar procedure.

\section{Derivation of (\ref{eq:expression of slip length by Hansen et al}) from the linearized fluctuating dynamics}
\label{sec:Derivation of bHTD from the linearized fluctuating dynamics}
In this Appendix, we show a proof of (\ref{eq:expression of slip length by Hansen et al}) by employing the model in Sect.~\ref{sec:model: linearized fluctuating hydrodynamics}.

In our model, the average velocity of the particles in the slab, $\hat{u}^x_{slab}(\Gamma)$, is replaced with the fluid velocity near the wall, $\tilde{\mathcal{V}}^x(+0,t)$. Based on this, we transform (\ref{eq:expression of slip length by Hansen et al}) to the expression with respect to frequency. By recalling that
\begin{eqnarray}
\lim_{\omega \to 0} A(\omega) = \int_0^{\infty} ds A(s),
\end{eqnarray} 
where $A(t)$ is any function and $A(\omega)$ is the Fourier transform of $A(t)$, (\ref{eq:expression of slip length by Hansen et al}) is rewritten as
\begin{eqnarray}
\frac{\eta}{b_A^{HTD}} = \frac{\lim_{\omega \to 0}D_{vA}(+0,\omega) }{\lim_{\omega \to 0}D_{vv}(+0,\omega)},
\label{eq:slip length of HTD frequency expression}
\end{eqnarray}
where $D_{vA}(z,\omega)$ is defined by (\ref{eq:C23}) and $D_{vv}(z,\omega)$ is given by
\begin{eqnarray}
D_{vv}(z,\omega) = \int_0^{\infty} dt \langle \tilde{\mathcal{V}}^{x}(z,t) \tilde{\mathcal{V}}^{x}(z,0) \rangle e^{i\omega t}.
\end{eqnarray}

In Appendix~\ref{sec:Sketch of the derivation of (153) and (154)}, $\lim_{\omega \to 0}D_{vA}(+0,\omega)$ is calculated as
\begin{eqnarray}
\lim_{\omega \to 0}\frac{D_{vA}(+0,\omega)}{k_B T A_{xy}} = \frac{\mathcal{L}+\mathcal{B}_B}{\mathcal{L}+\mathcal{B}_A+\mathcal{B}_B},
\label{eq:DVA omega 0 dammy}
\end{eqnarray}
where we have ignored the terms proportional to $\delta(0)$. Then, we calculate $D_{vv}(z,\omega)$ in $\omega \to 0$. The leading contribution of the Green function in $\omega \to 0$ is expressed as
\begin{eqnarray}
G(z,z_2;\omega) \to \frac{(\mathcal{L}+\mathcal{B}_B-z)(\mathcal{B}_A+z_2)}{\eta(\mathcal{L}+\mathcal{B}_A+\mathcal{B}_B)}
\label{eq:G omega 0 ACp}
\end{eqnarray}
for $z>z_2$, and
\begin{eqnarray}
G(z,z_2;\omega) \to \frac{(\mathcal{L}+\mathcal{B}_B-z_2)(\mathcal{B}_A+z)}{\eta(\mathcal{L}+\mathcal{B}_A+\mathcal{B}_B)}
\label{eq:G omega 0 ACm}
\end{eqnarray}
for $z<z_2$. By recalling that $D_{vv}(z,\omega)$ is given by (\ref{eq:velocity-velocity correlation function}), we substitute (\ref{eq:G omega 0 ACp}) and (\ref{eq:G omega 0 ACm}) into (\ref{eq:velocity-velocity correlation function}). This yields
\begin{eqnarray}
\lim_{\omega \to 0}  \frac{D_{vv}(+0,\omega)}{k_B T A_{xy}} = \frac{\mathcal{B}_A(\mathcal{L}+\mathcal{B}_B)}{\eta(\mathcal{L}+\mathcal{B}_A+\mathcal{B}_B)},
\label{eq:DVV omega 0}
\end{eqnarray}
where we have ignored the terms proportional to $\delta(0)$. Substituting (\ref{eq:DVA omega 0 dammy}) and (\ref{eq:DVV omega 0}) into (\ref{eq:slip length of HTD frequency expression}) produces
\begin{eqnarray}
b^{HTD}_A = \mathcal{B}_A.
\end{eqnarray}
This is what we want to prove.

\bibliographystyle{spmpscinat_unsort}
\bibliography{references}

\end{document}